\documentclass[12pt]{article}
\usepackage{hyperref}
\usepackage{epsfig}
\usepackage{float}

\usepackage{caption}

\usepackage{amsmath}
\usepackage{amssymb}
\usepackage{graphicx}
\newlength{\abstractwidth}
\renewcommand{\thefootnote}{\fnsymbol{footnote}}
\renewcommand{\thanks}[1]{\footnote{#1}} % Use this for footnotes
\newcommand{\starttext}{
\setcounter{footnote}{0}
\renewcommand{\thefootnote}{\arabic{footnote}}}

\newcommand{\be}{\begin{equation}}
\newcommand{\bea}{\begin{eqnarray}}
\newcommand{\eea}{\end{eqnarray}}
\newcommand{\beq}{\begin{equation}}
\newcommand{\ee}{\end{equation}}

\def\eq{&=&}
\def\d{\partial}

\def\la{\langle}
\def\ra{\rangle}

\def\simleq{\; \raise0.3ex\hbox{$<$\kern-0.75em
\raise-1.1ex\hbox{$\sim$}}\; }
\def\simgeq{\; \raise0.3ex\hbox{$>$\kern-0.75em
\raise-1.1ex\hbox{$\sim$}}\; }

\def\bi{\begin{itemize}}
\def\ei{\end{itemize}}

\def\S{Schwarzschild}

\def\CA{{\cal{A}}}

\def\CC{{\cal{C}}}

\def\CQ{{\cal{Q}}}

\def\CA{{\cal{A}}}
\def\t{\tau}

\def\Tr{\bf Tr \it}

\def\bn{\bigskip \noindent}

\def\suk{SU(2^K)}
\def\Suk{$SU(2^K)$}

  \def\l{l_{ads}}
  \def\kl{k-local}
  \def\suk{SU(2^K)}

\makeatletter
\g@addto@macro\normalsize{%
  \setlength\abovedisplayskip{10pt}
  \setlength\belowdisplayskip{20pt}
  \setlength\abovedisplayshortskip{10pt}
  \setlength\belowdisplayshortskip{20pt}
}
\makeatother

\usepackage{color}

\begin{document}
  
\begin{titlepage}

\rightline{}
\bigskip
\bigskip\bigskip\bigskip\bigskip
\bigskip

\centerline{\Large \bf {Three Lectures on Complexity and Black Holes }}
\bn

\bigskip
\begin{center}
\bf   Leonard Susskind  \rm

\bigskip

 Stanford Institute for Theoretical Physics and Department of Physics, \\
Stanford University,
Stanford, CA 94305-4060, USA \\
\bigskip

%\vspace{1cm}
\end{center}

\begin{abstract}

This is the written version of three lectures on complexity and black holes
 given at PiTP 2018 summer program entitled "From Qubits to Spacetime." The first lecture describes  the meaning of quantum complexity, the analogy between entropy and complexity, and the second law of complexity.
 
 Lecture two reviews the connection between the second law of complexity and the interior of  black holes.  I discuss how firewalls are related to periods of non-increasing  complexity which typically only occur after an exponentially long time.
 
 The final lecture is about the thermodynamics of complexity, and ``uncomplexity" as a resource for doing computational work. I explain the remarkable power of ``one clean qubit," in both computational terms and in space-time terms.
 
 The lectures can also be found online at \url{https://static.ias.edu/pitp/2018/node/1796.html} .

\medskip
\noindent
\end{abstract}

\end{titlepage}

\starttext \baselineskip=17.63pt \setcounter{footnote}{0}
\tableofcontents

\part{Lecture I: Hilbert Space is Huge}

\section{Preface}

These lectures are a tale of two metrics on the same space---the space of states of a quantum system of $K$ qubits. One metric is familiar to you and the other probably very unfamiliar. They are extremely different measures of the distance between quantum states and they have different purposes.

The two metrics are totally dissimilar. One, the inner product metric, is small in the sense that no two points are more distant that $\pi/2.$ The other, relative complexity, is huge; almost all points are separated by a distance exponential in $K$. The inner product metric is positively curved like a sphere. The geometry of relative complexity is negatively curved, with the radius of curvature being much smaller than the maximum distance between points (the diameter of the space). Both are compact and homogeneous.

Although they both represent the distance between quantum states they are in a sense incommensurate: The distance between two points can be very small in one metric and exponentially large in the other. They represent very different relations between states.

Relative complexity should be of interest in quantum-computer science but has  not been of much use in the study of ordinary quantum systems, not even black holes if we are interested in the region outside the horizon. It is only when we ask about the interior of a black hole, the region behind the horizon, that relative complexity takes its place as an essential quantity.

You might ask why put so much effort into understanding what in practice  can never be seen, i.e., the black hole interior. The answer is two-fold: First Einstein's general theory of relativity predicts and describes the region behind the horizon and if we want to understand quantum gravity we have to take account of this fact. We will not have understood how the full geometry of GR emerges from quantum mechanics without following space into the black hole interior.

The second point is that in a de Sitter cosmology like ours everyone is behind someone else's horizon. We won't understand anything without understanding horizons.

\bn

These three lectures cover a certain aspect of complexity and black holes that I find most interesting, namely the relation to the second law of thermodynamics. Quantum complexity is not entropy, but there are definite parallels between the two. In fact complexity is the entropy of an ``auxiliary" classical system (called $\CA$) which has an exponentially large number of degrees of freedom. The system $\CA$  actually represents the motion of the state vector in Hilbert space. If the number of qubits describing the quantum system is $K$ then the number of classical degrees of freedom describing $\CA$
is $4^K$. The auxiliary system is chaotic has its own statistical thermodynamics, including a ``second law of complexity." This second law  plays the central role in the emergence of the spacetime behind the horizon, and is the main subject of these lectures.

\bn

Because of the limitations of time I have had to sacrifice two important subjects which I would like to have covered. The action of a Wheeler DeWitt patch is probably our best bet for  a precise geometric quantity to relate to complexity. I describe it very briefly in Lecture III.
On the pure quantum side our best bet for a definition of complexity may come from
Nielsen's approach, in which the geometry of Hilbert space is modified to a right-invariant Riemannian metric space.  I  mention the Nielsen approach in the conclusion but only briefly. Had I more time, these are the two things I would have included.

\bn

\section{How Huge?}
Let's consider the space of states of $K$ qubits and make a simple estimate of its size. By size I don't mean the dimensionality of the space defined by the number of mutually orthogonal vectors. I mean something more like the total number of unit vectors. It's of course infinite but we can regulate it.

The dimension of the Hilbert space is $2^K$ and a general vector has the form,

$$|\Psi\ra  = \sum_1^{2^K} \alpha_i |i\ra$$

The $\alpha$ are arbitrary complex numbers. At the moment we won't worry about normalizing $|\Psi\ra$ or dividing out the overall phase. Now let's regulate the infinities by restricting each 
$\alpha_i$ to be one of $m$ values. The total number of states is,

\be
\#states = m^{2^K} = \exp{(2^K \log{m})}
\label{states}
\ee
For $K=4, \ m=4$ the number of states is $4,294,967,296.$

The logarithm of the number of states is more manageable,
\be
\log{(\#states})=2^K \log{m} \approx 22.
\label{logstates}
\ee

There are two interesting things about \ref{logstates}. The first is how strongly it depends on $K,$ namely it grows as $2^K$. The second is how weakly it depends on the regulator parameter $m$. We'll see this trend many times  in what follows.

\section{Volume of CP(N)}

To do any kind of rigorous counting of points in a continuous space  we have to coarse-grain the space. An example is counting states in classical statistical mechanics where we coarse grain, replacing points by little balls of radius $\epsilon.$ This allows us to count states, for example to define entropy,
\be
S=\log{(\#states)}
\label{entropy}
\ee
 at the cost an additive term, $\Delta\log{\epsilon}$, $\Delta$ being the dimension of the phase space. In order to count states in $CP(2^K-1)$ or unitary operators in $\suk$ we have to do something similar.

 The space of normalized states with phase modded out is the projective space $CP{(2^K-1)}$.
  Let's calculate its volume using the usual Fubini-Study metric. The answer is not hard to guess: it is the volume of the sphere of the same dimension divided by $2\pi.$ The volume of $CP(N)$ is, 
 \be 
V[CP(N)] = \frac{\pi^N}{N!} \approx \left(  \frac{e \pi}{N}   \right)^N
\ee

As I said, in order to count states we need to regulate the geometry. The simplest way to do that is to replace points by small balls of radius $\epsilon.$
The volume of an epsilon-ball of dimension $2N$ is
\be 
V[B(2N)] =\frac{\pi^N}{N!} \epsilon^{2N} \approx  \left(  \frac{e \pi}{N}   \right)^N\epsilon^{2N}.
\ee

The obvious thing to do is to divide the volume of $CP(N) $ by the volume of an epsilon-ball. The result is very simple. The
number of an epsilon-balls in $CP(N)$ is  $\epsilon^{-2N}.$

Next we replace $N$ by $2^K-1$. We identify the number of states with
 the number of epsilon-balls in $CP(2^K).$ 
\be 
\#states =\left( \frac{1}{\epsilon}\right)^{2(2^K -1)}
\ee
or
\be 
\log{(\#states}) =  2(2^K-1) \log{(1/\epsilon})
\label{balls-in-CP}
\ee

For $K=4$ and $\epsilon = 1/2$ this gives $1,073,741,824$ $\epsilon$-balls.

Equation \ref{balls-in-CP} shows the same pattern as  \ref{logstates}. It grows rapidly with $K$ like $2^K$, and is only logarithmically sensitive to the cutoff parameter.

\section{Relative Complexity}

The enormity of the space of states of quantum systems begs the question of how to measure the distance between states. Are some states close to each other and others enormously far? The usual metric in state space (by which I mean $CP(N)$) is the inner-product metric. It's also called the Fubini-Study (FS) metric. The FS distance between two vectors $|A\ra$ and $|B\ra$ is,
\be 
d_{fs}(A,B) = \arccos{|\la A |B\ra|}.
\ee
It varies from $d_{fs}=0$ when $|A\ra = |B\ra$ to $\pi/2$ when $|A\ra$ and $ |B\ra$ are orthogonal. The space with the FS metric is ``small" in the sense that the furthest distance between points is $\pi/2.$

The FS metric has its purposes but it fails to capture something important. Imagine an intricate piece of machinery with a microscopic switch---a control qubit---that can be flipped on or off. Compare the distance between these two states, 
$$
|\bf{on}\ra     \ \ \   \ \ \   |\bf{off}\ra
$$

 with the distance between two other states: a brand new machine,  and  a pile of rust,
$$
|\bf{new}\ra        \ \ \  \ \ \   |\bf{rust \ pile}\ra
$$

  The FS distance between states $|\bf{new}\ra $ and $|\bf{rust \ pile}\ra$ is exactly the same as the distance between $|\bf{on}\ra $ and $|\bf{off}\ra $, namely $\pi/2$. If the only criterion was the FS distance it would be just as easy to:
%  $|\bf{rust \ pile}\ra $
%
\begin{enumerate}
\item Make a transition from $|\bf{new}\ra $ to 
$|\bf{rust \ pile}\ra $ as to make a transition from $|\bf{on}\ra $ to $|\bf{off}\ra $.
\item Create a coherent superposition of $|\bf{new}\ra $ and $|\bf{rust \ pile}\ra$, or a coherent superposition of $|\bf{on}\ra $ and $|\bf{off}\ra $.
\item Do a measurement that would be sensitive to the relative phases of such superpositions in the two cases.
\end{enumerate}

But of course this is not true. It is much easier  make a superposition of $|\bf{on}\ra $ and $|\bf{off}\ra $ (it just involves superposing two single-qubit states) than of $|\bf{new}\ra $ and $|\bf{rust \ pile}\ra $. Evidently the FS metric fails to capture the difference between these two cases.

If it were equally possible to physically apply \it any \rm operator to a system, then it would be no harder to superpose the machine and the rust pile, than the on-off states. But in the real world an operator that takes a machine to a rust pile is much more \it complex \rm than an operator that flips a qubit. In practice, and maybe in principle, we carry out very complex operations as sequences of simple operations. Once a set of  ``allowable" simple operations has been identified, a new measure of distance between states becomes possible. It can be defined as:

\bn
\it The minimal number of simple operations needed to go from 
$|A\ra$    to $|B \ra $.\rm  

\bn
In (quantum)information theory the simple operations are called gates and new distance measure is called relative complexity:

\bn \it

The relative complexity of $|A\ra$ and $|B\ra $  is the minimum number of gates required to go from $|A\ra$ to $|B\ra $.
\rm

\bn
One might object that relative complexity is a somewhat subjective idea since it depends on a choice of simple operations. But in a qubit context a simple operation is one that involves a small number of qubits. For example we might allow all one and two-qubit gates  as simple operations. Relative complexity would then be fairly well defined\footnote{There are still details associated with how precisely a product of gates must approximate a target state.} But then a skeptic may say: Yes, but what happens to your definition of complexity if I decide to allow three qubit gates?

I expect (but cannot prove) that the answer is: Not much happens at least for large complexity.  There are two kinds of ambiguities to worry about---additive and multiplicative. The additive ambiguities are associated with precision---how close do we have to get to the target to declare success? If we are required to get within inner-product distance $\epsilon,$ then there is an additive term in the complexity,  logarithmic in $\epsilon.$  It is closely related to the logarithmic ambiguity in  \ref{balls-in-CP}.

There is also a multiplicative ambiguity when we go from allowing only one and two qubit gates to allowing three qubit gates. That is because a three qubit gate can be approximated by some number of one and two qubit gates. That ambiguity  can be accounted for as long as the complexity is not too small. 

A basic assumption is that there a set of definitions of quantum complexity that exhibit a universal behavior up to additive and multiplicative ambiguities. 
The important rule is that the allowable gates must be restricted to be \kl \ with $k$ much smaller than the number of qubits comprising the system. What \kl \ means is that the gate involves no more than $k$ qubits.

We'll have more to say about relative complexity but let's first discuss the dual role of unitary operators.

\section{Dual Role of Unitaries}

Complexity can be defined for states\footnote{We continue to work with systems of $K$ qubits. The space of states is $2^K$ dimensional.}, but in many ways it is more natural to start with the complexity of operations that you can do to a system. That means the complexity of unitary operators. Unitary matrices have a dual role in quantum information theory. First of all they are the matrix representation of unitary operators,  
\be 
U = \sum U_{ij}|i\ra \la j|
\ee

Secondly they can be used to represent
maximally entangled states of $2K$ qubits,
\be 
|\Psi\ra = \sum U_{ij}|i\ra |\bar{j}\ra
\ee
\footnote{The bar in $\bar{J}$  indicates time reversal or in relativistic quantum field theory the CPT operation.}
In asking about the complexity of unitary operators we are also asking about the complexity of maximally entangled states\footnote{In the case of maximally entangled systems I mean a restricted form of complexity in which gates are only permitted to act on each side of the entangled system separately, thus preserving the maximal entanglement.},

The space of (special ) unitary operators is $\suk.$ It is much bigger than the space of states $CP(2^K-1)$. Let's begin with a crude estimate of its size.
A unitary matrix in $SU(N)$ has $(N^2-1)$ real parameters. If each parameter can take on $m$ values the number of unitary operators is
$$
\#unitaries = m^{(N^2-1)} 
$$
Setting $N= 2^K,$
\bea  
\#unitaries \eq m^{(4^K-1)} \cr \cr
\log{(\#unitaries)} &\approx& 4^K \log{m}
\label{log-Nunit}
\eea
This is the same pattern that we saw for states except that $2^K$ is replaced by $4^K.$

\section{Volume of $SU(2^K)$}

Let's make a more refined calculation of the number of operators in $SU(N)$ by dividing its volume by the volume of an epsilon ball of the same dimensionality (the dimension of $SU(N)$ is $N^2-1.$).
The volume of $SU(N)$ is \  \url{https://arxiv.org/abs/math-ph/0210033} 
\be 
V[SU(N)] = \frac{2\pi^{\frac{(N+2)(N-1)}{2}}}{1!2!3!....(N-1)!}
\ee
The volume of an epsilon-ball of dimension $N^2-1$ is
$$\frac{\pi^\frac{N^2-1}{2}}{  \left( \frac{     N^2-1}{2}  \right)!         }$$

\bn
Using Stirling's formula, and identifying the number of unitary operatos with the number  of epsilon-balls in $SU(N)$ 
\bea   
\#unitaries  &\approx& 
\left( \frac{N}{\epsilon^2} \right)^{\frac{N^2}{2}} \cr \cr
&=&  \left(  \frac{2^K}{\epsilon^2} \right)^{4^K/2}
\label{V-in-e-balls}
\eea
Taking the logarithm,
\be 
\log{(\#unitaries)} \approx  \frac{4^K}{2}K\log{2} + 4^K\log{\frac{1}{\epsilon}}
\label{log-balls}
\ee
which is comparable to \ref{log-Nunit}.
Again, we see the strong exponential dependence on $K$ and the weak logarithmic dependence on $\epsilon.$ The $\log{\frac{1}{\epsilon}}$ term  is multiplied by the dimension of the space.

\section{Exploring $\suk $}

We've seen that the space of unitary operators is  gigantic. Now I want to discuss how  to move through it. I've already hinted that we don't make big complexity jumps, but instead move in little steps called gates. A sequence of gates is called a circuit although it has nothing to do with periodicity. It's just a name.

\bn
\bf Definition: \rm

A \kl \ gate is a nontrivial k-qubit unitary operator chosen from some allowed universal gate set. We assume that if $g$ is in allowed set, so is $g^{\dag} $. 

\begin{figure}[H]
\begin{center}
\includegraphics[scale=.3]{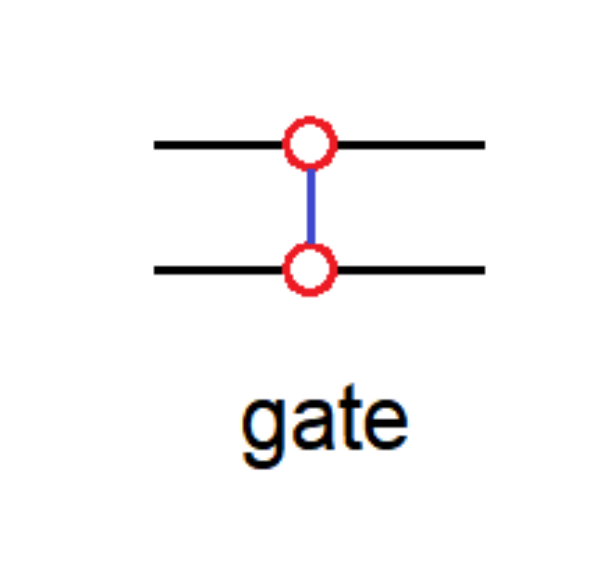}
\caption{Time flows from left to right. A gate acts on an incoming state of two-qubits to give an outgoing state.}
\label{gate}
\end{center}
\end{figure}

Figure \ref{gate} is a schematic representation of a 2-qubit gate.

\begin{figure}[H]
\begin{center}
\includegraphics[scale=.3]{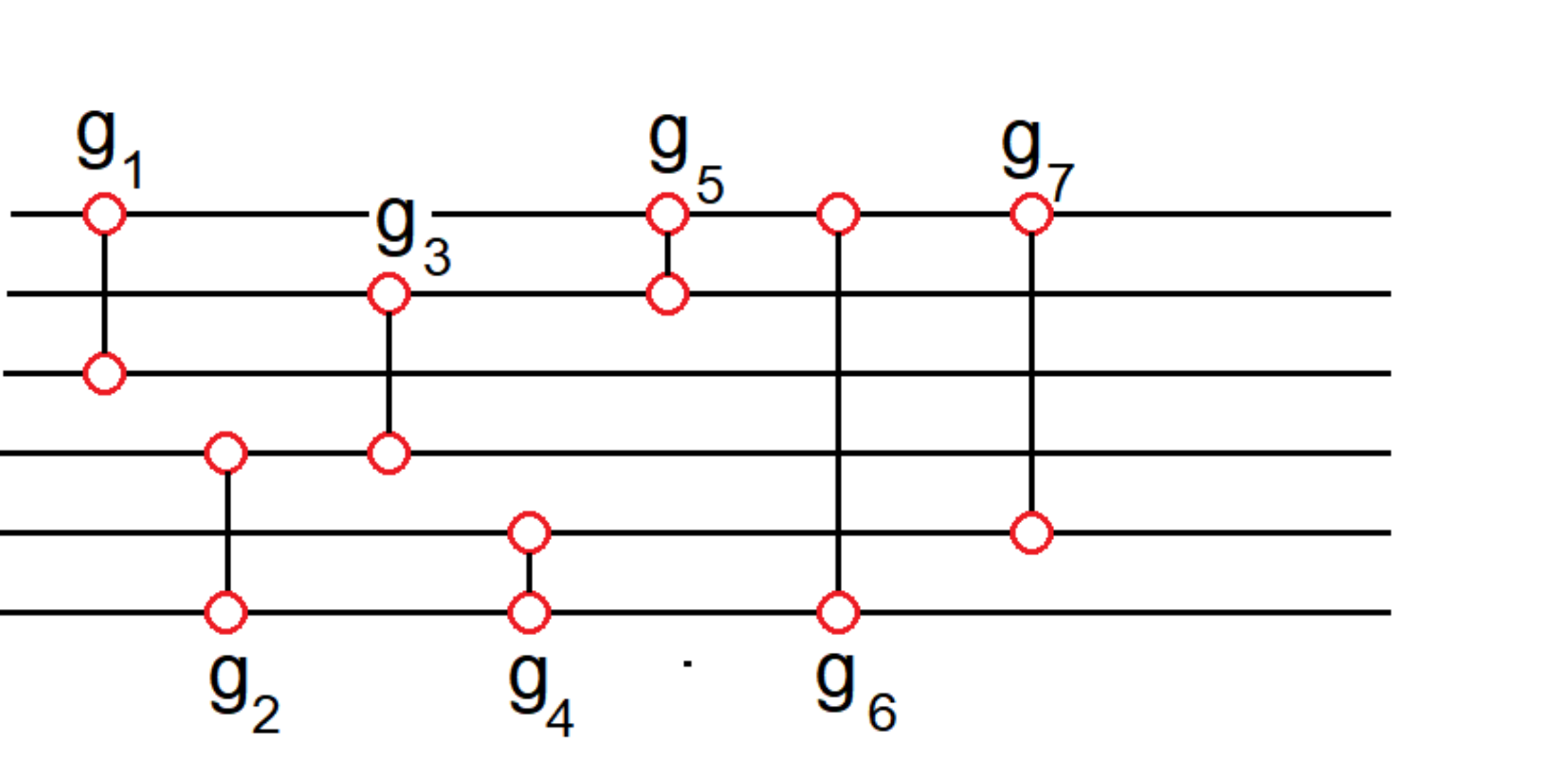}
\caption{k-local all-to-all circuit}
\label{circuit}
\end{center}
\end{figure}

\bn
Gates can be assembled into quantum circuits in the manner shown in figure \ref{circuit}.

\begin{figure}[H]
\begin{center}
\includegraphics[scale=.3]{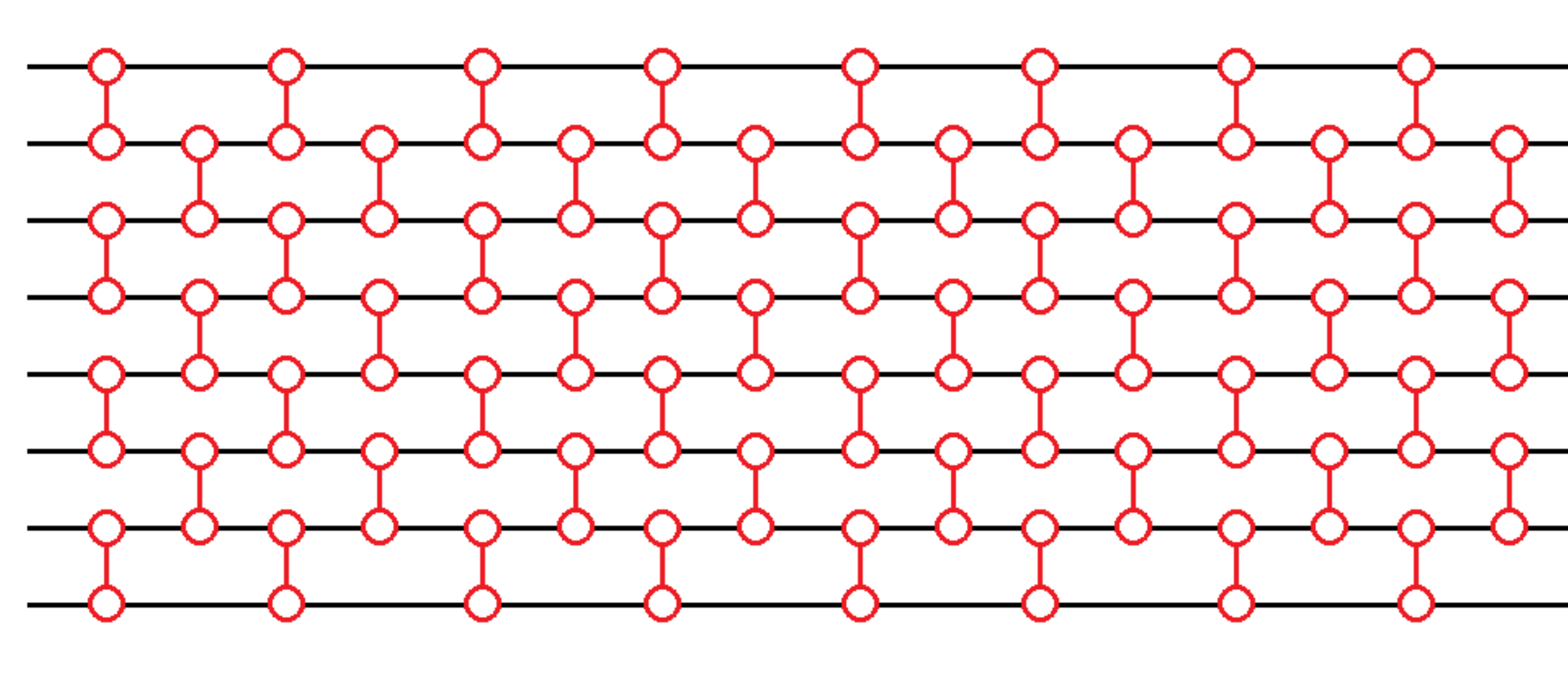}
\caption{Spatially local circuit}
\label{s-local}
\end{center}
\end{figure}

A \kl \ circuit is one made of \kl \ gates. k-locality is not the same as spatial locality although spatially local circuits are special cases of \kl \ circuits. In a spatially local circuit the qubits are arranged on a spatial lattice of some dimensionality. Gates are permitted only between near neigbors on the lattice. An example of a \kl  \ and spatially local circuit is shown in figure \ref{s-local}.
Such circuits would be valuable for simulating condensed matter systems but they are not the kind of circuits we will be interested in.

\bf Definition \rm
A \kl \ \it all-to-all\rm \ circuit is one which is \kl \ but permits any group of $k$ qubits to interact. Figure \ref{circuit} is 2-local \ and all-to-all.

\begin{figure}[H]
\begin{center}
\includegraphics[scale=.3]{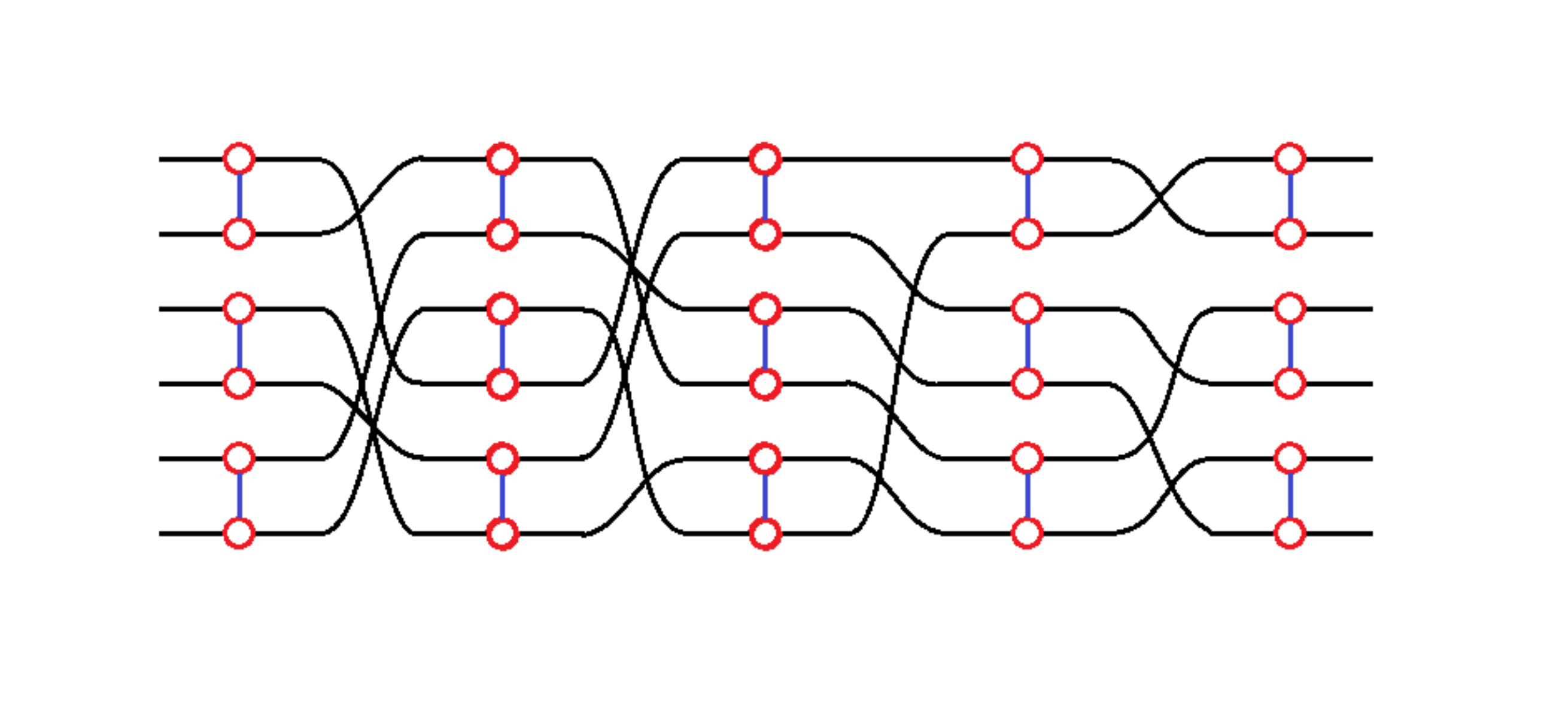}
\caption{Standard circuit architecture. Note that in each step the 
gates all commute because they act on non-overlapping qubits.}
\label{standard-circuit}
\end{center}
\end{figure}

In  figure \ref{circuit} the gates act in series, one gate at a time, but it is more efficient to allow parallel action of gates. The standard \kl \ all-to-all 
circuit architecture is shown in figure \ref{standard-circuit}.
The rule is that at each step the qubits are grouped into pairs and each pair interacts by means of a two-qubit gate.

We imagine that such a circuit is equipped with a clock, and with each tick of the clock $K/2$ gates act. We'll call this a ``step." The number of steps in a circuit is called its depth $D$. The depth of the circuit in figure \ref{standard-circuit} is $D=5.$  The number of gates in the circuit is $\frac{KD}{2}$.

Between steps the qubits may be permuted so that any pair can potentially interact. A circuit of this type can be called \kl \ (2-local in figure \ref{standard-circuit}) and all-to-all. The meaning of all-to-all is that any pair of qubits can potentially interact. Note that in a given time-step the different gates commute because they act on non-overlapping qubit pairs .

If we measure time $\tau$ in units of steps then the number of gates that act per unit time is $K/2.$

\bn

A given circuit prepares a particular unitary operator $U$.
Preparing $U$ by a series of steps can be viewed as a discrete motion of a fictitious classical particle, called the auxiliary system $\CA$ in Brown-Susskind
\url{https://arxiv.org/pdf/1701.01107.pdf}.

The auxiliary system represents the motion of $U(t)$ on $SU(N)$ as time unfolds.  The classical particle  starts at  the identity operator $I$ and ends at $U$.
\begin{figure}[H]
\begin{center}
\includegraphics[scale=.3]{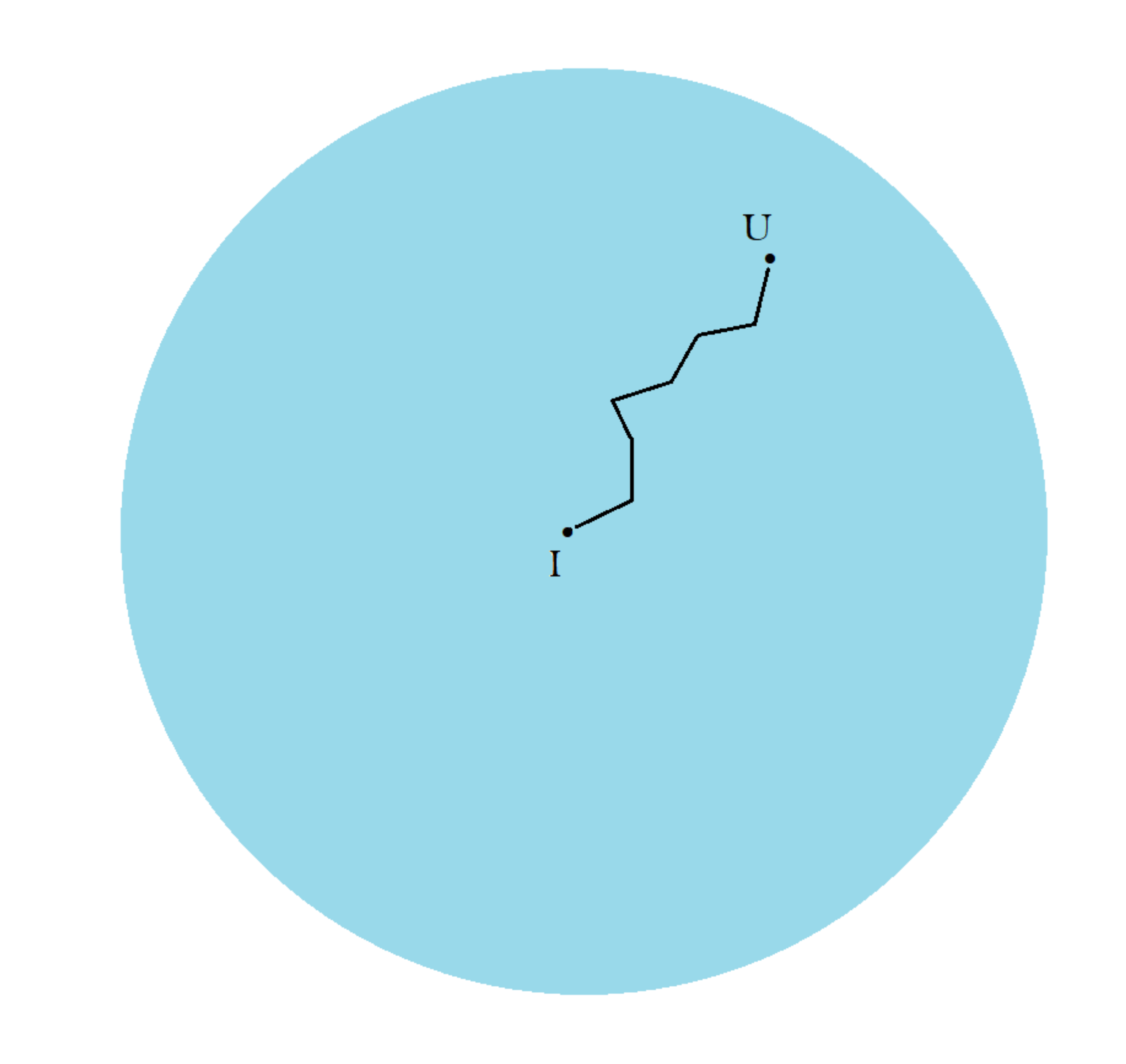}
\caption{The auxiliary system: The evolution of the unitary operator $U(t)$ can be thought of as the motion of a classical particle moving through the group space \Suk.  }
\label{motion}
\end{center}
\end{figure}

\subsection{Relative Complexity of Unitaries}

The standard inner product metric for unitaries is similar to the inner product metric for states. In fact if we think of unitaries as the wave functions of maximally entangled systems then it is the inner product for such states. The inner product distance between $U$ and $V$ is,
\be 
d(UV) = \arccos \   |\Tr U^{\dag}V|
\ee
where $\Tr$ means the normalized trace defined so that $\Tr I =1.$
For the same reasons that I discussed earlier, the inner product distance is not a useful measure of how difficult it is to go from $U$ to $V$ by small steps. A much better measure is relative complexity.\\

\bn
Given two  unitaries the relative complexity $\CC(U,V)$ is defined  as the minimum number of gates (in the allowed gate set)  satisfying,
\be 
U = g_n g_{n-1}....g_1 V
\label{uggggv}
\ee
to within tolerance $\epsilon.$ The relative complexity of $U$ and the identity may be defined to be the complexity of $U$.
\be 
\CC(U) \equiv \CC(U,I).
\ee

\begin{figure}[H]
\begin{center}
\includegraphics[scale=.3]{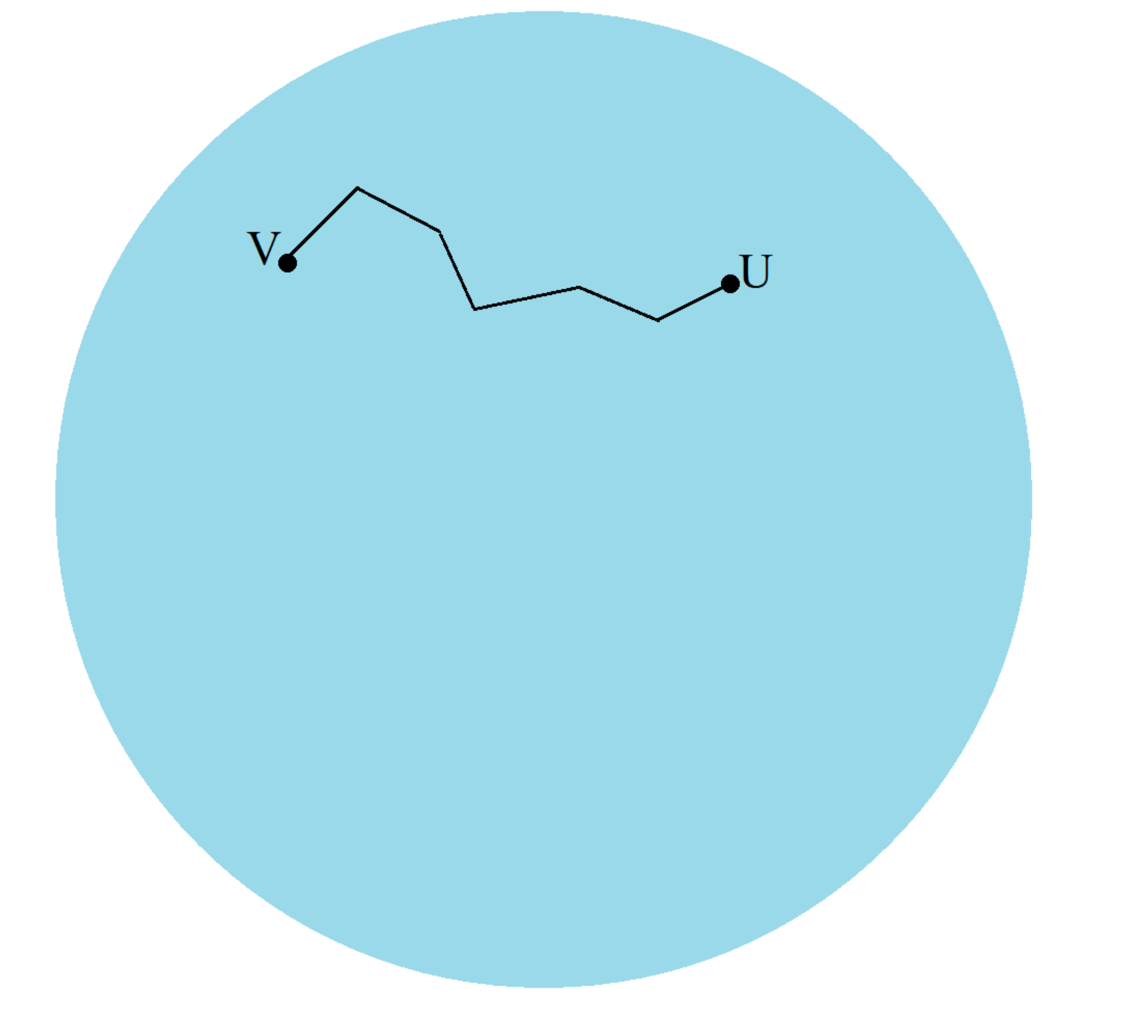}
\caption{Relative-C}
\label{relative-C}
\end{center}
\end{figure}

Figure \ref{relative-C} shows how the relative complexity can be thought of in terms of a discrete curve from $V$ to $U$.

\bn
Because the curve defining $\CC(U,V)$ is the shortest such path, we can think of it as a geodesic, but NOT a geodesic with respect to the inner product metric. It is a geodesic with respect to  relative complexity. The geometry defined by relative complexity is very different from the geometry defined by the inner product distance.

\bn
\bf Relative complexity is a metric \rm

\bn

Relative complexity satisfies the four defining conditions to be a metric.

\begin{enumerate}
\item $\CC \geq 0.$
\item $\CC(U,V)= 0 \ \rm iff \it \ U=V $
\item  $\CC(U,V) = \CC(V,U)$
\item  $\CC(U,V) \leq \CC(V,W) + \CC(W,V)$ \ \ \ \rm (Triangle \ Inequality)
\end{enumerate}

In fact it is a particular kind of metric called  \it right-invariant.\rm  

\bn
Suppose that 
\be 
U=(g_n g_{n-1}.......g_1) \  V
\label{U=ggggV2}
\ee
Then for any $W$ it follows that,
\be 
UW=( g_n g_{n-1}.......g_1) \  VW.
\label{U=ggggV3}
\ee
In other words the relative complexity of $U$ and $V$ is the same as that of $UW$ and $VW.$ 
\be 
\CC(U,V) = \CC(UW,VW) \ \ \ \ \ \ \rm (all  \it W)
\ee
This is what it means for $\CC$ to be  right-invariant.

\bn
On the other hand if we multiply from the left,
\be
WU=(W g_n g_{n-1}.......g_1W^{\dag}) \  WV.
\label{U=ggggV4}
\ee
But $(W g_n g_{n-1}.......g_1W^{\dag})$ 
generally is not a product of $n$ allowed gates. Therefore $\CC$ is not left-invariant\footnote{The usual inner-product metric is both left and right invariant. It is called bi-invariant.}.

%\bn
%Relative complexity and the inner-product metric define two %inequivalent  geometries on the same space, \Suk. 

\bn
It should be obvious from these remarks that quantum complexity is really a branch of geometry---right-invariant geometry\footnote{See Dowling, Nielsen  \url{https://arxiv.org/pdf/quant-ph/0701004.pdf} }

Most of the mathematical literature on the subject is about left-invariant geometry but of course this is just a matter of convention.

\subsection{Complexity is Discontinuous}
\bn
In describing relative complexity  I've  suppressed issues having to do with coarse-graining by epsilon balls. For example in \ref{uggggv}, \ref{U=ggggV2},  \ref{U=ggggV3}, and \ref{U=ggggV4}
I  ignored the phrase, \it to within a tolerance epsilon.\rm \ In fact as epsilon becomes small it takes an ever increasing number of gates to achieve that tolerance. The increase is only logarithmic in $\frac{1}{\epsilon}$ but nevertheless, with the present definition there is not a formal limit of relative complexity, or of the geometry of relative complexity.  One can only define a sequence of geometries as $\epsilon $ decreases.

The relation between the familiar inner-product metric and the relative complexity metric becomes wildly discontinuous as $\epsilon\to 0.$ For example two points can be close in complexity space, e.g., $|\bf {on}\ra$ and $|\bf {off}\ra$, and be maximally distant in the inner product metric. Similarly two states can be close in inner product and far in complexity. When this happens the two states have all expectation values close to one another, which means they yeild almost identical results for any experiment. Nevertheless 
 making a transition between them requires many gates.

It is an interesting question why a physicist would ever be concerned with  distinctions which have essentially no effect on expectation values. The answer is that if it were not for black holes, most likely no physicist would be.

\section{Graph Theory Perspective}

Circuits can be usefully described using graph theory\footnote{See also Henry Lin, ``Caley Graphs and Complexity Geometry",  \url{https://arxiv.org/pdf/1808.06620.pdf}.}. I'll call a graph  describing a circuit a \it circuit graph.\rm

Let's begin at the identity and act with a circuit of depth one, in other words a circuit of a single step with $K/2$ gates. For simplicity let's assume the allowed gate set is a single non-symmetric two-qubit gate. A choice must be made of how the $K$ qubits are paired. Each pairing will lead to a different unitary. I'll call the number of possible choices $d$ (an odd choice of notation but it corresponds with standard graph-theory notation). It's an easy combinatoric exercise to compute $d$,
\be 
d \sim \frac{K!}{\left(  K/2 \right)!}  \sim \left(  \frac{2K}{e}   \right)^{\frac{K}{2}}
\ee

Let's visualize this by a decision-tree embedded in $\suk$. At the central vertex is the identity operator. The branches (or edges) correspond to the $d$ choices. 

\begin{figure}[H]
\begin{center}
\includegraphics[scale=.3]{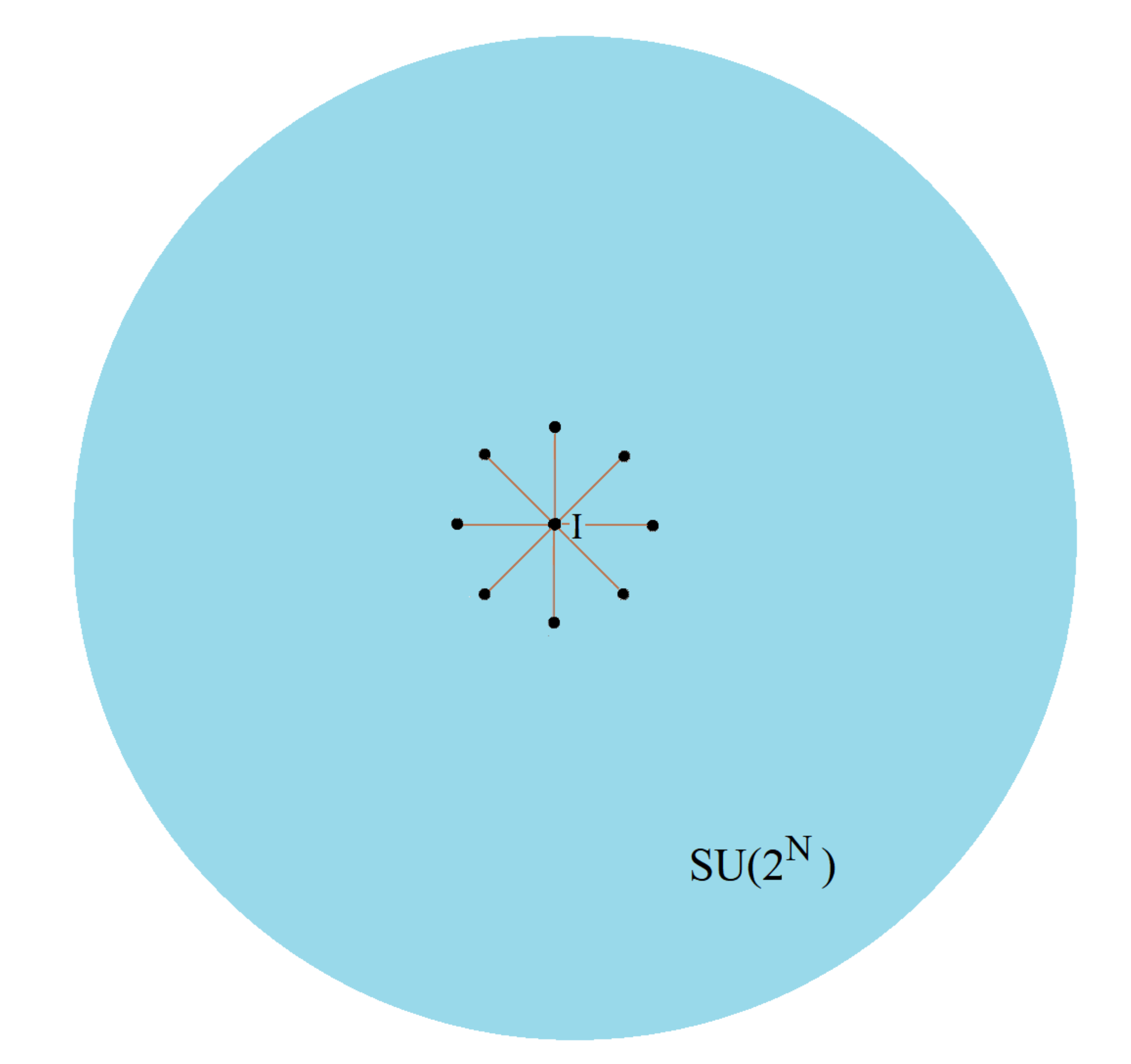}
\caption{Vertex with $d=8$.}
\label{spider}
\end{center}
\end{figure}
The number of branches or edges $d$ at a vertex is called the degree of that vertex. In our construction every internal vertex has degree $\frac{K!}{\left(  K/2 \right)!}$. A graph in which every vertex has degree $d$ is called d-regular. But so far the circuit graph is not d-regular because the outer boundary vertices have a single edge.
The leaves of the graph are unitary operators in $\suk.$

\bn

\begin{figure}[H]
\begin{center}
\includegraphics[scale=.3]{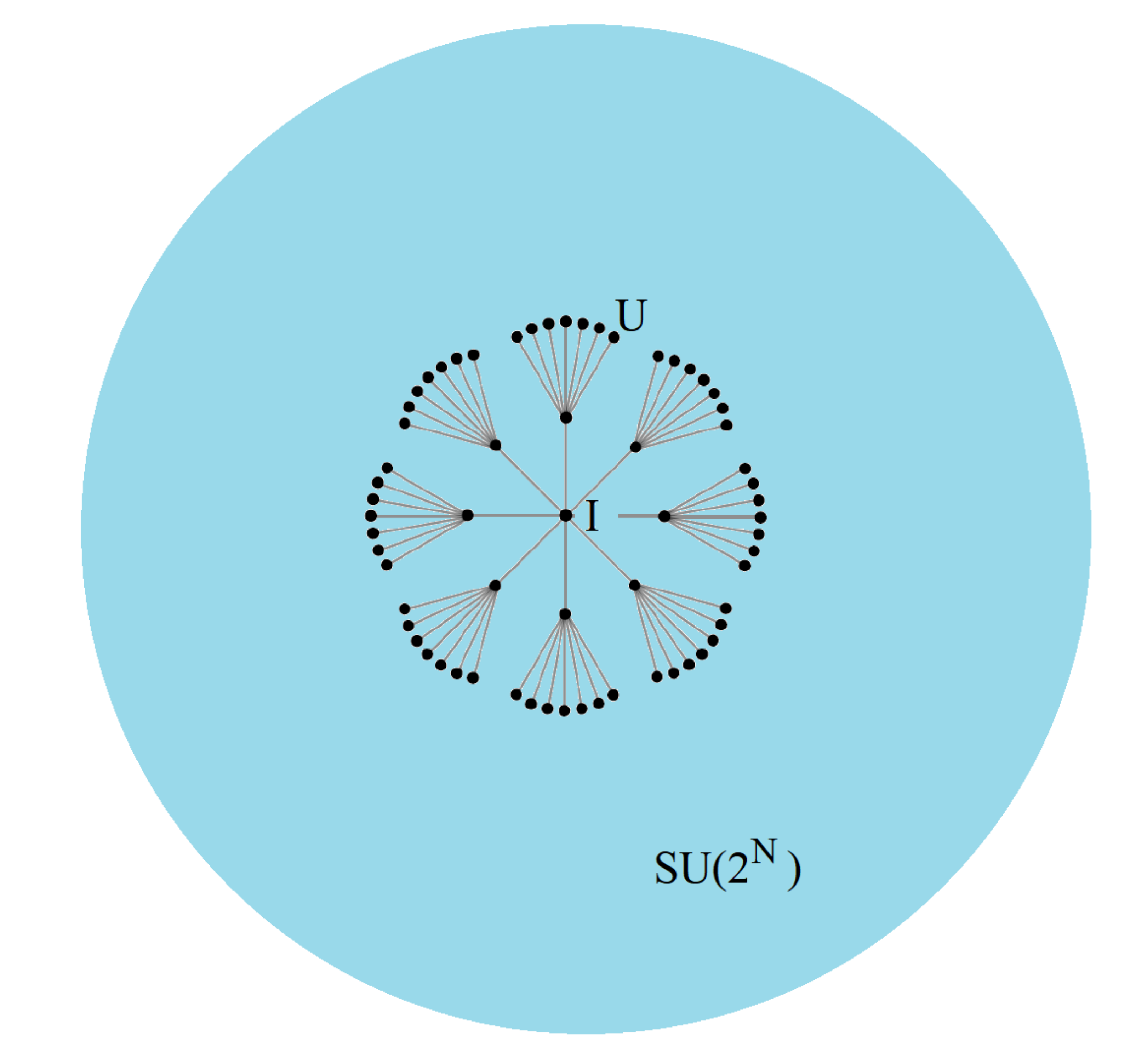}
\caption{Regular tree with degree 8, depth 2.}
\label{ssspider}
\end{center}
\end{figure}

Now let's grow the circuit depth by adding another step. I'm going to make one restriction of a technical nature. For large $K$ it makes no difference but it simplifies things. I will assume that when I add a step the choice of pairings is not the same as in the previous step. This  implies that the next layer of the tree has 
$$\frac{K!}{\left(  K/2 \right)!}-1$$
new branches. This is shown in figure \ref{ssspider}
Now comes an important assumption which underlies everything I will say. Each endpoint (leaf) of the tree represents a unitary operator. I will assume that the probability of two leaves being the same unitary is vanishingly small. Basically the reason is the extremely high dimension of the space of unitaries. If two leaves represent the same unitary to within $\epsilon$ I will say that a collision occurred. My provisional assumption is that collisions are so rare that they can be ignored.

This no-collision assumption  must eventually break down as the tree grows outward and we will calculate when that happens. But for now let's ignore collisions.

Assuming no-collisions, the number of unitaries that have been reached at depth $D$  is,
\be 
\#unitaries=d^D \approx \left(  \frac{2K}{e}   \right)^{\frac{DK}{2}}
\ee

The number of gates in a circuit of depth $D$ is $DK/2$. Assuming 
no-collisions, the path to each leaf is the minimal path, implying that the number of gates is the complexity. Thus we can write,

\bea 
\#unitaries \it \eq \left(  \frac{2K}{e}   \right)^{\CC} \cr \cr
 &\sim & e^{\CC \log{K}}
 \label{N=expC}
 \eea
Let me rephrase this formula in the following way:

\bn
\it The sub-volume of $\suk$ that corresponds to unitaries of complexity $\CC$ grows exponentially with $\CC.$ \rm

\bn
It says a number of things. First of all if we think of complexity as the distance from the origin (the unit operator ) then it says that the volume grows exponentially with radius---a characteristic of hyperbolic spaces of negative curvature. This may seem surprising: with the usual bi-invariant metric $\suk$ is positively curved. But we are talking about a different metric, relative complexity. This negative curvature is a symptom of chaos and in my opinion it, not fast scrambling, is the general signature of quantum chaos.

Equation \ref{N=expC} suggests something else. In classical statistical mechanics the volume of phase space associated with an ensemble of states of a given entropy is exponential in the entropy. The exponential growth of volume in $\suk$ associated with a given complexity is the basis for a deep analogy between complexity and entropy, including a \it Second Law of Complexity. \rm \  It can be summarized by the slogan:  Complexity is the entropy of the auxiliary system $\CA.$

\bn

\begin{figure}[H]
\begin{center}
\includegraphics[scale=.2]{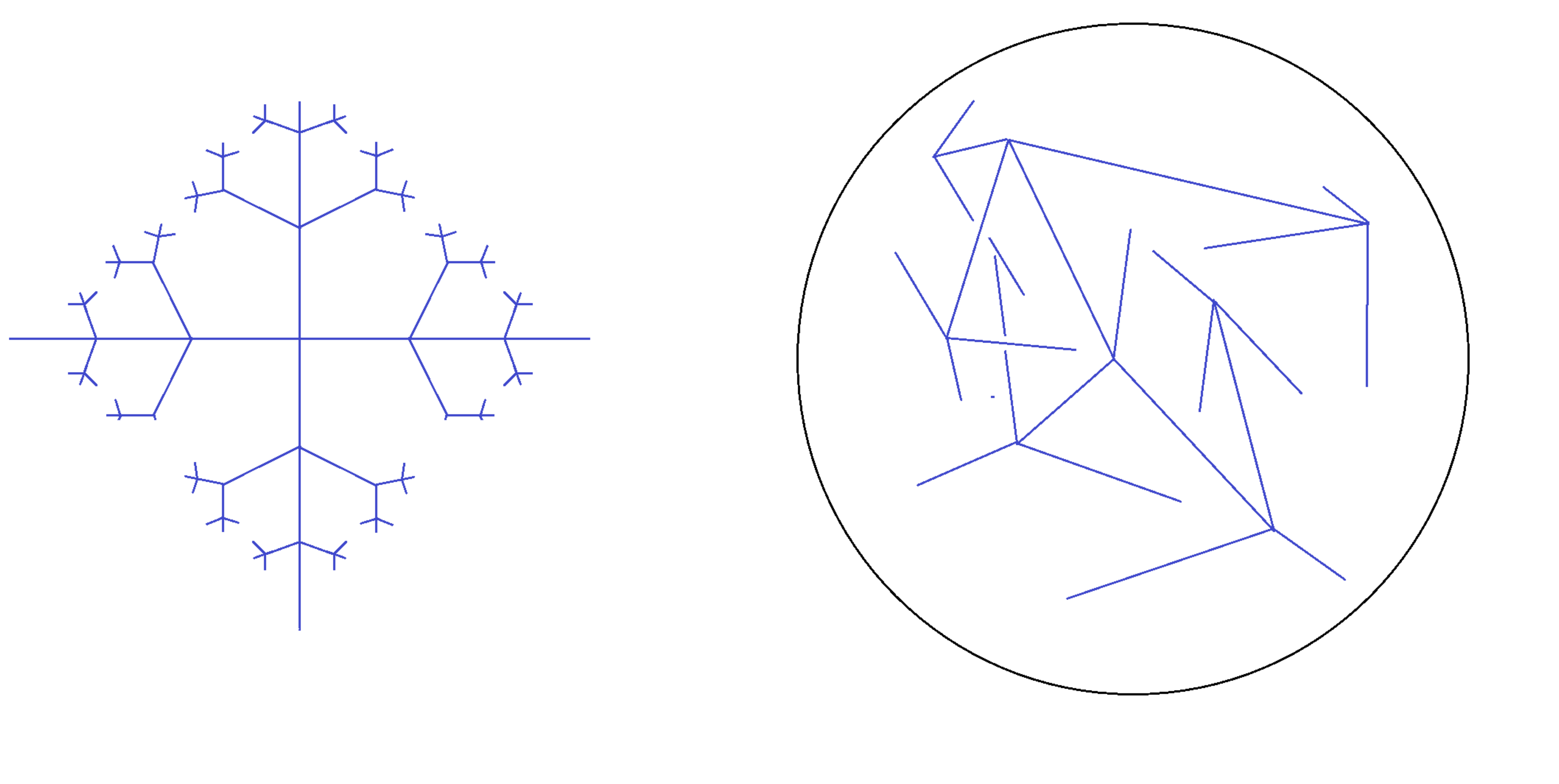}
\caption{Two views of the circuit tree. On the left is an abstract decision tree. The complexity of the unitary leaf grows linearly  with the number of steps from the identity at the center. Relative complexity is proportional to the number of steps between leaves. On the right the tree is embedded in \Suk \ with the usual inner product metric.}
\label{two-views}
\end{center}
\end{figure}

One more point about the circuit graph and its embedding in $SU(2^K)$. Figures \ref{spider} and \ref{ssspider} are very schematic. It is difficult to accurately convey the actual properties of the embedding with pictures given the fact that the space is extremely high dimensional. One thing to keep in mind is that with the usual bi-invariant metric, \Suk \ is small in the sense that the largest distance is $\pi/2.$ A single one-step circuit will move $U$ about that distance. So although the tree abstractly grows outward from the center as in the left panel of figure \ref{two-views}, the embedding in \Suk \ looks more like the right panel. 
The epsilon balls get filled in by the growing circuit in a very fractal manner which is related to the fact mentioned earlier that complexity is discontinuous as $\epsilon\to 0.$

\bn

\subsection{Collisions and Loops}

\begin{figure}[H]
\begin{center}
\includegraphics[scale=.4]{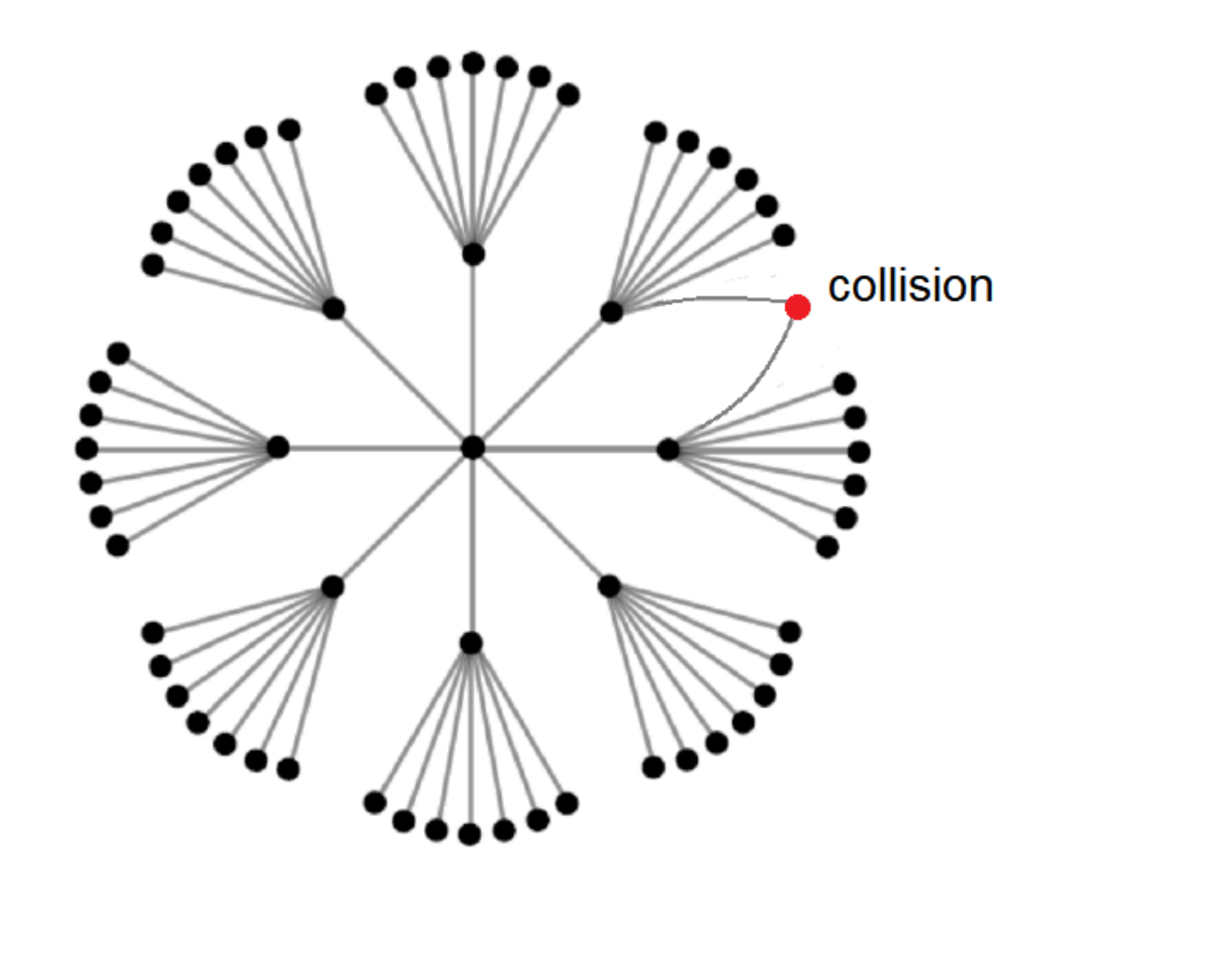}
\caption{Two paths collide on the same epsilon-ball.}
\label{collision}
\end{center}
\end{figure}

Each vertex of the tree represents a unitary operator and for that reason we can think of the tree as being embedded in $SU(2^K).$ 
Suppose a collision does occur. This means that two leaves of the tree are located at the same point is $\suk.$ We can represent this by drawing the two leaves as a single leaf as in figure \ref{collision}. 
The figure illustrates the fact that collisions induce loops in the graph. This allows us to translate the rarity of collisions into graph-theoretic terms. No collisions at all would mean no loops, i.e., exact tree-ness. But as we will see, eventually for very large depth, collisions must happen. This means that very large loops must occur. The correct formulation, which we will come back to, is that small loops must be absent or very rare. 

\bn

 Since the number of epsilon-regulated unitaries is finite we will eventually
 run out of room on $\suk$.  That happens when the number of leaves (as given by \ref{N=expC}) is equal to the total number of epsilon-balls as given in
  \ref{V-in-e-balls}. This determines the maximum possible complexity.

\be 
\left(  \frac{2K}{e}   \right)^{\CC_{max}} = \left(  \frac{2^K}{\epsilon^2} \right)^{4^K/2}
\ee
or

\be 
\CC_{max} = 4^K \left[\frac{1}{2} +\frac{|\log{\epsilon}|}{\log{K}}
\right]
\ee
Again, strong dependence on $K$, weak dependence on $\epsilon.$
\bn
Roughly  $$\CC_{max} \sim 4^K.$$

Apart from a factor of $K$ which is swamped by the exponential, this is also the depth at which collisions must occur. In other words it is the maximum radius at which the tree stops being tree-like. Finally it is also the largest distance  between nodes of the tree---the diameter of the graph.

We can now state the no-collision assumption more precisely. 

\bn

\it 

Loops smaller than $4^K$ are absent or very rare.\rm  

\bn
In graph-theoretic terms, the girth of the graph is $\sim 4^K.$

\bn
Another point follows from the fact that the total number of unitaries is $\sim   e^{4^K}$. We may identify this with the number of vertices in the graph. This implies that the diameter of the graph is logarithmic in the number of vertices.

\bn

\subsubsection*{The breakdown of no-collisions}
\begin{figure}[H]
\begin{center}
\includegraphics[scale=.3]{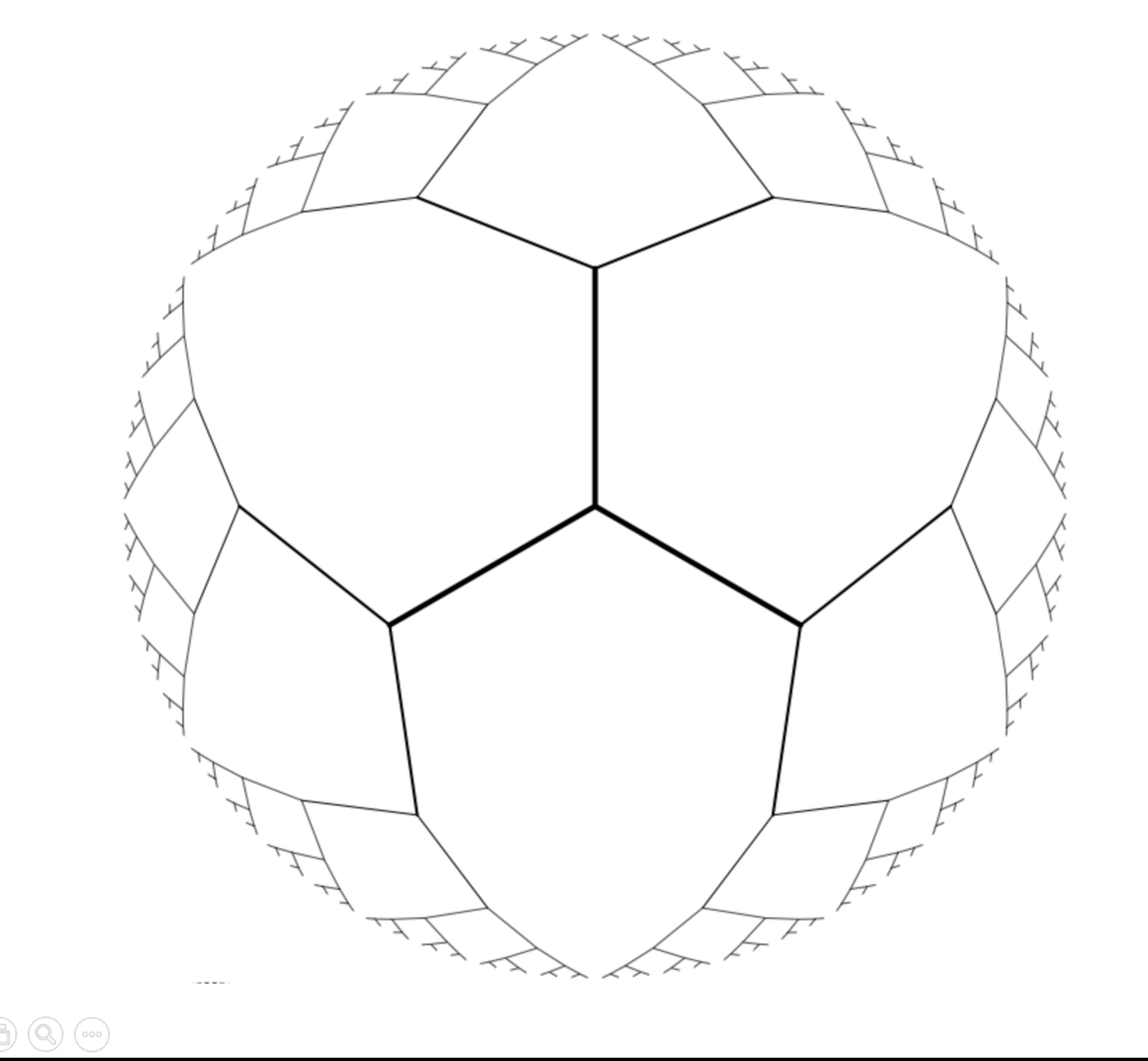}
\caption{Regular tree with degree 3}
\label{bethe}
\end{center}
\end{figure}

Let's consider what happens when the tree-ness breaks down.
Up to that point the graph is a d-regular tree similar to figure \ref{bethe} except with much higher degree.

\begin{figure}[H]
\begin{center}
\includegraphics[scale=.3]{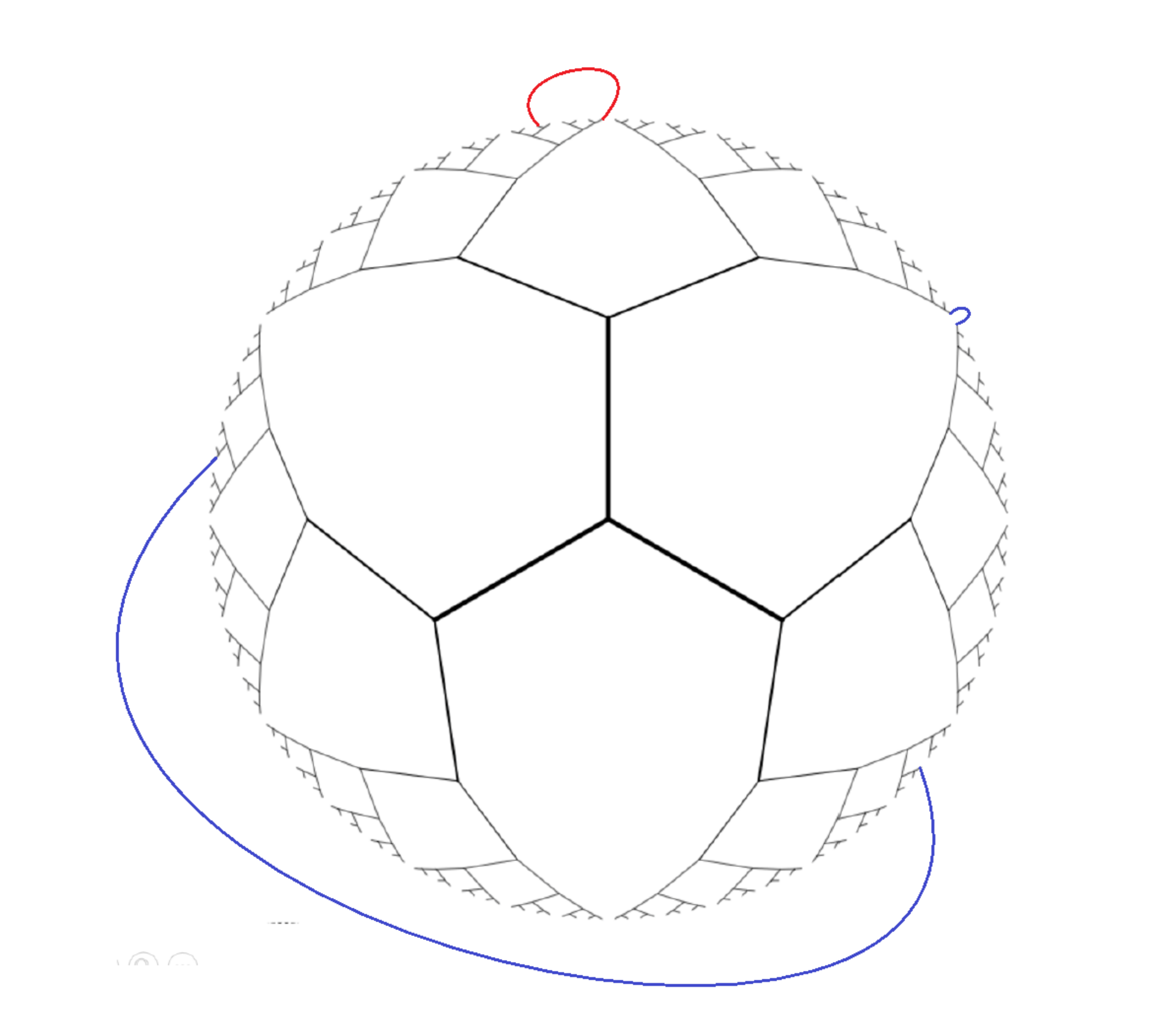}
\caption{Graph ceases being a tree and doubles back. Since a group-space is homogeneous the structure must look the same from every vertex. The shortest loops (girth of graph) is of order $4^K.$ Thus the red loop is too short. The loops must look more like the blue loop.}
\label{bethe-2}
\end{center}
\end{figure}

\bn
But once we reach $\CC_{max}$ the graph can not continue to grow. Collisions occur and loops must form. The graph must double back on itself and revisit previously visited epsilon-balls\footnote{If we follow an epsilon ball from the identity it will not in general perfectly coincide with an epsilon ball after executing a loop. There is a bit of sloppiness but it doesn't seem to be important. Note that if $\epsilon $  is decreased the maximum complexity increases and the graph becomes bigger.}. We show a couple of possible loops that might form in figure \ref{bethe-2}.
Now we know that any loop which passes through the central node must be very big, namely of length $\sim 4^K.$  But because the space $\suk$ is a group space, every point on the graph is the same as every other point. Thus it must be that loops passing through any point must be equally long. It is clear from the figure that loop containing the red segment is much shorter and should not occur, but the loop containing the blue segment is long and may occur.

\bn
Let me summarize the conjectured properties of circuit graphs generated by iterating one-step circuits: 

\begin{enumerate}
\item The degree is the same for all vertices and is given by
\be 
\rm degree \it = \frac{K!}{\left(  K/2 \right)!}  \sim \left(  \frac{2K}{e}   \right)^{\frac{K}{2}}.
\ee
This is much smaller than the number of vertices. The graph can be said to be sparse.
\item The number of vertices in the graph is of order $e^{4^K}.$
\item  The greatest distance between vertices (diameter)  is $\CC_{max} \sim 4^K$. The diameter is therefore logarithmic in the number of vertices.
\item Loops of length less than $4^K$ are rare or absent.
\item The graph is homogenous and from any point looks tree-like out to distances of order the diameter.
\end{enumerate}

These properties are very familiar to graph theorists. They are the properties of a good \it expander \rm graph.  Graphs of this type are discrete analogs of finiite negatively curved geometries such as the hyperbolic plane, with identifications that render it compact\footnote{The relation between complexity and such geometries was described in Brown, Susskind and Zhao, \url{https://arxiv.org/pdf/1608.02612.pdf}.}

\begin{figure}[H]
\begin{center}
\includegraphics[scale=.2]{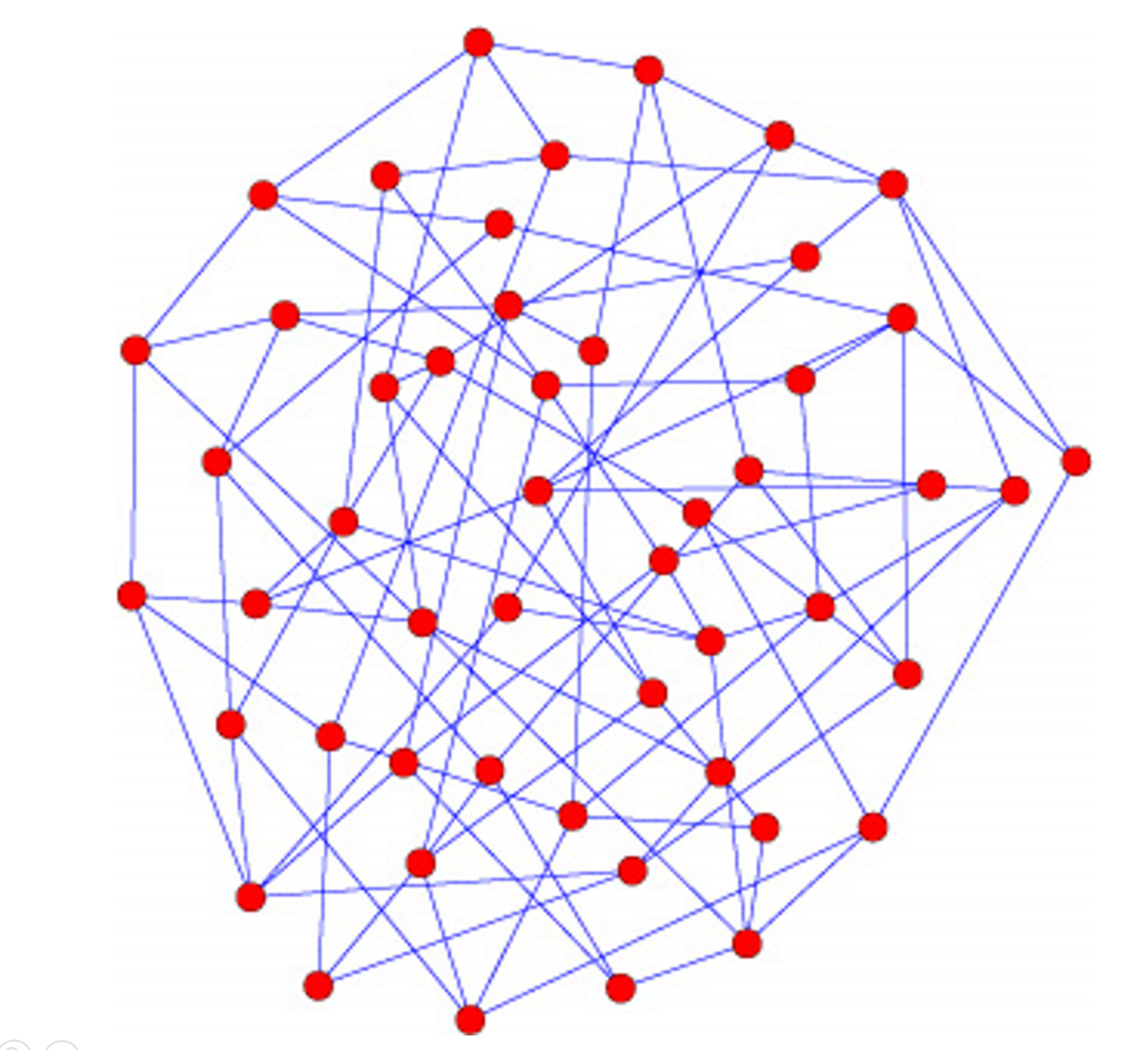}
\caption{50-vertex, (almost) regular $d=4$ expander. Generated by some AI optimization  protocol.} 
\label{expander}
\end{center}
\end{figure}
\begin{figure}[H]
\begin{center}
\includegraphics[scale=.2]{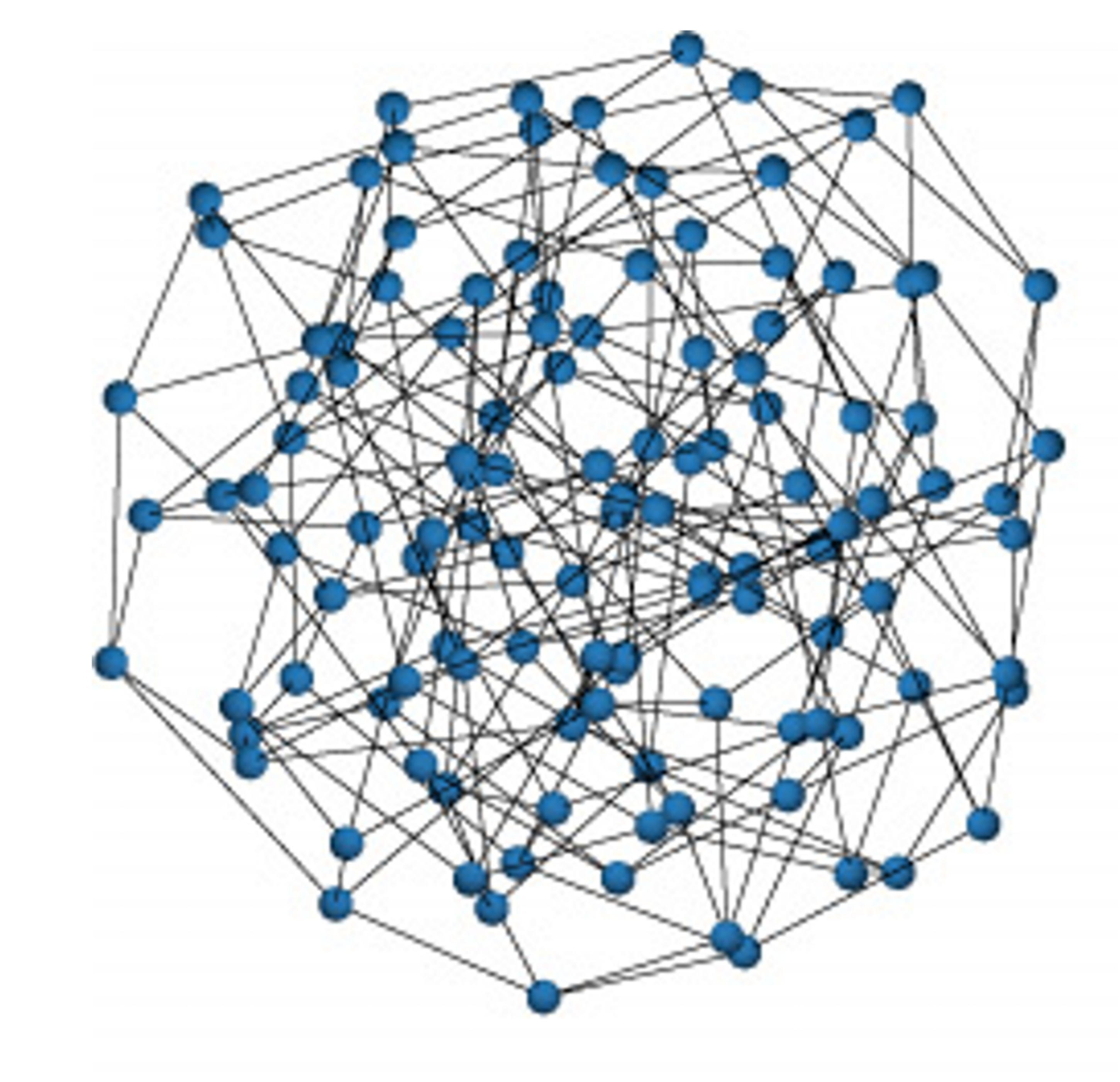}
\caption{120 vertex, regular  $d=4$ Ramanujan-expander. Quoting Donetti, et al.:
``Some remarks on its topological properties are in order. The average distance 3.714 is relatively small (i.e. one
reaches any of the 120 nodes starting from any arbitrary origin in less than four steps, on average."  Note: $\log{120}=4.78749174278$. Not too different from  3.714.  The authors then go on to say, ``The clustering coefficient vanishes,
reflecting the absence of short loops (triangles), and the minimum loop size is large, equal to 6, and identical for
all the nodes. In a nutshell: the network homogeneity is remarkable; all nodes look alike, forming rather intricate,
decentralized structure, with delta-peak distributed topological properties."}   
\label{expander2}
\end{center}
\end{figure}

\bn

\section{The Second Law of Quantum Complexity}

 I'm not sure why complexity theorists never remarked on the similarity of quantum complexity and classical entropy, or the existence of a second law of complexity\footnote{There is a very interesting paper by Zurek  in which he argues that that the classical entopy of a system is the ensemble average of the Kolmogorov algorithmic complexity of states in phase space. [W. Zurek,  Phys.\ Rev.\ A {\bf 40}, 4731 (1989)] The idea is very similar to the connection between quantum-computational-complexity and entropy described by Brown et.al in  \url{https://arxiv.org/pdf/1701.0110.pdf}}. 
 Just to be clear, although ensemble-averaged complexity is a kind of entropy it is NOT conventional entropy. The conventional entropy of a system of $K$ qubits is bounded by $K\log{2}$, the logarithm of the maximum number of mutually orthogonal vectors in the Hilbert space. By contrast the maximum  complexity is exponential in $K$.  Maximum complexity $\sim 4^K\log{1/\epsilon}$, the number of $\epsilon$-balls in $\suk.$ 

The quantity $4^K \log{}1/\epsilon$ 
does have an interpretation in terms of the entropy of a classical \it auxiliary \rm system 
$\CA$ associated with the quantum system of $K$ qubits. We may think of it as the maximum entropy of a classical system with $4^K$ classical degrees of freedom.  The auxiliary  system is just the classical $4^K$ collection of coordinates\footnote{Stictly speaking, $4^K-1$.} that describe the evolving time-evolution operator $U(t).$ Figure \ref{motion} illustrates the auxiliary system.  Quantities describing the auxiliary system will carry a subscript $\CA.$

For now let us consider a simplified version of the quantum evolution of a system at high temperature. We envision an
ensemble of fictitious particles moving on $\suk$. The particles are  random walkers which all start at the origin of $\suk$ (the identitiy operator) at some initial time. The dynamics is discrete: at each step the position is updated by applying a depth-one circuit. This means the particles execute random walks on the graph
that I just explained.

At each vertex the decision for the next step is made randomly. I will allow the possibility of back-steps along the previous edge. 
Initially the probability is concentrated at the origin and the entropy of the fictitious system is zero.

After one time-step the particle is at the first level of the tree on one of the leaves, as in figure \ref{spider}. The fictitious entropy is $\log{d}$ and the complexity is $\frac{K}{2}.$ In the next step the particle has a probability $\sim 1/d$ to back-track, but for large $d$ that is negligible. With probability close to one the particle moves outward to the next level where the complexity is 
$K.$ and the auxiliary  entropy $S_{\CA}$ is $\log{d^2} = 2\log{d}.$
\bn

After $n$ steps the particle, with high probability, will be at the $n^{th}$ level, the auxiliary entropy will be $S_{\CA} = n\log{d}$,  and the complexity will be
$nK/2.$
Evidently the complexity and fictitious entropy of the auxiliary system  are related,
\be 
S_{\CA} \approx  \CC  \  \log{K}
\label{CsimS}
\ee

This identification is dependent on the negative curvature and high dimensionality of complexity space. These two ingredients are what insure that collisions are rare and that we can identify the level of the tree with the minimum distance from the origin at $I$. In other words, up to a factor $K/2$ we may identify the depth with complexity.

Of course it is not rigorously true that there are no collisions. It's just that collisions are rare for sub-exponential time. We can be reasonably sure that almost all leaves  (vertices) have complexity proportional to their level, but it is much harder to know that a given leaf has not had collisions in the  past. It's for this reason that we identify the entropy $S_{\CA}$ with the ensemble averaged complexity.

The second law of complexity is just the second law of thermodynamics---the overwhelming statistical likelihood that entropy will increase---applied to the ensemble average of complexity. The reason why complexity almost always increases when it is less than maximum is the same as why classical entropy  almost always increases when it is less than maximum---the number of states exponentially increases with increasing entropy/complexity.

Let us follow a particular member of the ensemble. As long as it is not an exponential number of steps from $I$ the complexity will simply reflect the exponentially growing number of states as we mover outward from the origin. It will with very high probability increase linearly with time. 
However once $\CC \sim 4^K$ the particle will have reached the maximum distance on the graph and the complexity will stop increasing. Complexity equilibrium will have been achieved. The number of states with  maximum complexity is so vast, that the particles will get lost among them and remain at maximum complexity for a recurrence time. The recurrence time for the classical system $\CA$  will be $t_{recur} = \exp{S_{\CA}}$ which is doubly exponential in $K,$.
\be
t_{recur} \sim e^{4^K}
\ee

Thus we expect a singly  exponential time $\sim 4^K$ during which complexity linearly grows, after which it remains approximately constant at its maximum. But then, on gigantically long time scales $\sim \exp\exp K$ it will recur to small values, and $U(t)$ will return to the neighborhood of the identity.

\begin{figure}[H]
\begin{center}
\includegraphics[scale=.3]{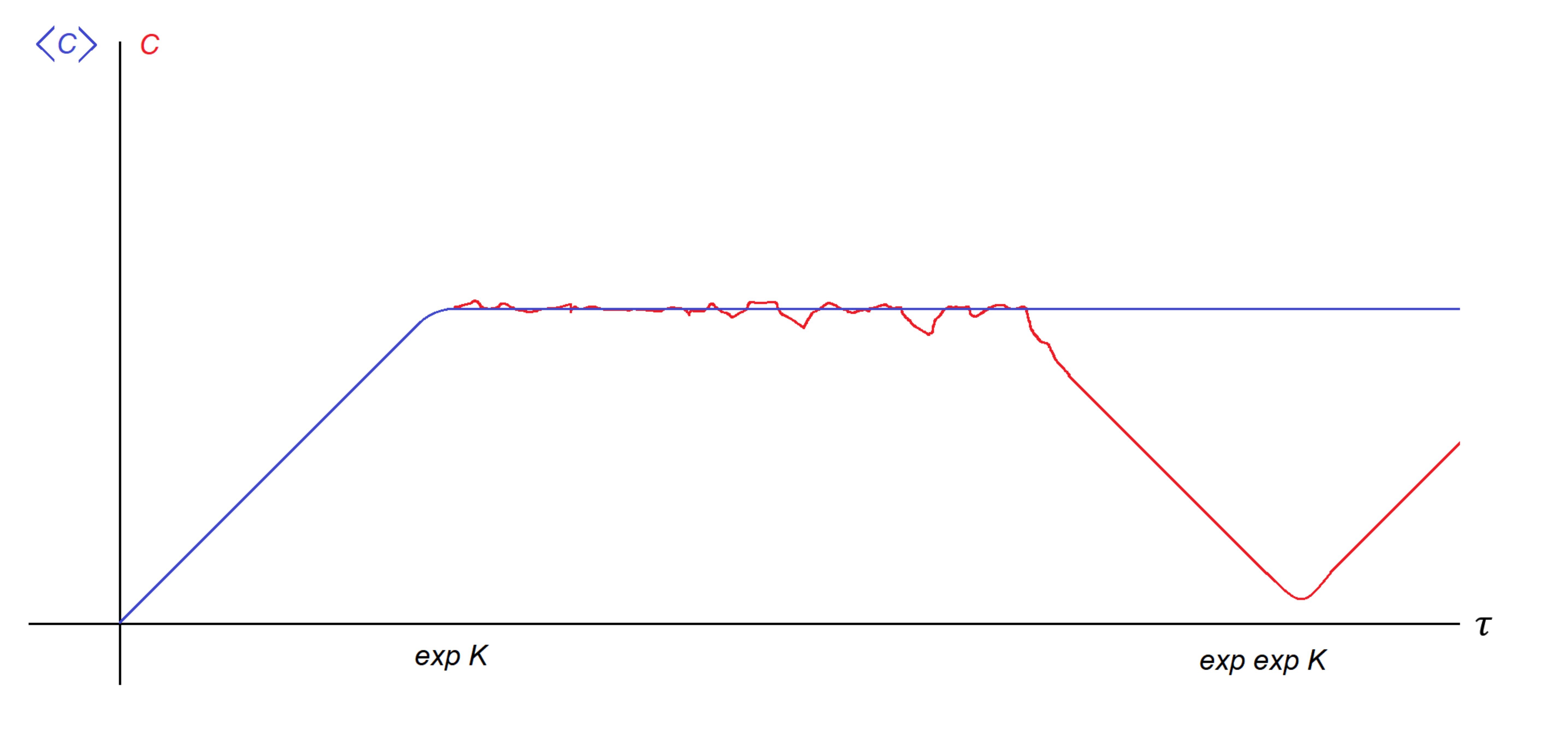}
\caption{Evolution of complexity with time. The ragged red curve is the evolution for a specific instance of an ensemble. The smooth curve is the ensemble average.}
\label{2nd-law}
\end{center}
\end{figure}

The transition from linear growth to complexity equilibrium is very sudden.

\begin{figure}[H]
\begin{center}
\includegraphics[scale=.3]{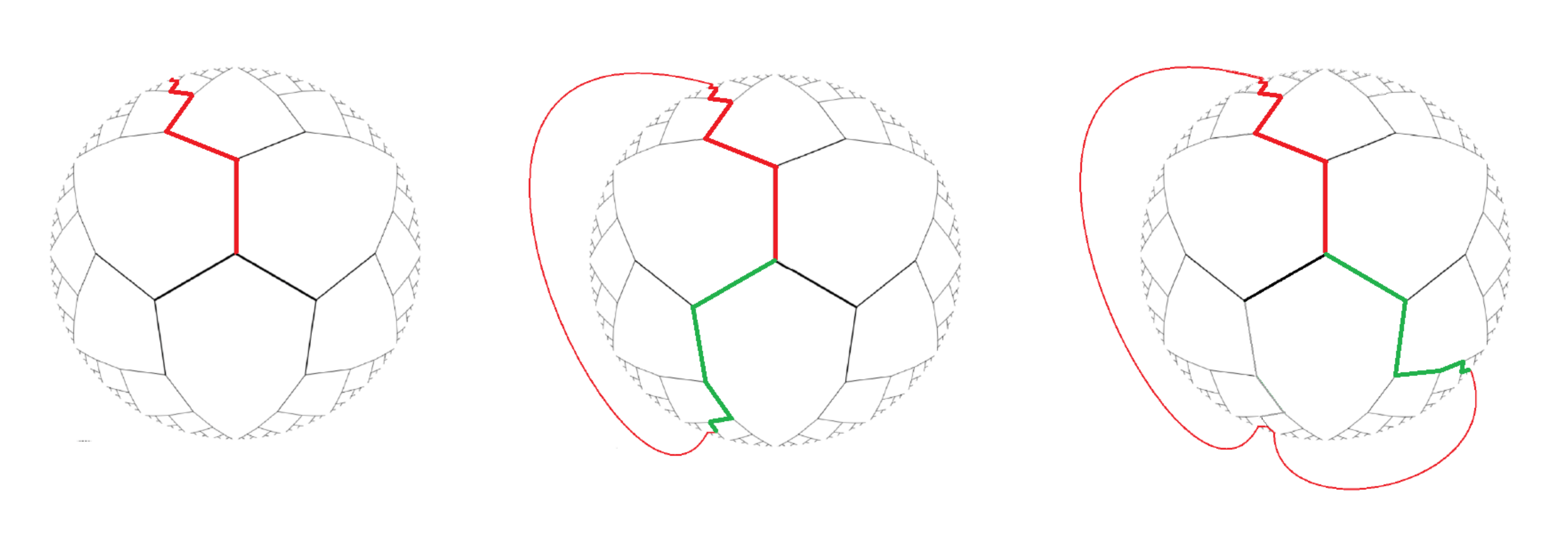}
\caption{The left panel shows a (red) trajectory beginning at the identity and proceeding outward along toward greater complexity. It eventually reaches maximum complexity. The middle panel follows the evolution as it doubles back to a slightly less-than-maximal complexity. At this point the minimal path jumps to the green path. In the third panel the trajectory continues to maximal complexity and then jumps again. }
\label{cut}
\end{center}
\end{figure}
\bn
It is interesting to see how this behavior can be understood in terms of circuit graphs. We can get an idea by looking at figure \ref{cut}. The figure shows the evolution of the $\CA$ auxiliary system moving according to some ``dynamical" rule that mocks up Hamiltonian evolution starting with $U(0) = I.$ The trajectory moves outward and until collisions occur the complexity increases linearly with time. The complexity is proportional to the graph distance from $I$. That's shown as  the red trajectory in the left panel.

Once the trajectory reaches maximum complexity it keeps going, but it has no choice but to visit previously visited sites. With overwhelming probability it will jump to another almost maximally complex state. That is shown in the middle panel. 

But now, there is a shorter path to the end of the red trajectory. It is shown in green. The complexity is \bf NOT \rm the length of the red trajectory but rather the length of the green trajectory. Thus the complexity does not increase and may even decrease a bit.

From there the dynamical red trajectory continues. With overwhelming likelihood it moves outward because the overwhelming number of branches reach outward. It soon reaches the next point where it has to loop around. A new green trajectory forms that is quite different than the previous one, but also has close-to-maximal complexity. That is the reason why the top of the curve in figure \ref{2nd-law} is a bit ragged.

Eventually the auxiliary particle will find its way back to low complexity and the cycle will repeat but this takes a quantum-recurrence time, $\exp \exp K$.

Note that the transition from linearly increasing complexity to equilibrium at the top of the curve is sudden. It's similar to a first order phase transition where another local minimum crosses over and becomes the global minimum.

\subsection{Hamiltonian Evolution}

In the live lectures at PiTP I ran out of time before I had a chance to discuss how complexity evolves when a system's dynamics is governed by a time independent \kl \ Hamiltonian. In that case the constraint of energy conservation restricts the motion of $U(t)$ to lie on a relatively low dimensional subspace of \Suk. Let's write $U(t)$ in the form,
\bea
U(t) \eq e^{\-iHt}  \cr \cr
\eq \sum_{i=1}^{2^K} |E_i\ra \la E_i| e^{-iE_i t}
\eea
For a given Hamiltonian the motion is restricted to a torus defined by the set of $2^K$ phases,
\be 
e^{i\theta_i} = e^{-iE_i t}.
\ee
In other words $U(t)$ moves on a $2^K$-dimensional torus embedded in the $(4^K-1)$-dimensional group \Suk. Although in itself a very large space the torus can only cover a tiny fraction of the full \Suk \ space. 

The first question is whether $U(t)$ fills the torus? Typically the answer is yes. If the energy levels are incommensurate, as will be the case if the system is chaotic, then the motion will be ergodic on the torus. The recurrence time for $U$ to return to the neighborhood of the identity  is the time for all the phases to get simultaneously close to $1$, and that takes time,
\be 
t_{recur}\sim \exp{2^K}.
\ee
This is in contrast to the recurrence time $\sim \exp{4^K}$  random (Brownian) circuits which fill the entire $4^K$-dimensional unitary group. Therefore it is not possible for $U(t)$ to visit any but an infinitesimal  fraction of the  maximally complex unitaries with $\CC\sim 4^K.$ 

On the other hand the unitaries of complexity $\leq 2^K$ comprise a set with volume comparable to the torus. It is therefore likely that for a Hamiltonian system the growth of complexity in figure \ref{2nd-law}  reaches $2^K$.

\bn

This raises an interesting question: In equation \ref{N=expC} I estimated the volume of the portion of \Suk \ with complexity $\CC$ and got,
$$e^{\CC \log{K}}.$$  That was the basis for the connection between complexity and the entropy of the auxiliary system, \ref{CsimS},
$$
S_{\CA} \approx \CC \log{K}.
$$

A similar question can be formulated for motion restricted to the torus: What is the volume of the torus occupied by unitaries of complexity $\CC?$ I expect that the answer is exponential,
$e^{ \alpha \CC}$,
 but with $\alpha$ parametrically smaller than $ \log {K}$. My guess is that $\alpha$ is independent of $K.$ In that case \ref{CsimS} would be replaced by,
 \be 
 S_{\CA} \approx \CC.
 \ee

\bn

This concludes lecture I. In the next lecture I will explain what   complexity and its second law has to do with black holes.

%%%%%%%%%%%%%%%%%%%%%%%%%%%%%%%%%%%%%%%%%%%%%%%%%%%%%%%%%%%%%%

%%%%%%%%%%%%%%%%%%%%%%%%%%%%%%%%%%%%%%%%%%%%%%%%%%%%%%%%%%%%%

%\begin{titlepage}

\rightline{}
\bigskip
\bigskip\bigskip\bigskip\bigskip
\bigskip

\part{Lecture II: Black Holes and the Second Law of Complexity}

\section{Preface}
When a star collapses a horizon forms and its area grows until it reaches its final value.  This is visible from outside the horizon and is known to be a manifestation of a statisitcal law, the second law of thermodynamics: Entropy increases until thermal equilibrium is reached.

There is another similar but less well-known phenomenon: the growth
of the spatial  volume behind the horizon of the black hole. 
At first sight this could  be another example of the increase of entropy, but more careful thought shows that this is not so. The growth of the interior continues long past the time when the black hole has come to thermal equilibrium. Something else---not entropy---increases. What that something else is should have been one of the deepest mysteries of black hole physics---if anyone had ever thought to ask about it.

Complexity is of course that something. Although complexity is most emphatically not ordinary entropy, the theory of complexity has a very strong thermodynamic flavor, as I began to explain in the last lecture. In this lecture we will see how another statistical law, ``The Second Law of Complexity" controls physics  behind the horizon in a manner very similar to the way the second law of thermodynamics controls the the dynamics of the horizon as seen from outside the black hole. 

\section{The Black Hole-Quantum Circuit Correspondence}

Complexity is mostly studied in computer science departments.  The language is that of \it bits,  qubits, \rm  and logical \it gates\rm, not Hamiltonians and quantum fields.
At the present time there is no completely consistent framework in which a black hole can be represented as a finite collection of qubits, the kind of system for which complexity is a reasonably well formulated concept.
I will not try to fill the gap in this lecture. What we will do is to to exhibit a \it pattern of similarities \rm between black hole phenomena and corresponding phenomena that occur in \kl \ all-to-all quantum circuits. The correspondences include:

\bi  
\item \ \ \ \ Black hole of entropy $S$  \ \ \ \ \ \  ------ \ \ \ \ Quantum circuit with $K\sim S$ qubits.
\item \ \ \ \  \ \ \ \ \ \ \ \ \kl \ gates  \ \ \ \ \ \ \ \ \ \ \ \ \  ------ \ \ \ \ \ \ \kl \ Hamiltonians

\item \ \ \ \ \ \ \ \ \ \ \ \ Clock time $\tau$   \ \ \ \ \ \ \ \ \ \ \ \ \  ------ \ \ \ \ \  Rindler time $\tau$
\ei

\bn
It is possible that all quantum systems can be described as systems of qubits, but often the description would involve many more qubits than are necessary to represent the phenomena at hand. A good rule of thumb is that in order to describe a system with entropy $S$ it is sufficient to use a number of qubits of order $S$. In that case the qubit system should be studied at temperatures high enough that every degree of freedom carries of order one unit of entropy.

To keep the discussion simple and concrete I will generally set $k=2$ but things are easily generalized to arbitrary $k$ as long as $k<<K.$

\subsection{Two Problems}
Let us consider two problems:

\begin{enumerate}
\item 
One evening before going to sleep Alice decided to record a single bit $x$ in  her quantum computer, so she put the computer into the state $|\Psi_0\ra = |x00...000\ra$ with $x$ either equal to $0$ or $1.$ Her intention was to come back the next morning and to retrieve $x$. But unfortunately she accidentally left the 
 computer running  all night, gate by gate executing some useless computation. By morning, because of the strong tendency for complexity to increase, the quantum computer had arrived at a  fairly  complex state $|\Psi \ra.$ Alice will need to do something pretty complicated if she wants to recover the value of $x$. One thing she can do is to run the computer in reverse for the same length of time as it ran overnight. 
 
 But it is possible that there is a shorter sequence of gates that will bring the computer back to the state $|\Psi_0\ra = |x00...000\ra$. The question is: What is the minimal number of gates that can accomplish this task? As you may guess, the answer is the relative complexity of $ |\Psi_0 \ra$ and $ |\Psi \ra$.

\item  The second problem involves a black hole and two observers, Alice and Bob. When the black hole was formed Bob accidently got caught behind the Horizon in a kind of one-sided wormhole\footnote{A ``bridge to nowhere," Susskind-Zhao 2014. . }. With (external) time, the one-sided wormhole grew, so that Bob receded further and further from the horizon. Alice knows how to extract Bob from behind the horizon, but the increased separation due to the wormhole growth makes it difficult.  Therefore Alice would like to manipulate the black hole, from the outside, so as to shorten the wormhole as much as possible before attempting the extraction. How hard is it to do this and how long must it take? The answer will depend on how much the interior of the black hole grew between   the time that Bob fell in and the time that Alice decides to extract Bob.

\end{enumerate}

These two problems seem unconnected but in fact they are the same. The basis for this claim is the duality between the growth of black hole interiors as governed by GR, and the growth of quantum complexity---the second law of complexity---that I described at the end of Lecture I.

\subsection{Circuits and Black Holes}

The properties of black holes suggest that a description in terms of qubits should have a particular form. First, how many qubits do we need? To fully and exactly describe the black hole and the infinite AdS-like space surrounding it we need an infinite number, but that's like saying that in order to exactly describe the solar system we need all of the degrees of freedom out to the cosmic horizon. Clearly that is overkill. 

On the other hand there is a minimum number needed to describe a system with entropy $S$. The maximum entropy of a system of $K$ qubits is $K.$ Therefore we need at least $S$ qubits. I will assume that the minimal description is not much bigger. Thus we will compare black holes of entropy $S$ with qubit systems of size $K\sim S.$

One might think these qubits are localized at definite locations on the horizon---perhaps a kind of lattice---and interact locally.  However known properties of black holes indicate that this is not correct. In particular the fast-scrambling property
implies that the interactions between qubits is \it \kl \ \rm  and \it all-to-all. \rm \  This means that the qubit Hamiltonian involves sums of terms, each involving $k$ qubits or fewer, and every pair of qubits interact in at least one of those terms. This implies that the qubits are completely de-localized.  In itself this is a surprising conclusion but it is not new. I will assume you are familiar with it from previous lectures. SYK and BFSS matrix theory are examples. In SYK the parameter $k$ is  called $q.$

\section{The Growth of Wormholes}

\begin{figure}[H]
\begin{center}
\includegraphics[scale=.3]{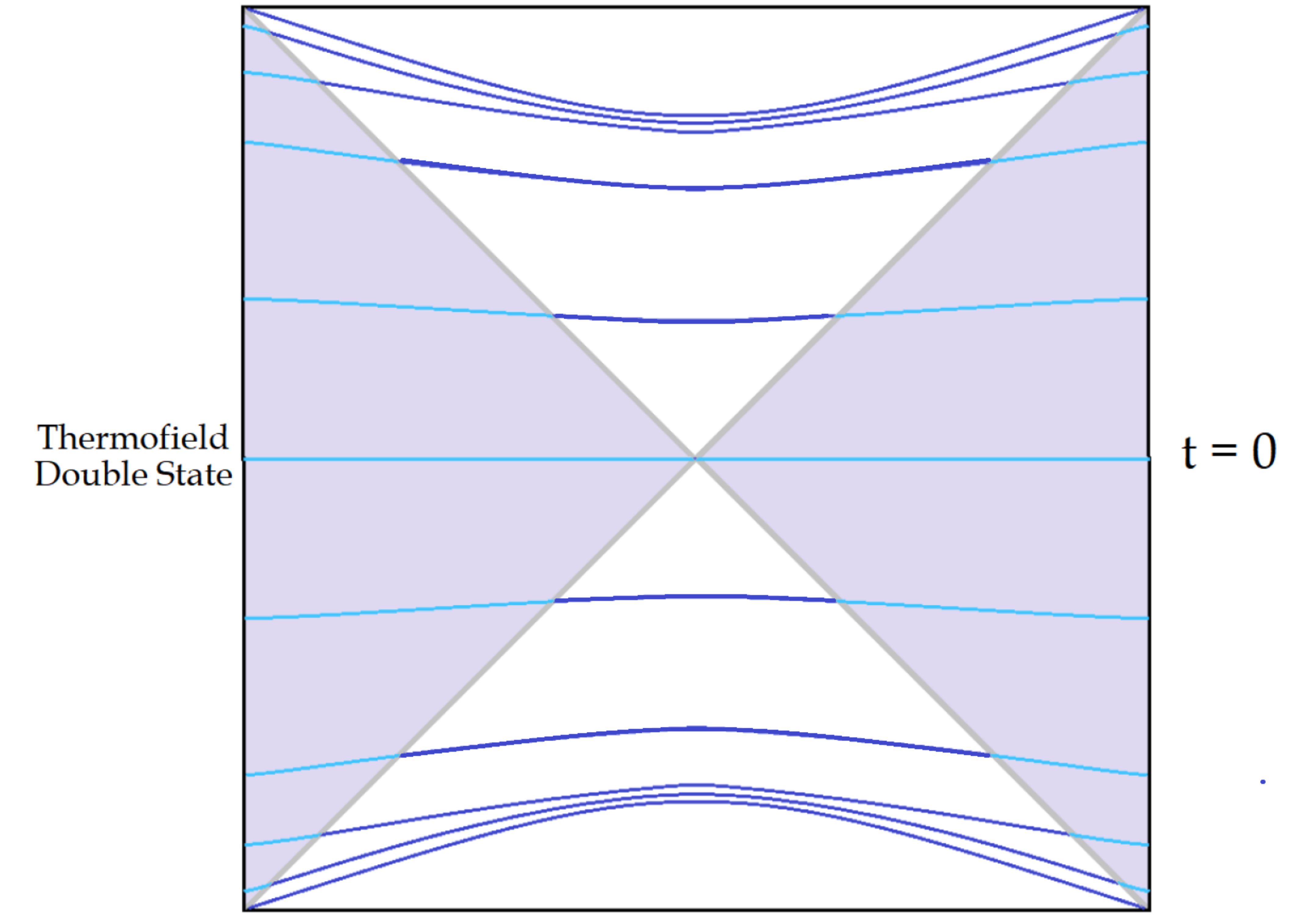}
\caption{Penrose diagram for an AdS eternal black hole. The diagram is foliated by maximal slices. The darker blue portions of the slices represent the wormhole behind the horizons.
The volume of the slices ``bounces," decreasing in the lower part of the diagram, and increasing in the upper part. }
\label{Foliated-BTZ2}
\end{center}
\end{figure}

Wormholes or Einstein-Rosen bridges grow with time.
To illustrate this fact we  examine the Penrose diagram \ref{Foliated-BTZ2} of an AdS ``eternal" black hole, or what is really two entangled black holes connected by an Einstein-Rosen bridge.

\bn
I've also drawn a space-like foliation that I'll define precisely in a moment.
One can see that the spatial size of each slice varies as we move up and down the diagram. In particular if we focus on the portion of the slices behind the horizons---the darker blue segments---it is evident that a kind of bounce occurs. The slice at $t=0$ has vanishing length, which then increases symmetrically, both into the future and the past. At first sight I found this very puzzling. It reminded me of something I had thought about long before while I was a student trying to understand the second law of thermodynamics.

Imagine a gas of a large number of classical molecules in a sealed box. At $t=0$ the molecules are all at rest and are confined to a extremely small volume near the center of the box. The entropy of this configuration is very small.
If we now evolve the state into the future the forces between molecules will act to spread their velocities and their positions. The  second law of thermodynamics will push the system to thermal equilibrium, filling the box uniformly, at least to the naked eye. The entropy will increase to the maximum value consistent with the energy. That much is obvious.

If, at $t=0$, instead of running the system into the future we run it back in time,  exactly the same thing will happen but in reverse. The entire history, past to future, will consist of a bounce. The gas will start in apparent equilibrium in the far past, spontaneously  begin to contract and collect near the center of the box, and then expand back out into the full volume. The entropy will decrease from its large initial equilibrium value to approximately zero, and then bounce back.

Why do we never see such 2nd-law-violating  phenomena? One answer is that they are very unstable;
 any minute ``butterfly" perturbation    will quickly reverse the decrease of entropy  and start it increasing. Nevertheless the unperturbed entropy-bounce is a genuine  solution of the classical equations of motion.

My first reaction  to the time-symmetrical  bounce in figure \ref{Foliated-BTZ2} was that it must be some similar phenomenon---a very fine-tuned entropy-bounce. But this cannot be right. The  black holes on both sides of the Penrose diagram are in perfect thermal equilibrium throughout the entire history. It cannot be entropy that bounced, nor can  the second law of thermodynamics be responsible for the  tendency for the wormhole to grow. These paradoxical  facts puzzled me for more than thirty years.

Before resolving the paradox I will define the slicing precisely. At each time (vertical height in the diagram) we  imagine anchoring a space-like surface on the two boundaries. The surface can then be interpolated  in a coordinate invariant way by choosing the slice of \it  maximum spatial volume.\rm  \ These maximal slices are the blue lines and the portion of them behind the horizons is the darker blue segments. We will identify this portion behind the horizons with the wormhole.
It is clear from figure \ref{Foliated-BTZ2} that the spatial volume of the wormhole bounces. It  goes to infinity as the anchoring time goes to $\pm \infty,$ and passes through zero at $t=0.$ 

Because the slices and their volumes are defined in a coordinate invariant manner, according to the standard AdS/CFT lore  there must be a holographic description of the wormhole volume: some property of the quantum state of the dual pair of CFTs at time $t$. Whatever that quantity is it should be minimal  in the thermofield double, and should increase into both the future and the past. Classically it should increase indefinitely but for reasons I will explain, quantum mechanics necessarily bounds the growth.  The 30 year puzzle for me was: What kind of quantity---if not entropy---can behave this way?

\begin{figure}[H]
\begin{center}
\includegraphics[scale=.3]{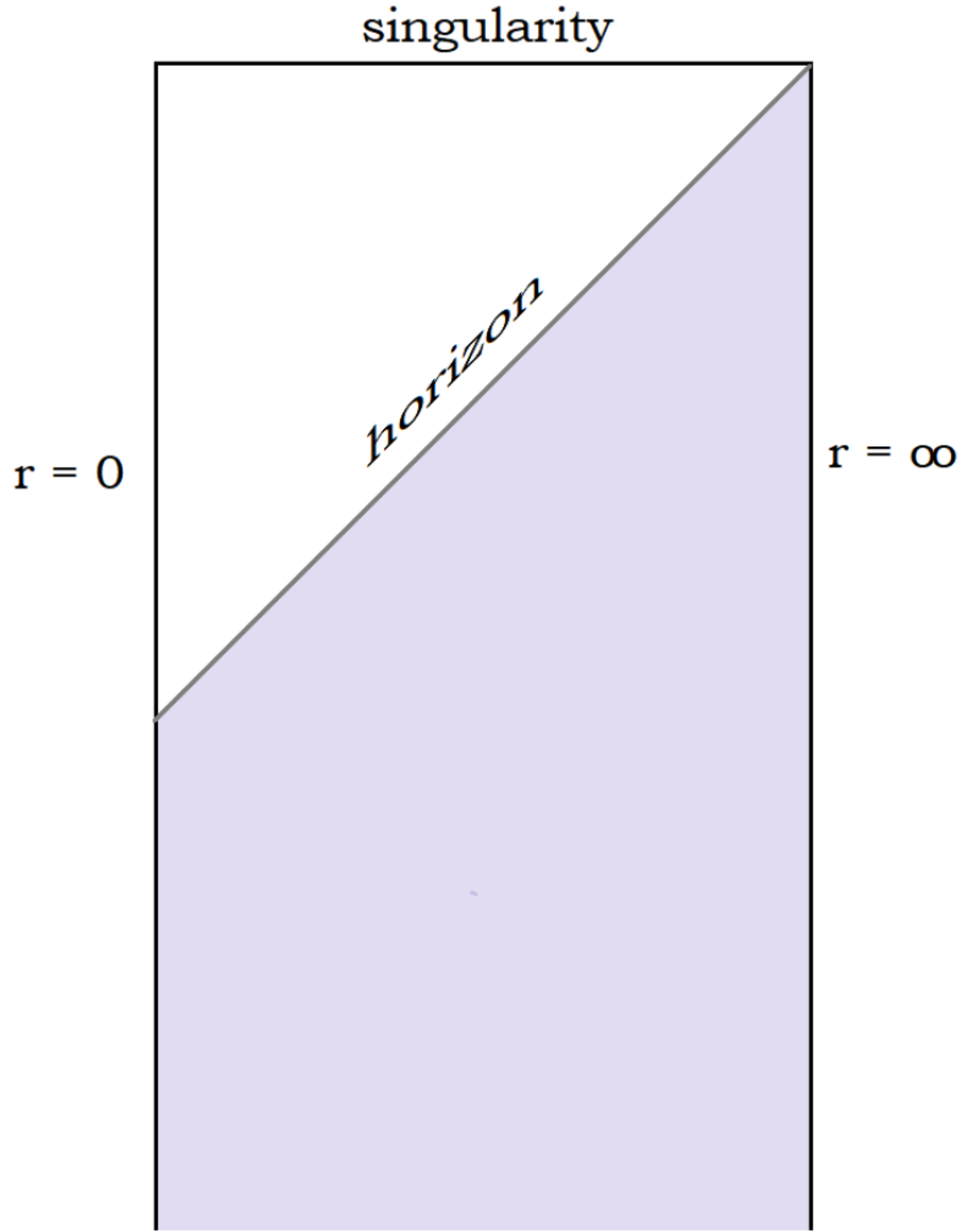}
\caption{One sided AdS black hole formed by collapse.}
\label{one-sided}
\end{center}
\end{figure}

This problem is not special to the two-sided black hole. It also occurs in the theory of a single un-entangled black hole. Figure \ref{one-sided} shows the Penrose diagram for a one-sided black AdS hole formed by collapse.

\begin{figure}[H]
\begin{center}
\includegraphics[scale=.3]{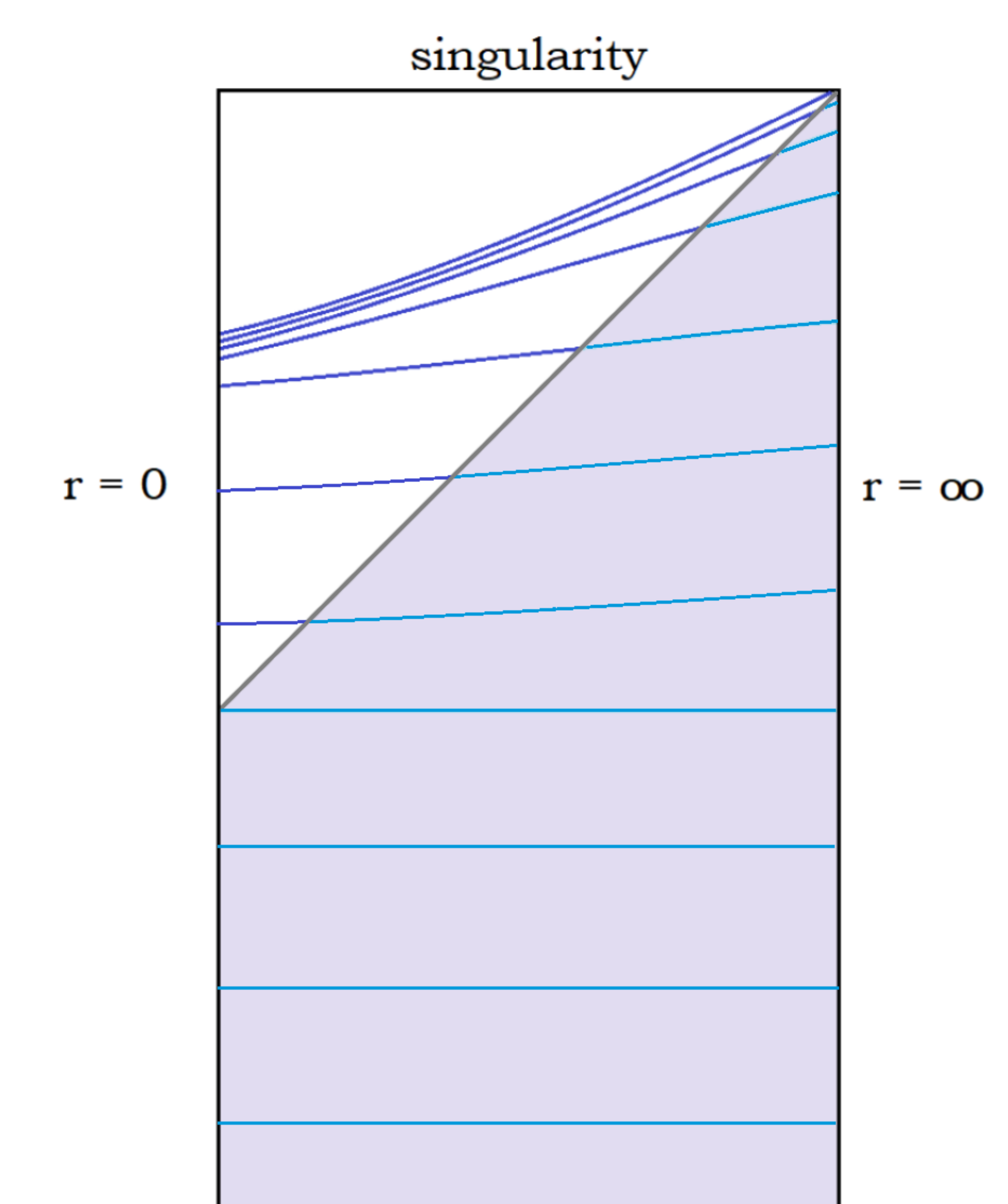}
\caption{One sided AdS black hole foliated by maximal slices. The dark portions of the slices represents the region behind the horizon, i.e., the bridge to nowhere.}
\label{one-sided-foli}
\end{center}
\end{figure}

In figure \ref{one-sided-foli} the same history is shown foliated by maximal slices anchored at the boundary. 

In the one-sided case there is no bounce, only  a growth that begins when the horizon forms. In the remote past the slices do not penetrate behind the horizon at all, so we assign them zero volume. At some time the maximal slice intersects the point at which the horizon forms at $r=0.$ From then on the portion of the maximal slice behind the horizon grows as  a function of anchoring time.

In this one-sided case the black hole is not in perfect thermal equilibrium throughout its history. The entropy starts zero or very small, and then  quickly saturates at the equilibrium value. This takes a very short time\footnote{About a millisecond for a solar mass black hole.}, but the increase of the volume of the dark blue slices behind the horizon goes on indefinitely, at least classically. Evidently some feature of the quantum state evolves long after the system has come to thermal equilibrium.

As a matter of terminology, the portion of each slice behind the horizon in figure \ref{Foliated-BTZ2} is the instantaneous Einstein-Rosen bridge or wormhole connecting the two horizons. The growth phenomenon we are interested in is the growth of the wormhole.  In figures \ref{one-sided} and \ref{one-sided-foli} there are not two sides to connect, but there is a growing region behind the horizon that Ying Zhao and I called the ``bridge-to-nowhere." Both the ERB and the bridge-to-nowhere grow in similar manner.

\begin{figure}[H]
\begin{center}
\includegraphics[scale=.2]{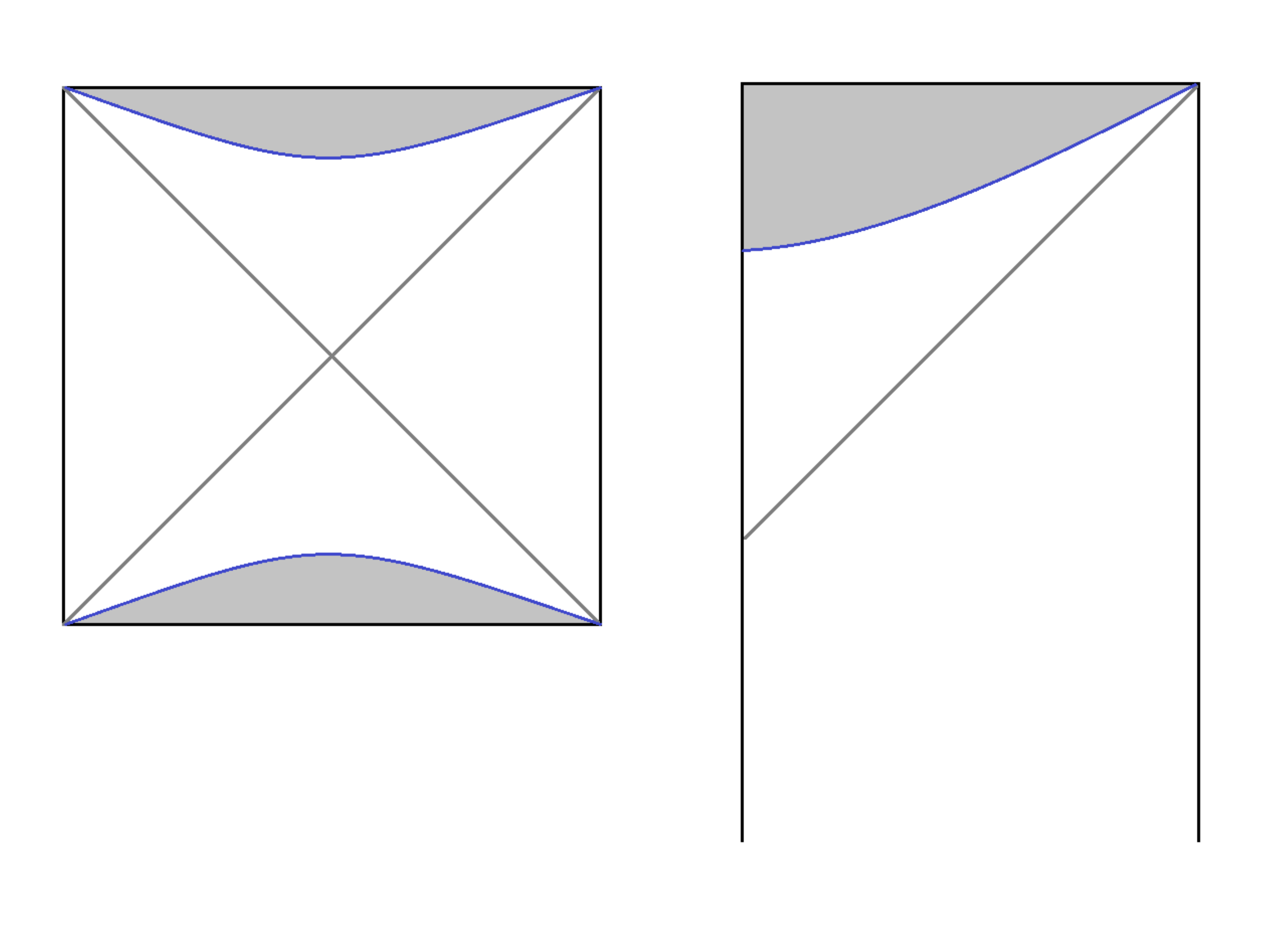}
\caption{The grey areas are not covered by the maximal-slice foliation.}
\label{no-fol}
\end{center}
\end{figure}

One point to note is that the maximal surfaces do not foliate the entire geometry. As the anchoring time goes to infinity the maximal surfaces tend to an asymptotic maximal surface that lies somewhat below the singularity. Figure \ref{no-fol} shows the regions that are not covered by the foliation.

To reiterate, the important virtue of maximal slicing is that it is defined in a coordinate-independent manner.

\subsection{Properties of Growth}

Let's review some facts about  the growth of wormhole volume that follow from the black hole geometry. First of all, except for an early transient, lasting about a thermal time, the growth  is linear with time. In the two-sided case the rate of growth of the volume satisfies,
\be 
\frac{dV(t)}{dt} \sim l_{ads}A T 
\label{Vdot}
\ee
where $t$ is the anchoring time, $A$ is the horizon area, $T$ is the black hole temperature, and $\l$ is the AdS length scale. This is not hard to show using the AdS black hole metric but an easier argument is just dimensional analysis. 
In the one-sided case the asymptotic  rate of growth is half that of the two-sided case.

Let's define a geometric quantity $C$ (yes, we will eventually identify it with complexity, but not yet),
\be 
C \equiv \frac{V}{G\l}
\label{CV}
\ee
Then \ref{Vdot} gives
\be 
\frac{dC}{dt}=\frac{A}{G}T
\ee
and using the Hawking-Bekenstein area-entropy relation
\be  
\frac{dC}{dt}=ST
\label{dCdt=ST1}
\ee
In a little while we will see that this is  the expected rate of growth of complexity for a circuit with $K\sim S$ qubits. 

\subsection{Rindler Time and CV}

\begin{figure}[H]
\begin{center}
\includegraphics[scale=.4]{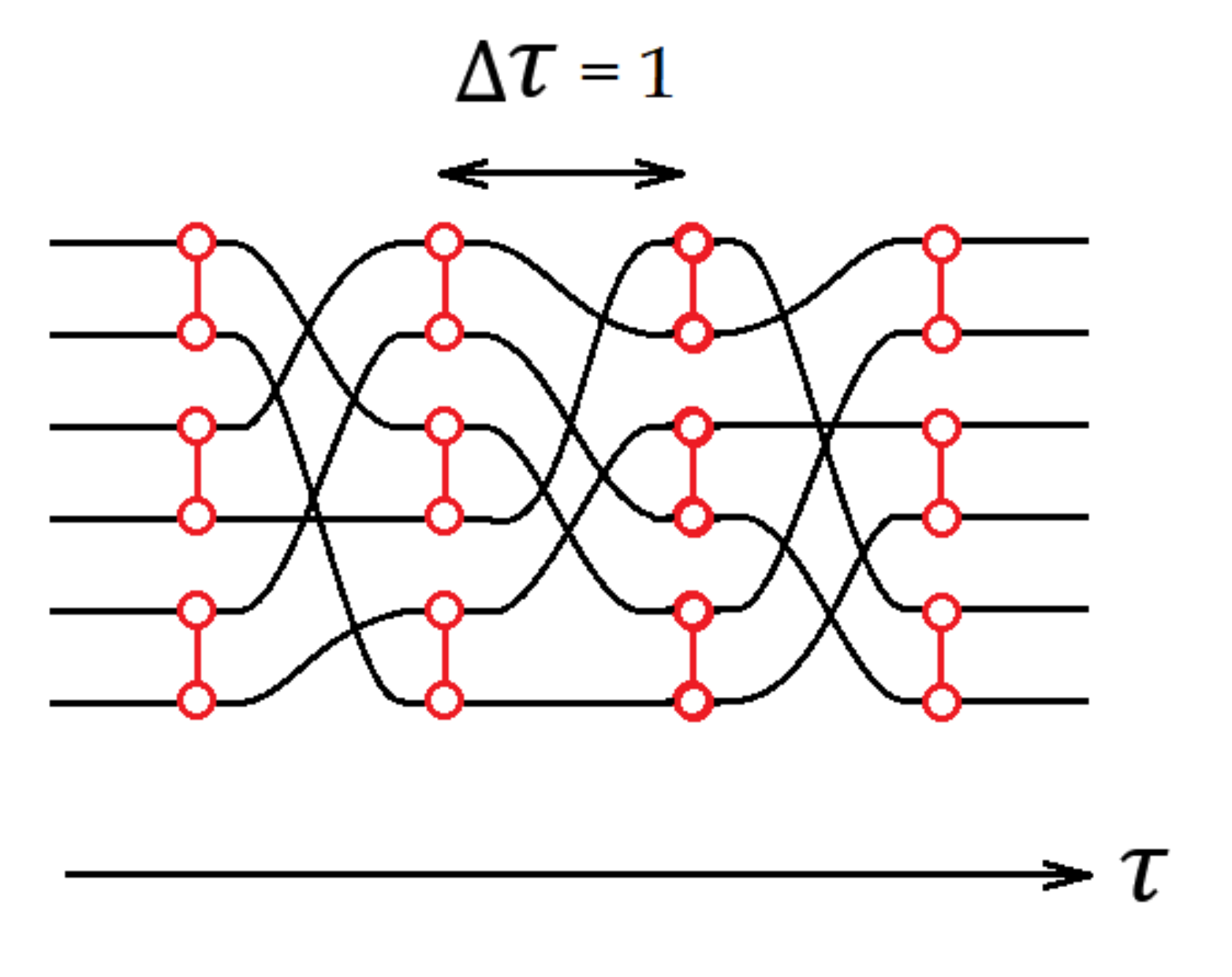}
\caption{}
\label{cirkuit}
\end{center}
\end{figure}

The standard architecture for quantum circuits (explained in Lecture I) is illustrated in figure \ref{cirkuit}.
We assume that the circuit comes equipped with a clock, and in each time-step $K/2$ gates act. For sub-exponential time the number of gates in a circuit corresponds to the complexity\footnote{This is the no-collisions assumption discussed at length in Lecture I. It is expected to be correct for sub-exponential time.} $\CC$ of the unitary operator prepared by the circuit satisfies,
\be 
\frac{d\CC}{\d\t} = \frac{K}{2}
\ee
or, assuming $K\sim S$,
\be 
\frac{d\CC}{\d\t} \sim S.
\label{dC-dtau=S}
\ee

If  black holes can be viewed as quantum circuits one may ask what the time-coordinate $\tau$ corresponds to on the black hole side. One hint is that since complexity and entropy are dimensionless, $\tau$ is also dimensionless. The ordinary Schwarzschild coordinate $t$ of a black hole has dimensions of time and therefore cannot be the same as $\t.$ There is however a universal dimensionless time for all non-extremal black holes, namely the Rindler time. The near-horizon geometry has the universal form,
\be 
ds^2 = -\rho^2 d\tau^2 + d\rho^2 +r(\rho)^2 d\Omega^2
\ee
where $\rho$ is proper distance from the horizon and $\tau $ is a hyperbolic boost angle. 
The natural candidate for the circuit time for a black hole is the Rindler time. For now let's accept this identification but 
later, in section \ref{Sec: Lyap and Rind} I will give a precise quantitative argument for identifying circuit time with Rindler time.

Thus I will therefore use the 
 same symbol $\t$ for circuit time, and for Rindler time. We may therefore assume that in the context of  black hole evolution the complexity grows according to \ref{dC-dtau=S} with $\tau$ being Rindler time.

An asymptotic   observer at infinity sees  rate of computation  red-shifted. It is a universal fact for all non-extremal black holes, that the asymptotic time $t$ is related to Rindler time by 
\be 
\t = \frac{2\pi t}{\beta} = 2\pi Tt.
\ee

It follows that the rate of computation as seen from infinity, is
\be 
\frac{d\CC}{dt}  
= ST.
\ee
Compare this with \ref{dCdt=ST1} and one sees that  the quantity $C$ is in fact the complexity $\CC.$ Accordingly \ref{CV} may be written in the form,
\be 
\CC =\frac{V}{G \l }.
\label{CV-dualtiy}
\ee
This formula connecting  complexity and the volume of the the wormhole is the called
 CV  correspondence. It is one of two versions of the complexity-geometry correspondence, the other being  complexity-action formulation. Time will not permit me to go over the CA correspondence, but there are many papers that you can consult.

\bn

Let me summarize and emphasize two points:
\begin{enumerate}
\item Circuit time, measured in steps, corresponds to the Rindler time of a black hole. 
Now repeat and memorize:

\it
Circuit time, measured in steps, corresponds to the Rindler time of a black hole. \rm

\bn 

\item Complexity is proportional to the wormhole volume through
\ref{CV-dualtiy}
\end{enumerate}

Of course this latter point is a conjecture; it was not derived from  first-principles. The argument is due to Sherlock Holmes; I quote the great detective:  ``When you have excluded the impossible whatever remains, however improbable, must be the truth." We  can and will subject item 2 to rigorous testing.

\section{Exponential Time Breakdown of GR}\label{expTbreakdown}
According to classical general relativity an Einstein-Rosen bridge (ERB), or  a one-sided bridge-to-nowhere, grows forever. It is interesting that this growth is bounded by quantum mechanics. To see why we  consider a diagnostic of the length of the wormhole, namely the equal time correlation between CFT fields outside the black hole horizons. If the two fields are on the same side then the correlation will be time-independent because the density matrix on each side is thermal, and therefore stationary. But if we consider correlations between opposite sides it is well known that the correlation is large at $t=0$ and decreases with the magnitude of $t$. The cause of the decrease is simple; the wormhole separating the two sides grows. Roughly speaking we expect the correlation to  exponentially decreases with the distance through the wormhole,
\be 
\la \phi_L \phi_R \ra \sim e^{-L(t)/l_{ads}}
\label{correlate}
\ee
where $L(t)$ is the length of the wormhole\footnote{This formula holds until $L/l_{ads}$ is of order the entropy. Beyond that the correlation is dominated by noise. } at time $t$.

Recall that in quantum mechanics there is a theorem called the quantum-recurrence theorem:

\bn

\bf Quantum Recurrence Theorem: \it
\bn

 The state of a (non-integrable) system with a finite density of states will be quasiperiodic  with a recurrence time doubly exponential in the entropy. This applies both to black holes in AdS and to generic states of systems of $K \sim S$ qubits. The quantum recurrence time is $ exp \ (exp \  S).$ Over this enormous time scale the state will quasiperiodically  return arbitrarily close to its initial value. \rm

\bn
It follows  that over such long time scales the two-sided correlation 
$\la\phi_L \phi_R \ra$ will  return to its initial TFD value, and from \ref{correlate}   the distance through the wormhole will also  return to its minimal value. The classical phenomenon  of eternal growth must break down quantum mechanically as a consequence of the finite entropy of black holes (recall that classically the entropy diverges like $1/\hbar$).

I believe we can do better and argue that classical geometry must break down at a singly exponential time $\sim \exp{S}$. The argument goes as follows:
As the wormhole grows the evolving quantum state must pass through a series of orthogonal states. The  time scale for the state to become orthogonal is the Anandan-Aharonov time $1/\Delta E.$ As an example, for a 4-dimensional \S \  black hole the uncertainty in the energy is the Planck mass. Thus the system must pass through orthogonal states at the rate of one per Planck time.

The problem is that there are only a limited number of available  mutually orthogonal states, the number being of order $\exp(S).$ After an exponential time we run out of orthogonal states. Subsequent states   must become  superpositions of the previous states, all of which have sub-exponential wormhole length. This implies that the expectation value of the length of the wormhole must stop growing when $t\sim \exp{S}.$ This behavior is distinctly non-classical. We'll return to this issue.

\subsection{C=V}
Let's come back now to the  puzzle:  what grows long after thermal equilibrium is attained? 
 I only know one thing that has all the right features: quantum complexity. As I explained in Lecture I, quantum complexity has many of the characteristics of entropy but on a grand exponential time scale. It may be thought of as the entropy of the classical auxiliary system $\CA$ that I defined in Lecture I.

 To relate complexity to the spatial volume of the wormhole or Bridge-to-Nowhere we go back to equation \ref{CV},
$$\CC=\frac{V}{Gl_{ads}}.$$
This is the complexity=volume duality proposal.  The mysterious bounce shown in figure \ref{Foliated-BTZ2} is evidently a complexity bounce, not an  ordinary thermal entropy bounce.

Now let's turn to the confirmation of the CV hypothesis, not by deriving it but by detailed comparison of the behavior of volume (as determined from GR), and complexity (from the study of quantum circuits).

\section{Precursors }

\subsection{The Epidemic Model}

The best source of confirmation of the $\CC=V$ hypothesis is to perturb the evolution of the black hole, and see if the correspondence  between GR and quantum circuits  continues to hold. For example we may allow some time-dependent perturbations to the CFT Hamiltonian and calculate the effect on the black hole geometry.  At the same time we may study the effects of time-dependent perturbations on the complexity of a quantum circuit. By comparing the two we can confirm the complexity-volume  correspondence. In this context so-called precursor operators are a useful tool.

 Let us consider an operator of the form,
\be 
W(t) = e^{iHt}We^{-iHt} =U^{\dag}(t)W U(t)
\ee
in the Schrodinger picture.
Here $H$ is some all-to-all  \kl \ Hamiltonian acting on a system of $K$ qubits and $W$ is some single qubit unitary operator. With no loss of generality we can take $W$ to act on the first qubit. The complexity of $W$ is $1$. We write that in the form,
\be 
\CC(0)=1.
\ee
The object of our investigation is  $
\CC(t)$---the complexity of $W(t)$ as a function of time. By definition
$
\CC(t)
$
is the \it minimum \rm number of gates that it takes to construct $W(t).$  

Let us first consider the analogous problem for a quantum circuit composed of gates. To that end we replace the evolution operator $e^{-iHt}$ by a product of gates with the usual parallel architecture illustrated in figure \ref{cirkuit}. I'll write it in the form,
\be 
U(t) = g_n g_{n-1}....g_1.
\ee
The precursor operator becomes
\be 
W(t) = g^{\dag}....g^{\dag}_{n-1}g^{\dag}_n \  W \   g_n g_{n-1}....g_1.
\label{W-circuit}
\ee
One might be tempted to ascribe a complexity of $(2n+1)$ to $W(t)$. This however is an overestimate. While it is true that the circuit in \ref{W-circuit} has $2n+1$ gates the definition of complexity requires us to find the \it minimal \rm circuit that prepares $W(t)$.

By assumption $W$ is a very small perturbation on the system of $K$ qubits in that it only affects the first qubit. For all other qubits $W$ acts as the unit operator. Let's first examine the case in which $W$ is just the unit operator, $W=1.$ In that case all the gates in \ref{W-circuit} cancel and the complexity is exactly zero.

\begin{figure}[H]
\begin{center}
\includegraphics[scale=.3]{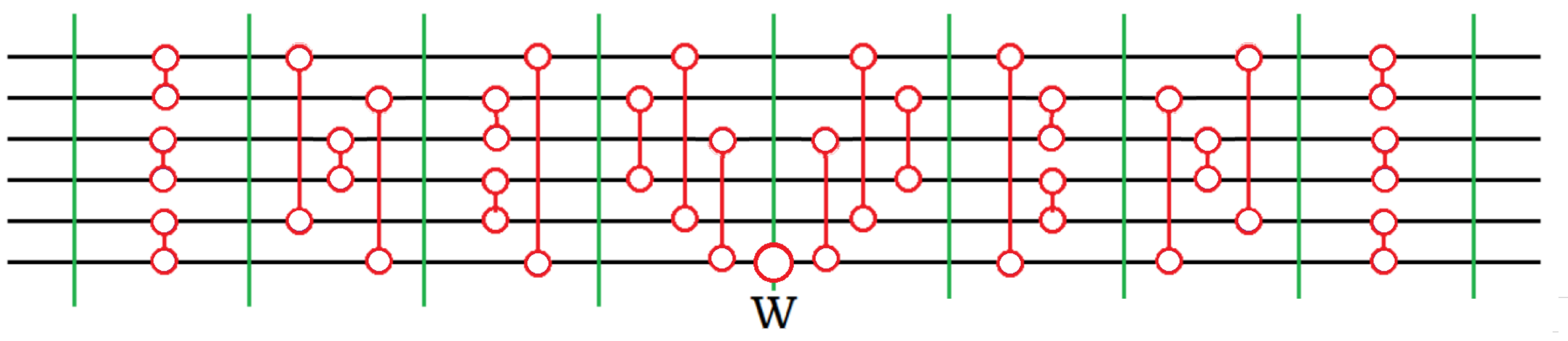}
\caption{Circuit  for a precursor}
\label{precursor-circuit3}
\end{center}
\end{figure}

\begin{figure}[H]
\begin{center}
\includegraphics[scale=.3]{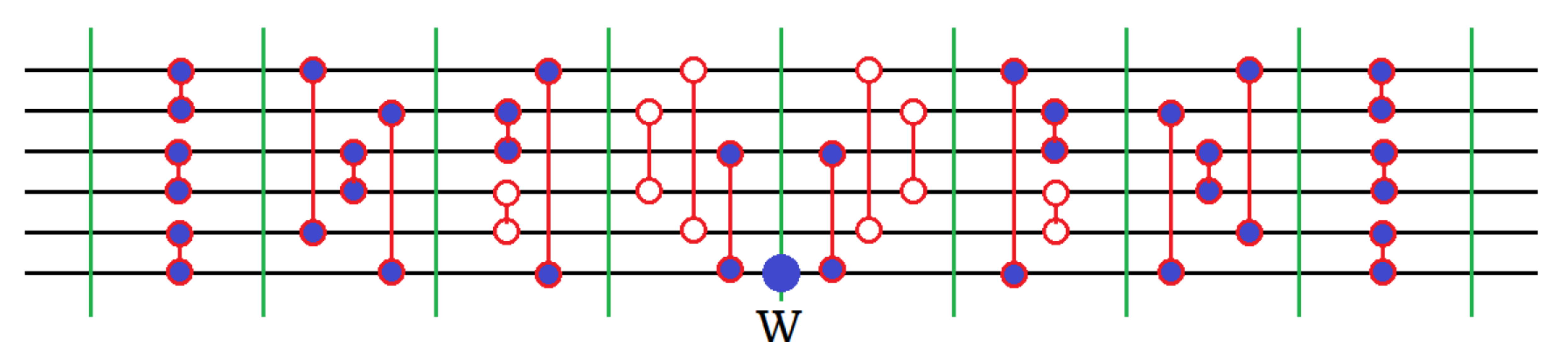}
\caption{Each infected (blue)  qubit  infects each each qubit that it interacts with. The epidemic grows until all qubits are infected.}
\label{precursor-circuit1}
\end{center}
\end{figure}

In the case where $W\neq 1$ many gates still cancel. To illustrate the cancelation consider the circuit in figure \ref{precursor-circuit3}. The operator $W$ acts between two sub-circuits which are mirror images of each other as in \ref{W-circuit}.
Start at $W$ and let us work our way toward the right. We will say that $W$ infects the qubit that it acts on (colors it blue).  Each  time an infected (blue) qubit interacts with an uninfected (white) qubit it infects it. The result is a growing epidemic in which eventually all qubits are infected. On the left side there is a mirror image epidemic. The growth of the epidemic is shown in figure \ref{precursor-circuit1}.

\begin{figure}[H]
\begin{center}
\includegraphics[scale=.3]{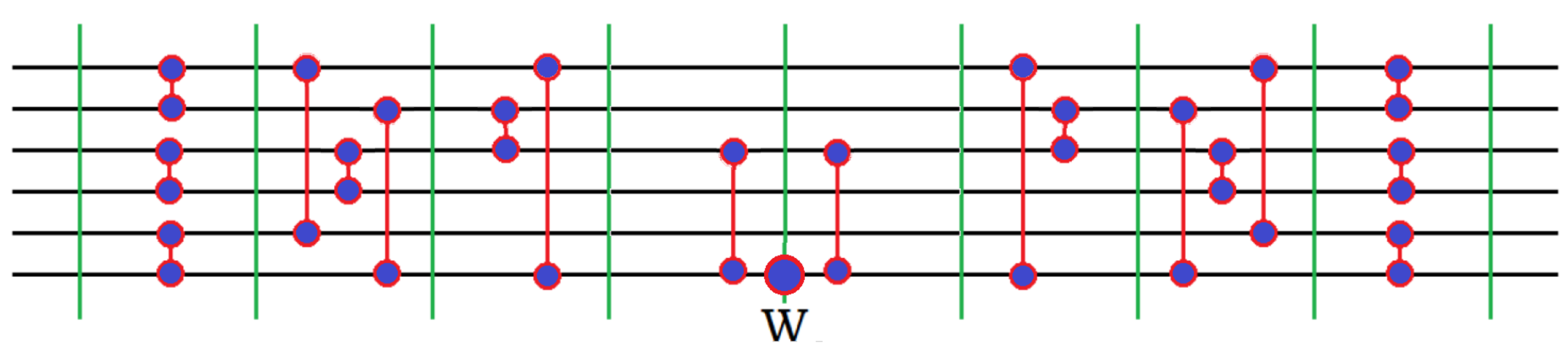}
\caption{Uninfected gates cancel with their mirror images.}
\label{precursor-circuit4}
\end{center}
\end{figure}

Notice  that any gate involving uninfected qubits can cancel a mirror image gate on the other side. The result, after canceling all un-infected gates, is shown in  figure \ref{precursor-circuit4}. 

Evidently the minimum  number of gates needed to prepare the precursor is less then the total number of gates in figure \ref{precursor-circuit3}. Let us quantitatively estimate it.

Suppose the circuit has evolved to the right by $n$ steps and at that stage the epidemic has \it size \rm $s(n).$ Then the average number of new infections that result in the next step is easy to compute. It is given by,
$$\Delta s = \frac{(K-s)s}{K-1}$$
The shift by one unit in the denominator can be ignored,
\be 
\Delta s = \frac{(K-s)s}{K}.
\label{difference equ}
\ee
We can replace this by a differential equation
\be 
\frac{ds}{d\tau} = \frac{(K-s)s}{K}
\ee
which is easily integrated,
\be 
s = \frac{Ke^{\tau}}{K + e^{\tau}}.
\label{logistic}
\ee

This ``epidemic function"\footnote{Also known as the logistic function.} governs the size of the precursor operator. 

Equation \ref{logistic} can written,
\be 
\frac{s}{K} = \frac{e^{\tau -\tau_*}}{1+e^{\tau -\tau_*}}
\ee
where $\tau_*$ is the \it scrambling  time \rm expressed in circuit time units.
\be 
\tau_* = \log{K}
\label{s-over-K}
\ee
The size $s(\t)$ has the following properties:
For small $\tau$ \ref{s-over-K} grows exponentially,
\be 
s(\tau) \approx e^{\tau}  \ \ \ \ \ \rm (small \ \tau)
\label{expgro}
\ee

\bn

For large $\tau$ the size saturates when all qubits become infected,
\be 
s(\tau) \to K.
\ee
and 
\be
\frac{s}{K}\to 1.
\ee
At the scrambling time a rather sharp crossover takes place and the size jumps rapidly to its saturation value $K$.

\subsection{Lyapunov and Rindler}\label{Sec: Lyap and Rind}

Let's come back to the relation between circuit clock time and Rindler time. You may think that the arguments were not very compelling  but equations \ref{difference equ} through \ref{expgro} provide a quantitative basis for the identification.
The exponential growth in \ref{expgro} is expected and should  be compared with the OTOC theory of scrambling, especially as applied to black holes.
There it is well known from the work of Kitaev and Maldacena, Shenker, and Stanford,  that in black hole physics size grows according to,
\be 
s(t) \approx e^{\frac{2\pi}{\beta}t}  \ \ \ \ 
\label{expgro-t}
\ee
The quantity $2\pi/\beta$ is called the quantum Lyapunov exponent and it is universal for large black holes as long as the \S \ radius is much  larger than the string scale.

Comparing  \ref{expgro-t} with \ref{expgro} we see that agreement indeed requires, 
\be 
\tau ={\frac{2\pi}{\beta}t}.
\ee

In other words circuit time is Rindler time.
  
\bn

\subsection*{Back to Size and Complexity}

Let us now consider the total number of  un-cancelled gates in figure \ref{precursor-circuit1}. This number is the complexity of the precursor at time $\tau$. 

$$
\#\rm{uncanceled \ gates} = \CC(\tau).
$$

\bn
The number of such gates on the right side at any discrete time $n$ is the sum of the sizes for all times less than or equal to $n$. In terms of the continuous time we can write,
\be 
\CC(\tau) = \int_0^{\tau} s(\tau')d\tau'
\ee
or using \ref{logistic},
\be  
\CC(\tau) = K\log(1+ e^{\tau-\tau_*}).
\label{compcircuit}
\ee

\begin{figure}[H]
\begin{center}
\includegraphics[scale=.3]{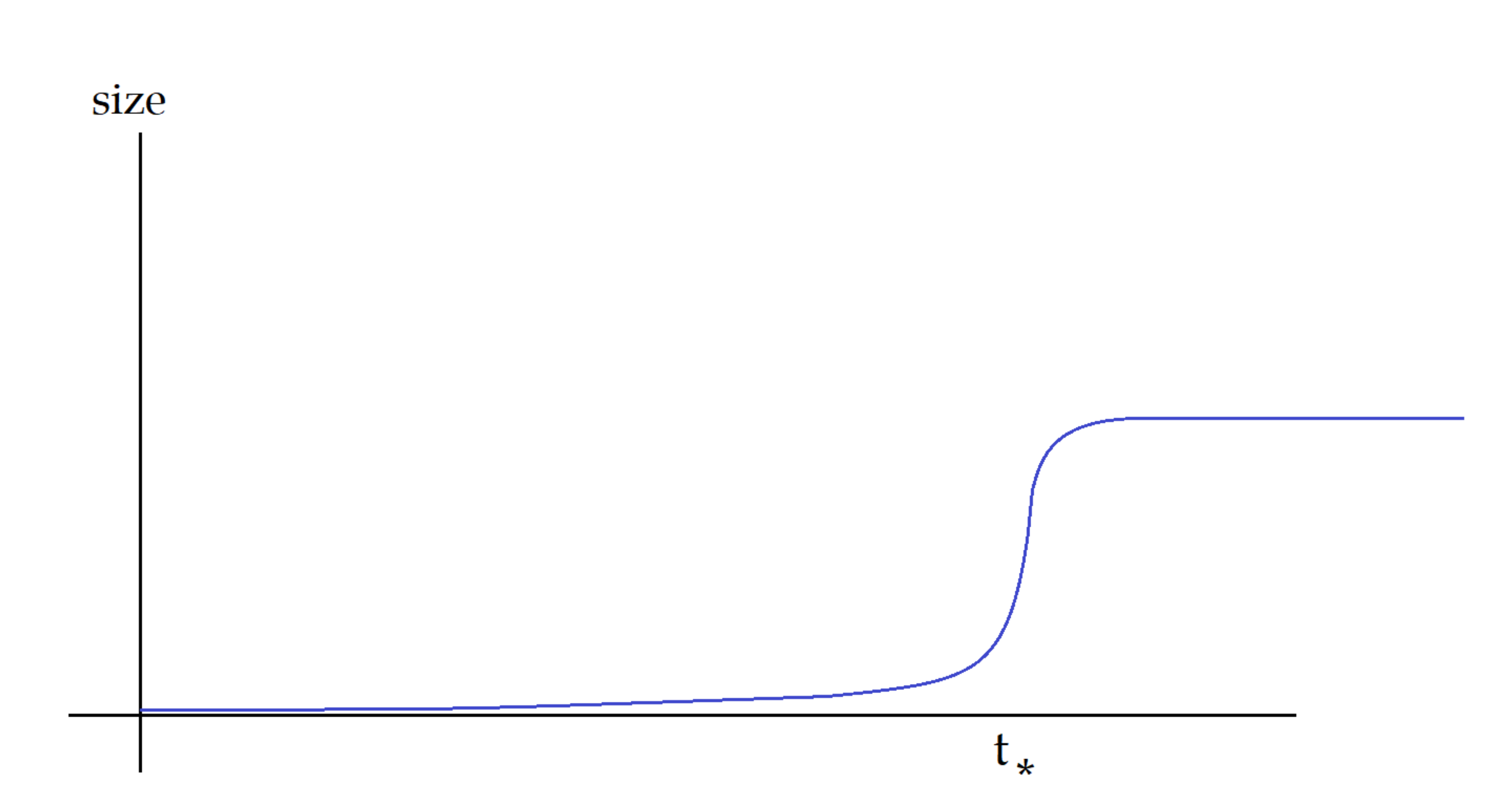}
\caption{size  as a function of time}
\label{size}
\end{center}
\end{figure}

\begin{figure}[H]
\begin{center}
\includegraphics[scale=.3]{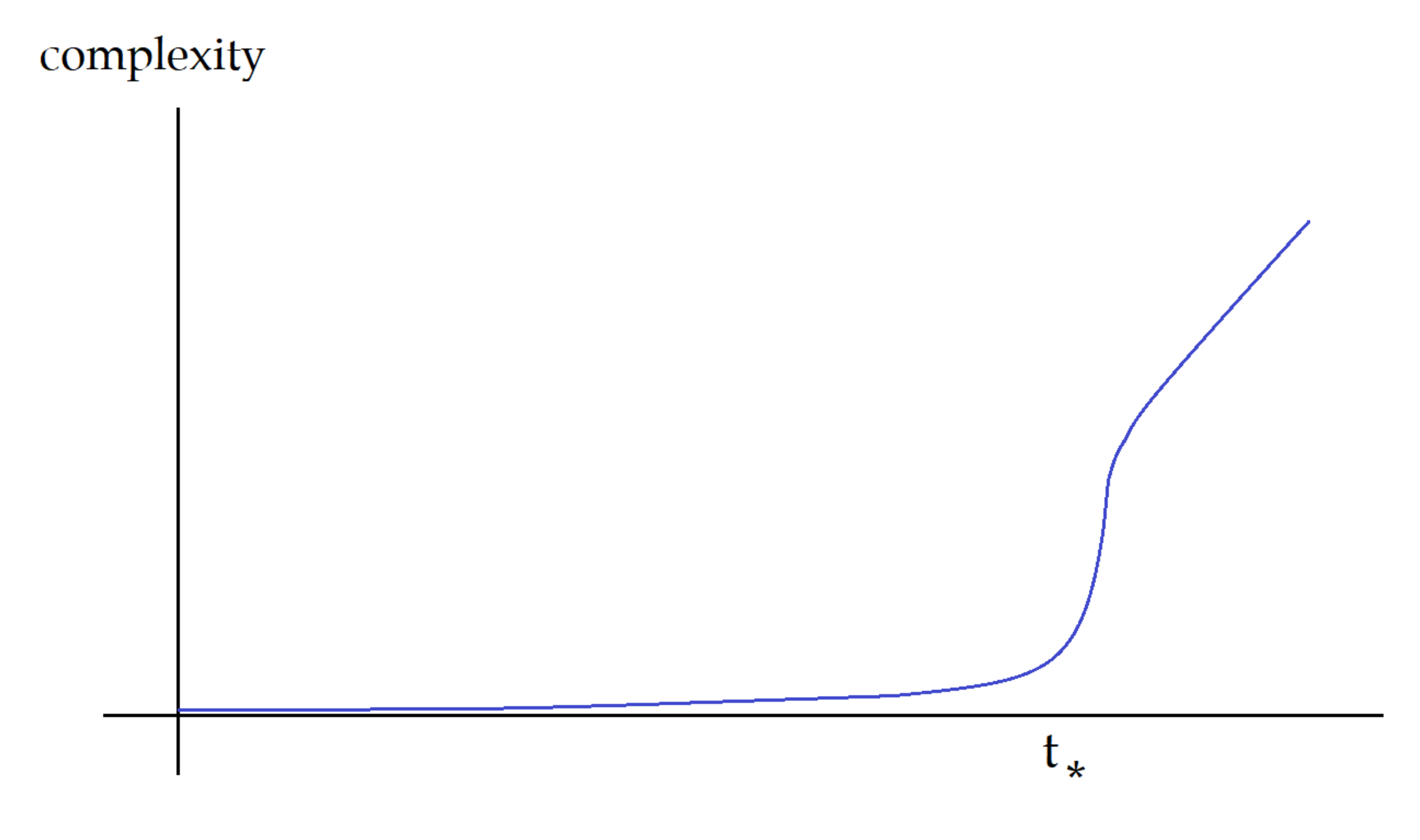}
\caption{ as a function of time}
\label{complexity}
\end{center}
\end{figure}

For large $K$ the size and complexity of the precursor are negligibly small (compared to $K$) until the scrambling time at which they quickly grow to $K$. After that the size is constant and the complexity grows linearly. The delay in the growth of complexity by $\tau_*$ is called the switchback effect 
(Stanford and Susskind, \url{https://arxiv.org/pdf/1406.2678.pdf}) 
 and is an important diagnostic that we will use later. The time-dependence of size and complexity are shown in figures \ref{size} and \ref{complexity}.


\bn

\bn

Now let us turn to precursors in black hole physics. We will see some remarkable correspondences between the geometry of black hole interiors and the equations that we have derived from the analysis of circuits.

\section{Precursors and Black Holes} 
In defining precursor operators in AdS/CFT it is tempting to identify the single qubit operator $W$ with a local low-dimension CFT field. However, local fields when they act inject infinite energy into the CFT system and this is not what we want. The remedy is to smear the field over a time interval. We would like the energy of the field to be such that it would increase the entropy of the black hole by one thermal unit. This means that the energy should be equal to the temperature. We therefore assume that $W_L$ is an operator that adds a thermal quantum to the left-side black hole. 

\begin{figure}[H]
\begin{center}
\includegraphics[scale=.3]{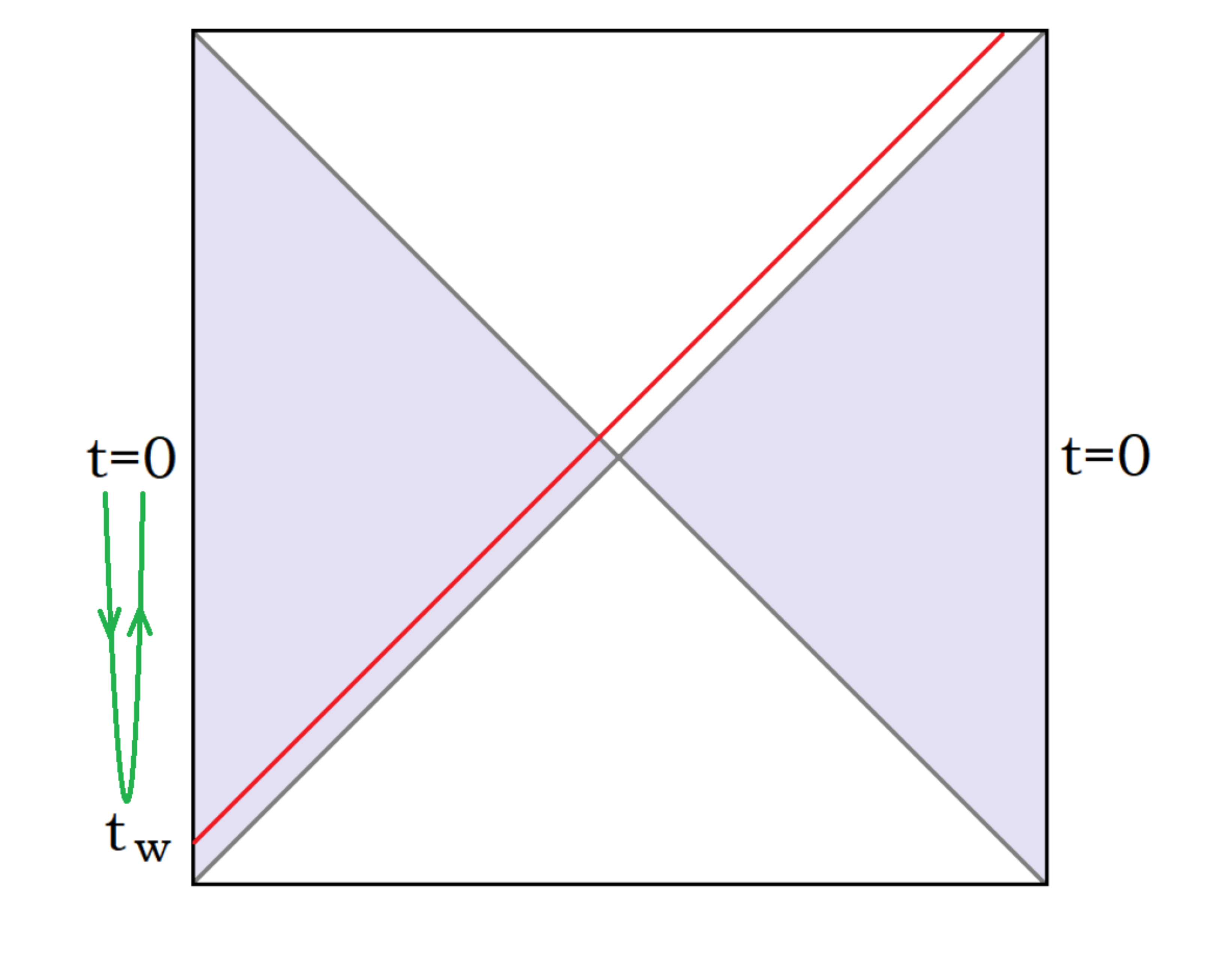}
\caption{}
\label{shock-wave}
\end{center}
\end{figure}

Now let us apply the precursor $ U^{\dag}(t_w) W_L \ U(t_w)$ to the thermofield-double state: 
\bea
|\Psi(t_w)\ra &\equiv&   U^{\dag}(t_w) W_L \ U(t_w) |TFD\ra \cr \cr
\eq W_L(t_w)|TFD\ra
\eea
The operator $W_L(t_L)$ is a Schrodinger picture operator  that acts at $t =0,$ but its effect is exactly the same as adding a thermal quantum to the left side at time  $t_w.$ We will be especially interested in the case of large negative $t_w.$ This shown in figure \ref{shock-wave}.

When $|t_w|>> t_*$ the infalling quantum created by $W_L$ undergoes a large blue shift as it falls to the horizon. The result is that for most of its world line it is an extremely energetic shock wave. This is despite the fact that the thermal quantum makes an extremely small change in the energy of the black hole. 

\begin{figure}[H]
\begin{center}
\includegraphics[scale=.3]{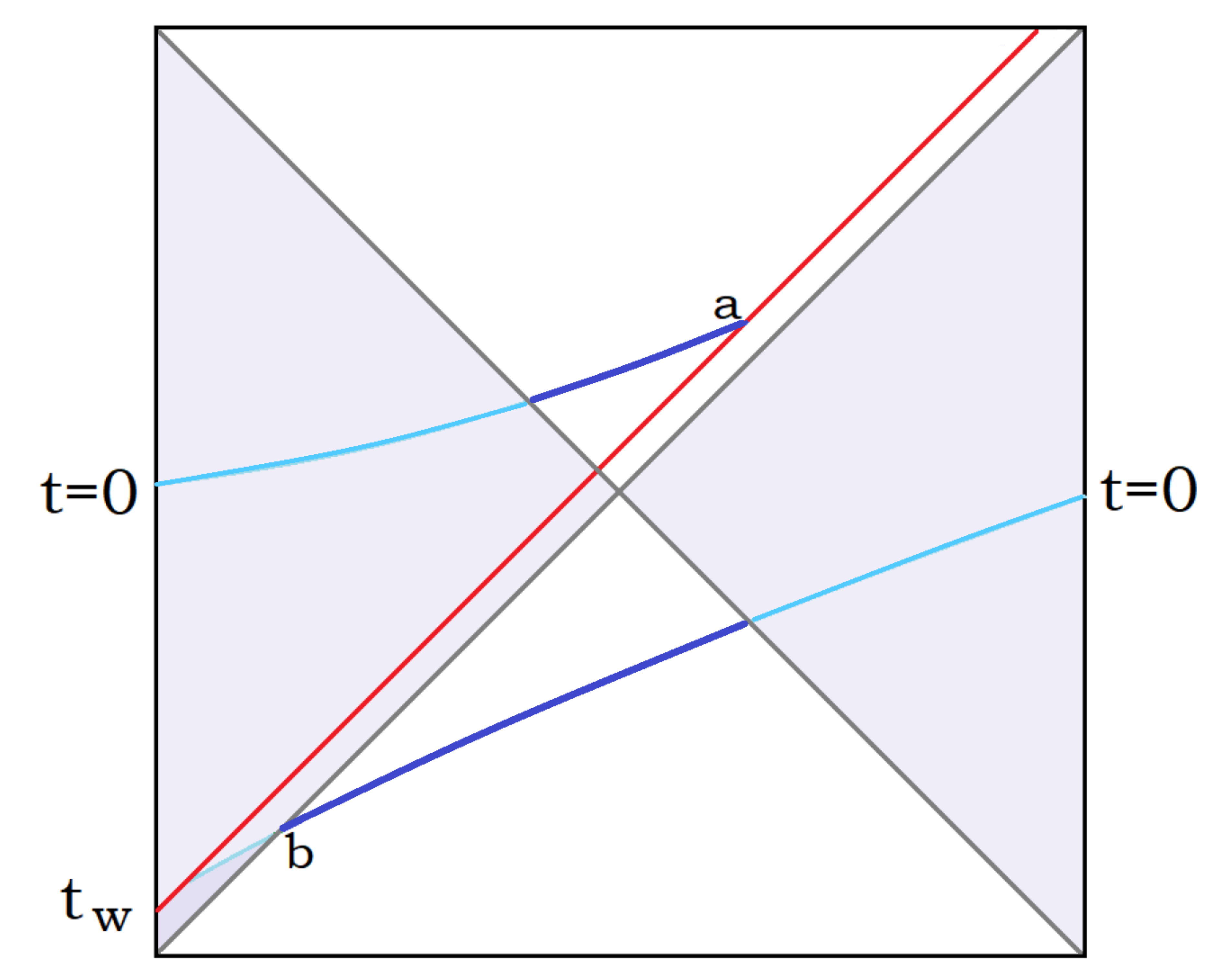}
\caption{}
\label{shock-wave-maxslice2}
\end{center}
\end{figure}

The shock wave has a large effect on the geometry but an easy way to understand it is through the Dray-'t Hooft effect ( T.~Dray and G.~'t Hooft,
  %``The Gravitational Shock Wave of a Massless Particle,''
  Nucl.\ Phys.\ B {\bf 253}, 173 (1985)) Instead of redrawing the Penrose diagram to reflect the gravitational back reaction on the shape of the diagram we can use the rule that any trajectory  crossing the shock wave gets displaced and jumps as shown in figure \ref{shock-wave-maxslice2} where we see the maximal slice anchored on both sides at $t=0$. The jump is from the point $a$ to the point $b$.

Without the shock wave the maximal slice at $t=0$ has no volume behind the horizon; it passes right through the bifurcate horizon. But as one can see from figure \ref{shock-wave-maxslice2} the shock wave causes the maximal slice to gain a significant volume as indicated by the dark blue surfaces.

In order to study the effects of such a shock wave I will give a general ``master formula" for the volume of the maximal slice as a function of the anchoring times $t_L, \ t_R$ and $t_w.$ The formula was derived by Shenker and Stanford 
\url{https://arxiv.org/pdf/1306.0622.pdf}
in another context but it was shown in Stanford-Susskind 
\url{https://arxiv.org/pdf/1406.2678.pdf}
 that it is a good approximation to the volume of maximal slices\footnote{It is a good approximation as long as the times are not exponentially large, at which the classical geometric description is expected to break down.}. It is a purely classical consequence  of general relativity,

\bn

\be 
\frac{V}{G\l} = S \ \log{\{
\cosh{\frac{\tau_L + \tau_R}{2}} 
+e^{(|\tau_w|-\tau_*) +\frac{1}{2}(\tau_L -\tau_R)}
\}}
\label{master-formula}
\ee

\bn
where $\tau$ is the dimensionless Rindler time, $\tau = 2\pi t/{\beta,}$ and $S$ is the entropy.

Let us consider the complexity $V/Gl_{ads}$ at $t_L = t_R =0.$ The answer should be the complexity of the precursor itself since the complexity of the TFD is zero. Using $S\approx K$ we find,
\be 
\CC(0,0) = K \log\left(1 + e^{(|\tau_w|-\tau_*)} \ \right),
\ee
in agreement with \ref{compcircuit}. 
This is a truly remarkable result that relates classical solutions of general relativity to properties of quantum circuit complexity.

The relation between complexity and volume is very general. It applies to any number of shock waves sent in at any times. It also applies to localized non-spherical shock waves,  D. Roberts, D. Stanford and L. Susskind,
\url{https://arxiv.org/pdf/1409.8180.pdf} .

\subsection{Instability of White Holes}

Earlier I talked about the meaning of the bounce in figure \ref{Foliated-BTZ2} . It resembles the entropy bounce of a confined gas if one constrains the gas to be in a very low entropy state at $t=0$. Let me give names to the various regions of the Penrose diagram. The left and right shaded regions are the left and right black hole exteriors. The upper un-shaded region is the black hole interior in which the complexity evolves in a manner consistent with the second law of complexity (complexity increases). The lower un-shaded region is the counter-intuitive region where complexity decreases. In general relativity it is the \it white hole \rm interior.

In the case of the confined gas, events in which the entropy decreases are never seen in the real world. Even if it were possible to set up the fine-tuned initial condition in the remote past, the bizarre behavior is extremely unstable. A tiny perturbation (the proverbial butterfly), by the time it spreads through the system will turn the evolution around so that the entropy will increase after a short time. 

The second law of complexity is patterned on the second law of thermodynamics and we should expect that the condition of decreasing complexity is unstable in the same way. Let us consider what happens if on the left side we perturb the white hole by adding a single thermal quantum at past time $t_w$. Using the master formula \ref{master-formula},  we vary $\tau_L$ while holding $\tau_R$ fixed.
In the white hole region where $(\tau_L+\tau_R)<<0$ we get, 
\be 
V \sim \log\left[ \frac{1}{2}e^{-\tau_L/2}  +e^{|\tau_w|-\tau_*+\tau_L/2}      \right] + const
\label{WH}
\ee
When $\tau_L$ is very negative the first term in the bracket dominates and $V$ decreases just as it would if no perturbation had been applied. The second term due to the perturbation increases with $\tau_L $. The crossover occurs at 

\be 
 \tau_L(crossover) = - \left(|\tau_w| -\tau_* \right)
\ee
Evidently the perturbation at $t_w$ initiates a process that results in a  reversal of the complexity-decrease after a scrambling time.  For large negative $\tau_L$ this is still deep in the original white hole region.
This shows that the decreasing complexity of the white hole behaves  exactly like the classical entropy of a fine-tuned entropy bounce, being unstable to tiny perturbations.

However I want to emphasize again that it is not an entropic  phenomenon in the usual sense. It is a complexity phenomenon.

\bn
\bf 
NOTE: THE LIVE VERSION OF LECTURE II ENDS HERE. IT CONTINUES IN LECTURE III.
\rm

\section{Complexity and Firewalls}

Let me
come now to the subject of firewalls. A firewall is supposed to be an extremely intense high energy shock wave propagating on or just behind the horizon. 
There are two paradoxes, both of which were claimed to be solved by firewalls. One is connected with entanglement and the other with ``typicality." The paradox of multiple entanglements is solved by the ER=EPR principle and I will not discuss it further. The ``Typical State Paradox" (D. Marolf and J. Polchinski, \url{https://arxiv.org/pdf/1307.4706.pdf}) 
is the one whose resolution has to do with complexity 
\url{https://arxiv.org/pdf/1507.02287.pdf}.

The typical state paradox can be easily explained in the case of the two-sided black hole. Imagine Bob on the right side and Alice on the left. Bob would like to jump into his black hole but without getting burned up at the horizon. The problem is that he has no idea what Alice might have done at her end. If one considers all the possible things she could have done  one may conclude that the situation is very dangerous. After all, she had all negative time, deep into the remote past, during which she might have perturbed her black hole. For example she might have dropped in one or more thermal photons deep in the exponential past. Almost anything she did in the remote past would have created  very intense shock waves or firewalls. So averaging over all possibilities, Bob should conclude that there is, with overwhelming likelihood, a firewall/shockwave at his horizon.

\begin{figure}[H]
\begin{center}
\includegraphics[scale=.3]{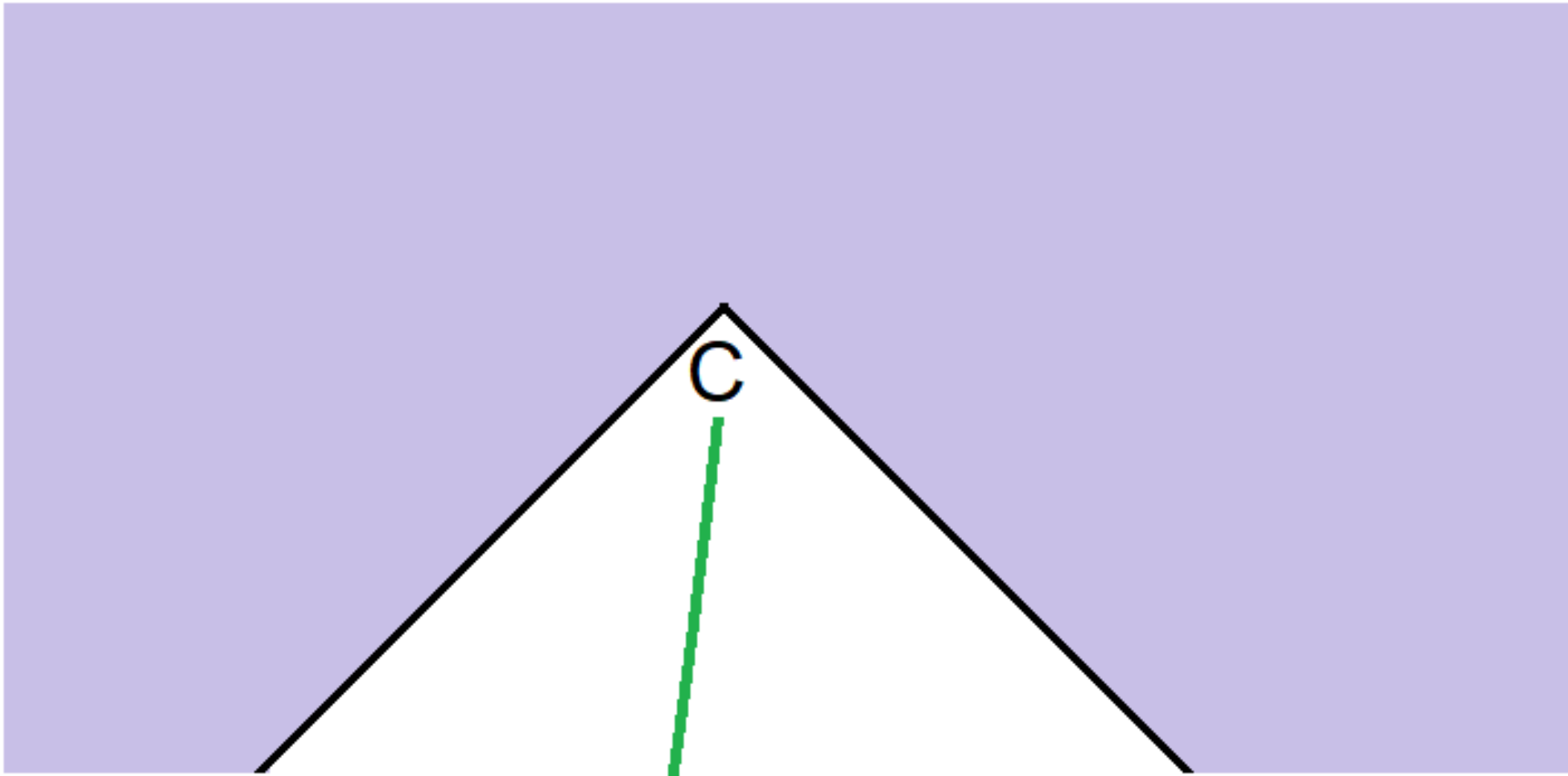}
\caption{}
\label{past}
\end{center}
\end{figure}

\begin{figure}[H]
\begin{center}
\includegraphics[scale=.3]{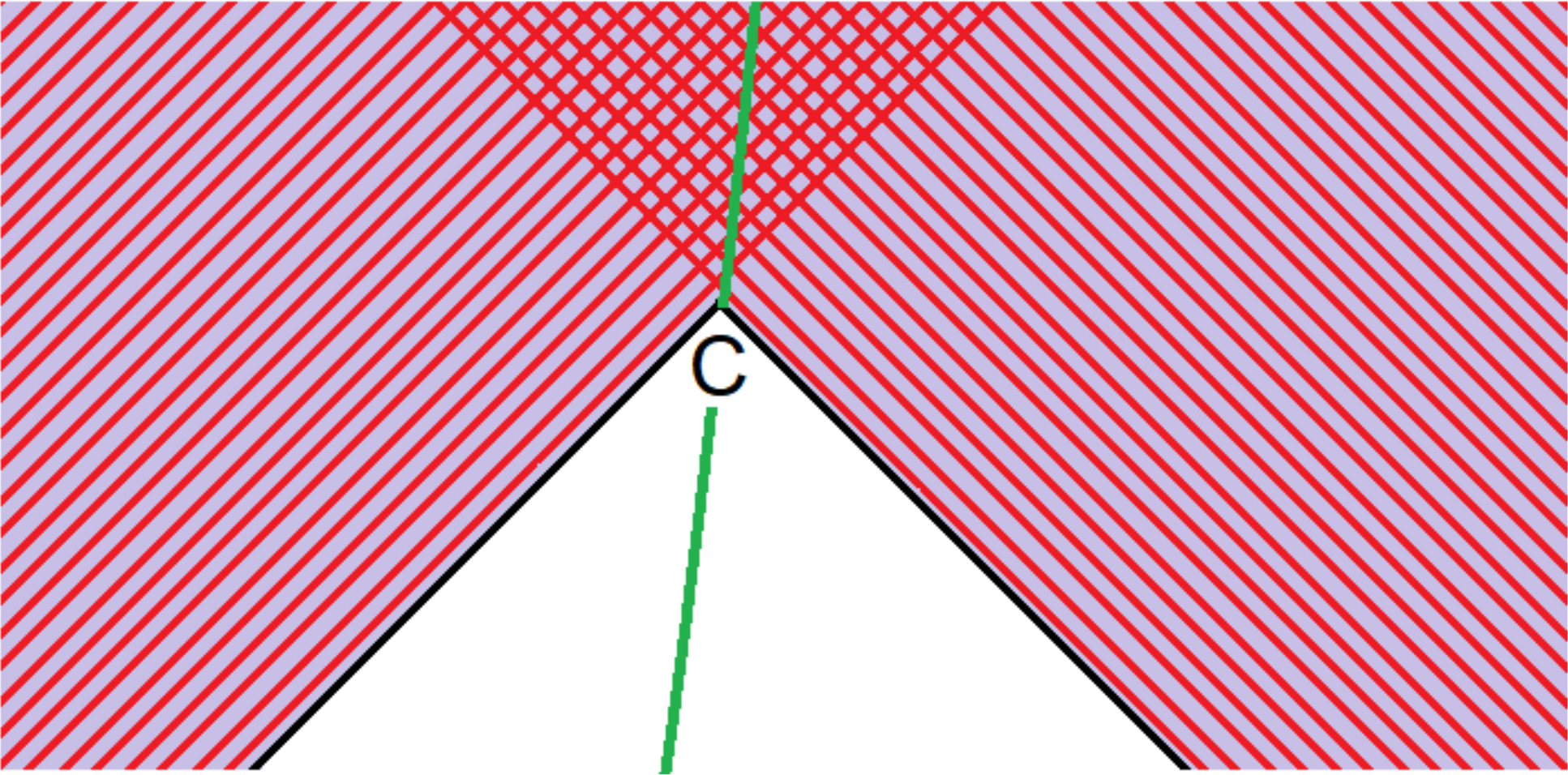}
\caption{}
\label{blast}
\end{center}
\end{figure}

That's the typical state argument. Before discussing its relation with complexity I want to compare it to a  similar cosmological argument. Consider Charlie whose world line arrives at spacetime point $C$ in figure \ref{past}.
He has a pretty good idea of what's in his past light-cone but in principle, he cannot know what is in the shaded region outside his causal past. Charlie would like to know what will happen to him in the next instant. So he averages over his ignorance, or to put it another way, he Haar-averages over all states that agree with his past experience. What he finds is very frightening. Almost all states consistent with Charlie's past contain a tremendous blast of super-high energy particles just outside his current past light-cone as in figure \ref{blast}.
\bn
So with virtual certainty he concludes that he will be burned to a crisp in the next nanosecond.

It's obvious what is wrong with this conclusion. Charlie needs a better cosmological theory of how his past is embedded in a larger spacetime. I won't go further into the details of cosmology here, but any reasonable theory of how Charlie got to his current location at $C$, the state of the world at that time is very exceptional---very un-Haar-typical.

The same thing is wrong with the the typical state theory of firewalls. I will explain why it is very likely that a Haar-random state of a condensed black-hole-like object \it does \rm have a firewall. But that ignores the question of how the black hole got there. We will see that any reasonable process of formation  of a black hole---stellar collapse is an example--- will give rise to a very atypical state.

\bn

Extremely energetic shock waves are certainly possible. Alice can create one on Bob's side by throwing in a thermal photon long in the past, at $t=t_w.$ To follow the complexity as a function of Bob's time we
return to the master formula \ref{master-formula}, this time   holding $\t_L$ fixed. We will  consider the complexity as a function of $\tau_R.$ In doing so we will see an surprising feature of  formula \ref{master-formula} . 
For $\tau_L = 0$,
\be 
V= const \log{  \{
\cosh\frac{\tau_R}{2} + e^{(|\tau_w|-\tau_*) - \frac{1}{•2}\tau_R}
\}
}
\label{tL=0}
\ee
Assume that $|\tau_w|>> \tau_*$ in which case the shock wave will be super-Planckian (recall the the energy of the shock wave grows exponentially with $|t_w|$). For $\tau_R$ not too large the second term dominates so that even at positive time the complexity \it decreases \rm as a function of $\tau_R.$ This is somewhat strange because  decreasing complexity violates the second law of complexity. Of course this is not really a law but a statistical tendency, but it may seem surprising that the mere act of throwing in a thermal quantum on the left can lead to decreasing complexity on the right. That seems too easy; violating the second law should be harder than that.

The point is that it \it is \rm hard---very hard. To create the shock wave we either had to start with the white hole in the remote past---an extremely fine tuned thing to do---or we have to apply a precursor which effectively means running the system backward for a long period of time. This is also very hard and unstable.
Evidently creating a firewall requires reversing the arrow of time and that is bound to be extraordinarily difficult.

The full history of the eternal black hole  generally has decreasing complexity in the white hole region. That is obvious from figure \ref{Foliated-BTZ2}, but the decrease is in the negative time region. In the shock wave case the decreasing  complexity  extends deep into the positive time region on Bob's side. 
It does eventually turn around when the two terms in \ref{tL=0} cross over at, $$t_R = |t_w| -t_*.$$  At that time the complexity begins to increase, but it only achieves its normal rate of increase far in the future  when $t = |t_w|.$

\begin{figure}[H]
\begin{center}
\includegraphics[scale=.3]{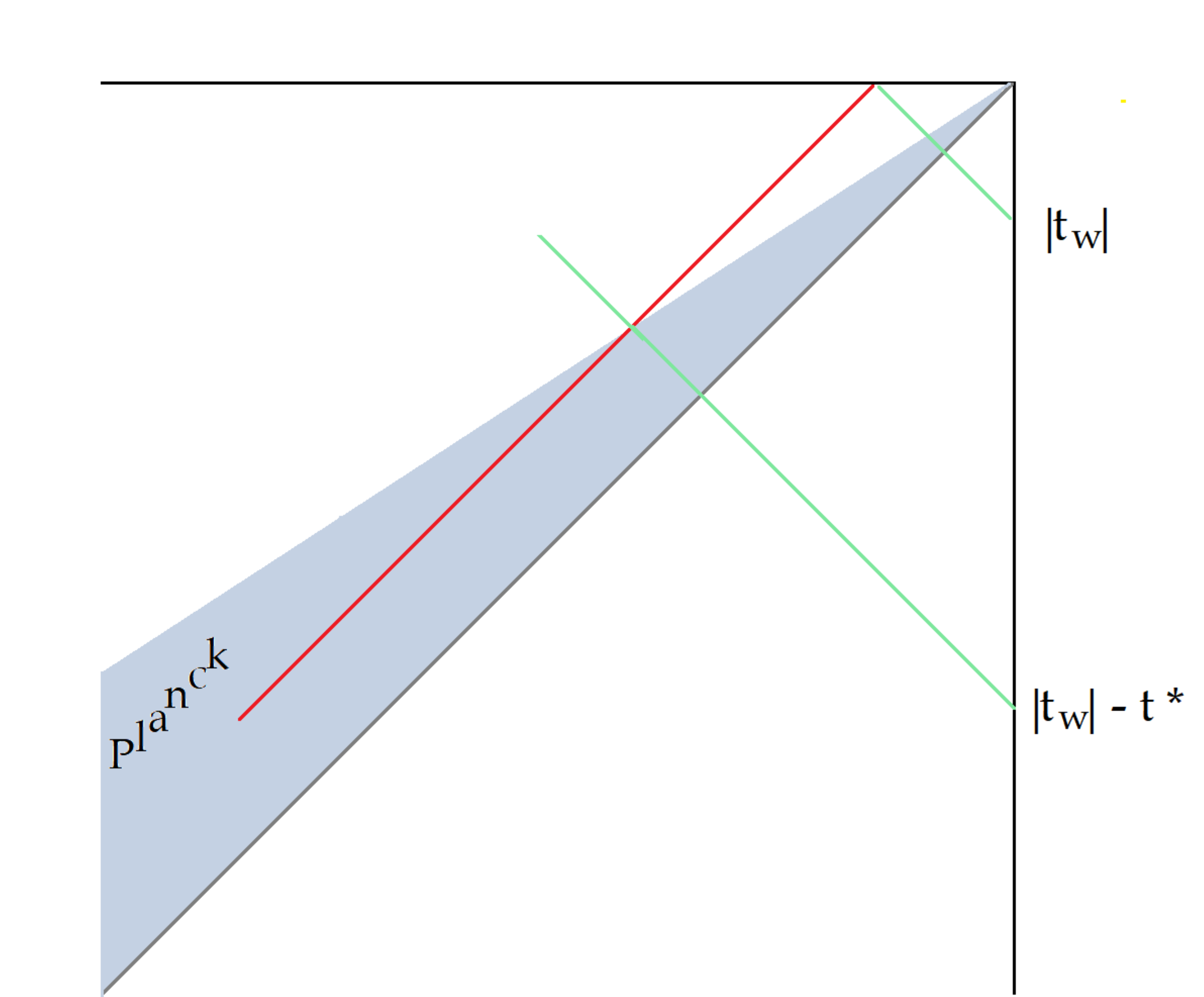}
\caption{An initially super-Planckian shockwave will eventually separate from the Planckian region and fall into the singularity. This occurs just at the Eddington-Finkelstein time when the complexity stops decreasing.}
\label{firewall}
\end{center}
\end{figure}

This behavior  is quite striking as can be seen from figure \ref{firewall}.
\bn
The grey region is within a Planck distance of the horizon as seen by an infalling observer. Any shockwave in that region is superplanckian. The red line is a shockwave which crosses out of the Planckian region and then falls into the singularity. The green lines are constant Edington-Finkelstein time slices. What we see is that the complexity decreases until the shockwave moves out of the Planckian region. As the shock wave approaches the singularity the complexity resumes its normal rate of growth, Zhao, \url{https://arxiv.org/pdf/1702.03957.pdf}.

The decrease of complexity is closely correlated with the location of the shock wave. One can see from figure \ref{shock-wave} that the more negative $t_w$ is, the closer to the right-side future horizon the shock wave will be. For very large $|\tau_w|$ the shock wave stays within a Planck distance of the horizon for a very long time before separating and falling into the singularity. In fact if we use Eddington-Finkelstein infalling coordinates for the right-side black hole, we find that the complexity turns around and starts to increase precisely when the shock wave separates from Planckian distance from the horizon.

\begin{figure}[H]
\begin{center}
\includegraphics[scale=.4]{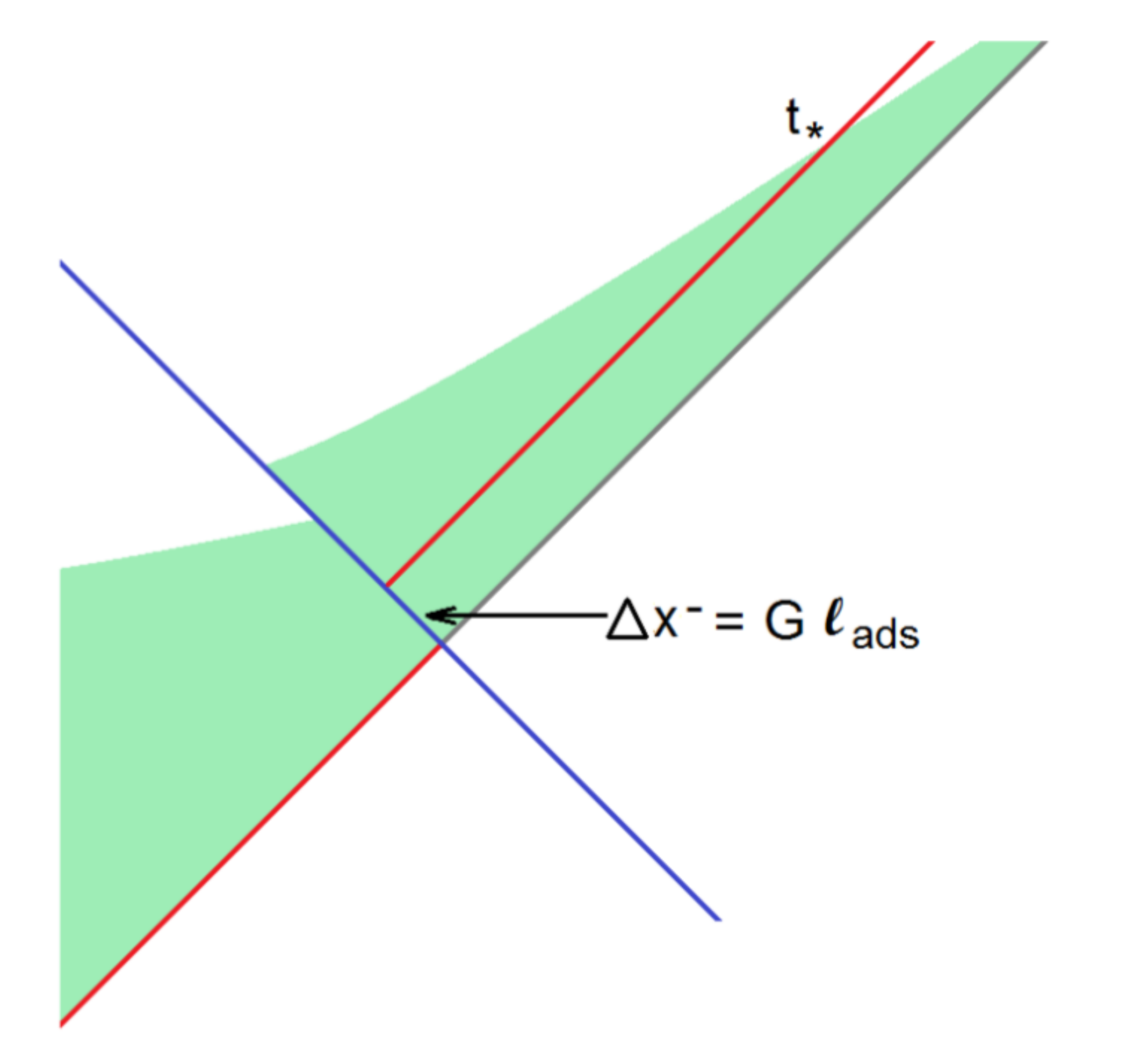}
\caption{An intially super-Planckian shockwave will be ``kicked" into the singularity by a single thermal quantum. The green region is the  Plankian region.}
\label{kick}
\end{center}
\end{figure}

Because the firewall is associated with decreasing complexity one might expect it to be unstable.  Let's suppose that $t_w$ is very far in the remote past and that Bob wants to jump into the black hole without being destroyed by the shock wave. All he has to do is throw in a single thermal photon. The result is shown in figure \ref{kick}.

\subsection{Firewalls are Fragile}\label{Sec: fragile FW}
The photon and the shock wave will undergo a high energy collision. On can think about the collision from the point of view of a frame in which the photon is high energy and the shock wave is low energy. In other words we can exchange the role of the photon and the shock wave. In that frame the infalling photon causes a Dray-'t Hooft shift of the original shock wave, kicking it out of the Planckian region and pushing it deeper into the black hole. The calculation is easy and one finds that no matter how high the energy of the original shock wave, it will fall into the singularity approximately a scrambling time after the photon is thrown in.
The lesson is that shock waves or firewalls are very fragile and can be destroyed by a single thermal photon. If Bob throws in the photon and waits a scrambling time he can be sure that the horizon will be transparent.

Lecture III is all about the counter-intuitive power of a single qubit---in this case a thermal photon---to accomplish what seem like impossible tasks. One of those tasks is destroying a firewall of tremendous energy\footnote{The energy of a shock wave grows exponentially with $|t_w|$. For reasons explained in Section \ref{expTbreakdown} \  the classical shock wave analysis holds up to exponential time $\exp S.$  A shock wave produced at that time has doubly exponential energy $\exp \exp S$ in the frame of an infalling observer.}
\bn

To get some sense for the orders of magnitude involved consider a solar mass black hole that has a super-Planckian shock wave behind the horizon. A single radio frequency photon with a wave length of a kilometer will destroy the firewall in a scrambling time which happens to be about a millisecond. A CMB microwave photon will destroy it even faster.

\bn

To summarize, firewalls are both very difficult to create and very fragile. To create one Alice must apply a precursor that effectively runs her black hole backward in time in order to create the shock wave. And to get rid of a firewall Bob only needs to drop in a thermal photon.

\subsection{What Happens After Exponential Time?}

Let's suppose that it is true that once complexity equilibrium is established, there is a high probability for a firewall. How stable is that firewall? The naive answer is that equilibrium is  stable and not likely to be significantly disturbed by butterflies. That is true for ordinary thermal equilibrium but complexity equilibrium is different.

Recall from Lecture I that the transition from growing complexity to equilibrium occurs abruptly when the complexity reaches its maximum $e^S$. Now suppose a thermal photon is  thrown in. It will increase the entropy by only one unit which seems negligible. But  the maximum complexity changes from  $e^S$ to $e^{S+\log{2}} = 2\times e^S.$
This means the black hole suddenly finds itself exponentially far from maximal complexity, with plenty of room for complexity to increase for another exponential time.

At exponentially large time, $t=\exp{K},$ the system reaches complexity-equilibrium and the complexity suddenly\footnote{At the end of the first lecture I explained why the transition is sudden.} stops growing. Something non-classical happens and the question is how does this affect an observer who falls through the (right) horizon? First let's ask the question for evolution of the exact TFD state.

The answer in this case is  that nothing special happens. The reason is that $|TFD\ra$ has a special symmetry, sometimes called time-translation symmetry.  But its not really time-translation; it's  boost symmetry in which $t_R$ and $t_L$ shift in opposite directions,
\bea
t_R &\to& t_R +c \cr \cr
t_L   &\to& t_L -c
\label{boost}
\eea
The transformation is generated by the difference of right and left Hamiltonians,
\be 
\rm Boost \ Generator \it \ = H_R-H_L
\ee

Let's call the operator that emits Bob from the boundary $B(t_R)$ and the time that he is emitted, $t_R = T$. The state after Bob has entered the geometry is
\be 
B(T) \ |TFD\ra.
\ee
The argument is  that a boost can transform this state  to
\be 
B(0) \ |TFD\ra,
\ee
so  that even if $T$ is exponentially large, it has no effect.

The thermofield double state is  very fine-tuned,  but we may ask what happens if we break the boost symmetry by a tiny amount. Again the answer is not much  if  $T$ is not very large, but if it is large all bets are off. 

\begin{figure}[H]
\begin{center}
\includegraphics[scale=.2]{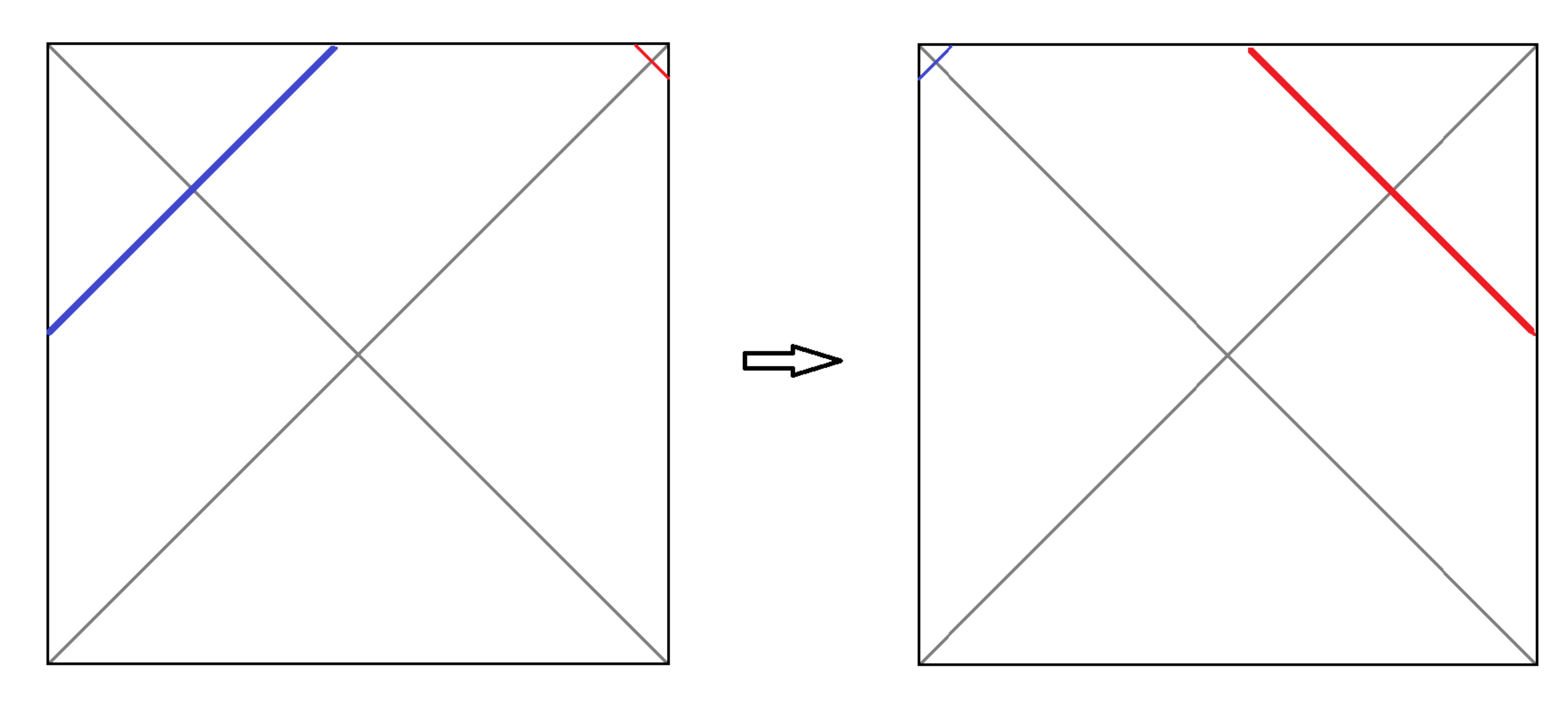}
\caption{}
\label{boost1}
\end{center}
\end{figure}

To break the symmetry let's suppose that Alice, on the left side, perturbs the system at $t_L=0$ with a thermal photon. She does this  by applying the operator $A(0)$. As before Bob jumps in at $t_R=T$. The state of the system becomes,
\be 
A(0)B(T) \ |TFD\ra
\ee
We can again transform Bob to $t_R=0$ but now Alice's perturbation is transformed to $t_L = T.$
\be 
A(0)B(T) \ |TFD\ra  \to  A(T)B(0) \ |TFD\ra
\label{ABTFD}
\ee
This is illustrated in figure \ref{boost1}.

We can write the rhs of  \ref{ABTFD} as
\be 
 A(T)B(0)  \ |TFD\ra = e^{-iH_LT}A(0)e^{iHT}B(0 ) \ |TFD\ra
 \label{ABTFD1}
\ee
This time we cannot eliminate Alice's perturbation, but we might think it is not important. It looks safely out of the way in the upper left corner of the diagram. To see that this argument is flawed we go to an extremely late  quantum recurrence time. The point of the quantum recurrence time is that $e^{-iHT}$ can be arbitrarily close to the unit operator $I$. In that case \ref{ABTFD1} becomes,
\be 
 A(T)B(0) \ |TFD\ra =  A(0)B(0) \ |TFD\ra
 \label{ABTFD2}
\ee
In fact if we make $T$ less than the recurrence time by twice the scrambling time the state becomes
\be 
 A(T)B(0) \ |TFD\ra =  A(-2t_*)B(0) \ |TFD\ra.
 \label{ABTFD3}
\ee

The state \ref{ABTFD3} is not benign; it is extremely dangerous to Bob. It contains a very high energy shock wave created by $A(-2t_*)$ which Bob will experience as a firewall.
This is an existence proof that if the boost symmetry is slightly broken then Bob's classical expectations of a transparent horizon will be wrong if he jumps in late enough at just the wrong time.

I don't know what happens at the earlier singly exponential time when complexity stops increasing. I'd like to show that firewalls have a high probability of occurrence but I can't. But I have also been unable to show that firewalls don't occur. The Marolf-Polchinski generic state argument suggests that they do and I will accept that.

\subsection{The Fragility of Complexity Equilibrium}

Let's  take the default position: If there is no reason to believe that  firewalls don't occur when the complexity stops increasing, then they probably do occur.
 How stable would such a firewall be? The naive answer is that equilibrium is  stable and not likely to be significantly disturbed by butterflies. 
 That is true for ordinary thermal equilibrium, say of a gas of $N$ classical molecules. 
 
 Thermal equilibrium is a state of maximum entropy in which the entropy is proportional to the number of molecules,
 \be 
 S_{max}\sim N.
 \ee
Now lets perturb the state by adding a single molecule. The result will be that the maximum entropy will be increased by one bit,
\be 
S_{max} \to N+1.
\ee

Although initially out of equilibrium, the new molecule will  come to equilibrium with the new entropy being $\sim N+1.$ This will typically take a very short time.

Let's compare this with what happens if we add a single qubit to a system of $K$ qubits in complexity equilibrium. The maximum complexity of the original $K$ qubit system is $2^K.$ Adding a qubit brings the maximum complexity to $2^{(K+1)}= 2\times 2^K$. This means that the new state is far out of complexity equilibrium; the complexity can increase for an additional amount $\sim 2^K.$
  
Now let's return to the black hole in complexity equilibrium at maximal complexity. If we drop in a single thermal photon the entropy increases by one bit, but the maximal complexity doubles.
The system has been kicked far out of complexity equilibrium by the addition of a thermal photon.
To return to equilibrium a minimum of $2^K$ gates must act, and this will take an exponential time. By dropping in the thermal photon we buy an additional exponential time during which the complexity increases. Complexity equilibrium is evidently very fragile.

There is another way to think about this. If indeed a black hole in complexity equilibrium has a firewall at its horizon, throwing in a thermal photon will   have the same effect as in section \ref{Sec: fragile FW}. The photon will kick the shock wave  and push it into the singularity. The horizon will be clear and the wormhole will be free to grow again for another exponential time.

\bn

\section{Do Typical States have Firewalls?}

Does a typical state of  black hole entangled with a second \it purifying \rm system have a firewall? The answer is subtle and depends on what we mean by typical. There are a number of different situations.

\subsection{AdS Black Holes}
First of all for black holes in AdS which cannot evaporate, but which are entangled through wormholes to other AdS black holes:

\bi 
\item
Almost all states of the entangled system are maximally complex. If we accept the default position---without an argument against a firewall, a given maximally complex state has a high likelihood of having a firewall---then almost all states are likely to have firewalls. To be firewall-free the state of the system much be exceptional and out of complexity equilibrium.
\item 
Exceptional states are ones with complexity much less than maximum.  For these states there is room for complexity to increase at a \it normal rate.\rm \ 
The fraction of states which are exceptional in this sense is very small. Over long time scales the system remains in complexity equilibrium for doubly exponential time-periods and only out of equilibrium for singly exponential periods. Thus the fraction of out of equilibrium states is incredibly small,
\be 
f \sim \frac{e^S}{e^{e^S}}.
\ee

\ei
Evidently the overwhelming majority of states may  have firewalls, just as  Marolf and Polchinski 	claimed 
\url{https://arxiv.org/pdf/1307.4706.pdf}.

\bn

But if we mean by typical, the kind of states that are typically created by ``natural" processes, such as stellar collapse or the collision of super energetic particles, then we come to the opposite conclusion. (By a natural process I mean one that takes a time  much less than $e^S.$ )

\bi 

\item Black holes (or anything else) that are formed by natural processes start our in  very low complexity states. This is obvious because to create a maximally complex state takes an exponential time. The initial complexity for such  a normally formed black hole is probably a good deal less than $S$, the entropy.

\item The second law of complexity implies that
 with overwhelming probability the complexity will increase for a very long time, $\sim \exp{S}$, before complexity equilibrium is established. 
 
 \item During normal periods of complexity increase the horizon is transparent with no firewall blocking entrance to the black hole.
 
\ei
 
Thus while it is likely that  Haar-typical states overwhelmingly have firewalls, naturally occurring states are exceptional. They start life with very low complexity and  second law of complexity ensures that their complexity grows so that they are firewall-free for an exponentially long time.

 \subsection{ Evaporating Black Holes}

 Finally there are black holes in asymptotically flat space that can evaporate. Such black holes may start out in low complexity pure states. As they evaporate and shrink they become entangled with their own evaporation products, so that by the Page time the entanglement is close to maximal. One might think that as the black hole becomes small the complexity must decrease,  leading to a firewall. This is not correct; complexity increases throughout the entire evaporation process.

 In the two-sided eternal AdS black hole, chaotic interactions  at both ends increase the complexity at a rate twice as fast as for a one-sided black hole. In the evaporating case the second side is replaced by the outgoing Hawking radiation.  Because it is free-streaming the radiation does not undergo chaotic interactions. It is as if the left-side (the radiation side) Hamiltonian were turned off. The wormhole still grows but only from the right side. At any given time the rate of growth is $ST,$ which for conventional \S \ black holes is proportional to the mass $M$,
 $$d\CC/dt \sim M.$$
 
 In the case of an evaporating black hole the mass is time dependent. In Planck units the mass evolves according to,
 \be 
 M(t)^3 = M_0^3 -t
 \ee 
 where $M_0$ is the initial mass of the black hole. 
 
 What goes to zero as the black hole shrinks is not the complexity but the rate of complexity growth.  The complexity growth rate satisfies,
\be 
d\CC/dt \sim M(t) = (M_0^3-t)^{1/3},
\ee
giving,
\be 
\CC(t) \sim   M_0^4 -  (M_0^3 -t)^{4/3} 
\ee
The complexity of the combined state of the black hole and the radiation monotonically grows until the black hole has evaporated. Thereafter it is constant.

But what is it that is complex at the end of the process? It can't be the black hole: it has disappeared.  Obviously the complexity resides in the  final pure state of the radiation. In Planck units the final complexity is of order $M_0^3$ or equivalently,
\be 
\CC_{final} \sim S^2.
\ee
where $S$ is the initial entropy of the black hole.
This is a rather complex state, but  very far from maximally complex.
 
At no point during the evaporation was the complexity decreasing. The wormhole or Einstein-Rosen bridge connecting the black hole to the radiation continues to grow despite the fact that the black hole shrinks, until the evaporation process has been completed. There is no reason to think that a firewall ever materializes.

\bn

This brings us to the end of Lecture II. In the next lecture we will study the thermodynamics of complexity.

%%%%%%%%%%%%%%%%%%%%%%%%%%%%%%%%%%%%%%%%%%%%%%%%%%%%%%%%%%%%%%

%%%%%%%%%%%%%%%%%%%%%%%%%%%%%%%%%%%%%%%%%%%%%%%%%%%%%%%%%%%%%

%\begin{titlepage}

\rightline{}
\bigskip
\bigskip\bigskip\bigskip\bigskip
\bigskip

\part{Lecture III: The Thermodynamics of Complexity}

\section{Preface}

This lecture is about the thermodynamics of complexity---thermodynamics as nineteenth century steam engineers would have understood the term. The question of that day was: How do you quantify the amount of mechanical work that can be extracted from a given  amount of energy? Entropy was of course the key concept. Entropy, or more properly, the lack of entropy is a \it resource \rm for doing work.

What exactly is a resource? There is whole subject called resource theory  (Coecke,  Fritz, Spekkens, 
\url{https://arxiv.org/pdf/1409.5531.pdf}
 that we won't have time to get into, but I'll give a quick definition. First of all a resource for the purpose of doing $X$ is something that you need in order to do $X$. It is necessary, but not generally sufficient.

Secondly, accomplishing $X$ expends part of the resource. The more X you do, the less X you can do without renewing the resource. There is much more to resource theory but that's all we'll need today.

It was Schrodinger who first introduced  \it negentropy \rm as a useful resource. He  defined it to be minus the entropy. We'll need to be a little more precise.
  Suppose we have a certain amount of energy available, and also suppose the entropy of the system is $S$. The system may be out of equilibrium, in which case $S$ will be less than the maximal entropy (consistent with that amount of energy). The negentropy is the difference between the maximum entropy and the actual entropy,

$$
\rm negentropy \it = S_{max} -S
$$
Negentropy is the room for entropy to grow.
Whenever negentropy is non-zero, some  amount of work can be extracted, but in doing the work one must expend negentropy. When all the negentropy is used up a system will be in thermal equilibrium, and no further work can be extracted.

I explained in Lecture I that complexity is a lot like entropy, not the usual entropy of a quantum system but the entropy of an auxiliary classical system. The auxiliary system describes the motion of a state-vector in Hilbert space, or the motion of the time-development operator $U(t)$ on \Suk. It is natural to ask whether there is a concept analogous to negentropy  that represents a resource, and for  what purpose  is it a resource?

\section{Negentropy}

\bn

Let's begin with some examples of negentropy. Consider a box of gas composed of $n$ classical particles. The microstate of the box is defined by $n$ points in the phase space of a single particle. The macrostate can be described thermodynamically in terms of a much smaller number of thermodynamic variables including entropy.
If the entropy is maximal (subject to some constraint) then the gas is in equilibrium and the negentropy is zero. No work can be extracted.

\bn

\begin{figure}[H]
\begin{center}
\includegraphics[scale=.2]{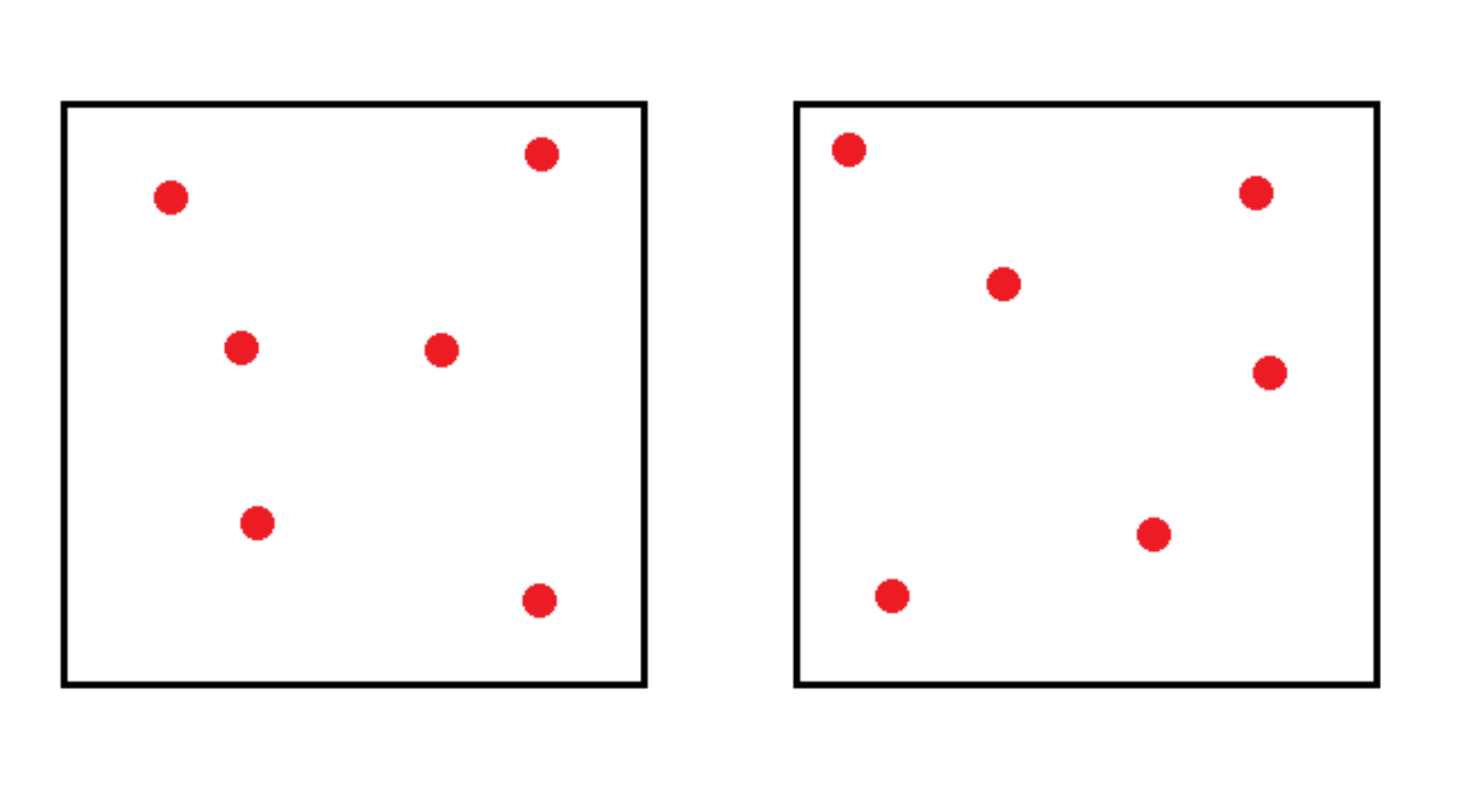}
\caption{}
\label{2box-uncor}
\end{center}
\end{figure}

Now suppose we have two thermodynamically identical boxes of gas, where each box is in  thermal equilibrium.  Let's also assume the microstates are uncorrelated except for the fact that they are macroscopically similar. See figure \ref{2box-uncor} 
The entropy of either box is $S_{max}$ and the total entropy is 
$2S_{max}$. Moreover the maximum total entropy is also 
$2S_{max}$. It follows that the negentropy is zero and no work can be be extracted.

\begin{figure}[H]
\begin{center}
\includegraphics[scale=.2]{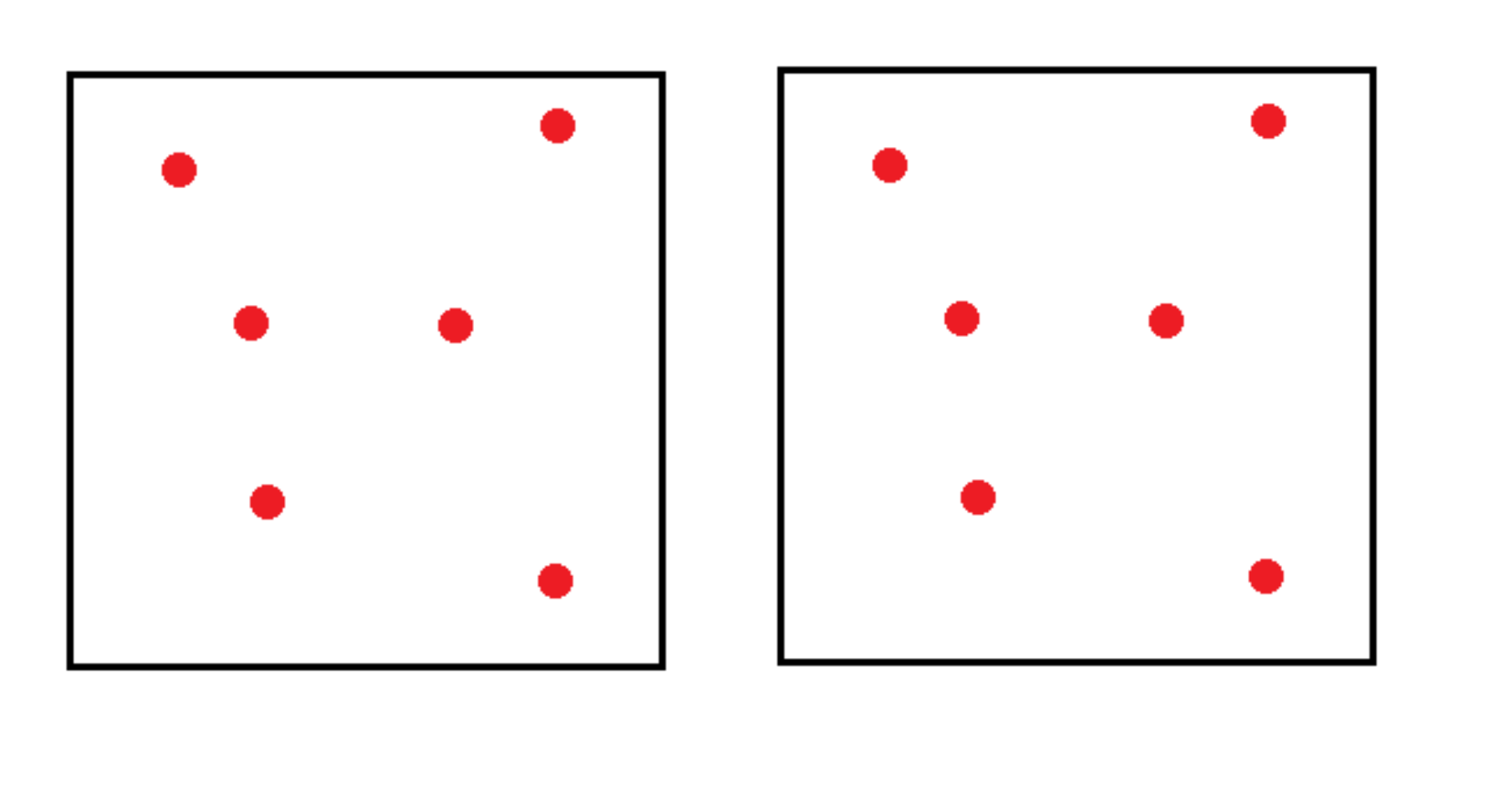}
\caption{}
\label{2box-cor}
\end{center}
\end{figure}

Now let's compare that with another example involving the same two boxes, each in equilibrium, but this time the microstates are \it identical \rm as illustrated in figure \ref{2box-cor} .
Each box is in equilibrium but the combined system has much less entropy than the previous example. In fact the total entropy is exactly the same as for the individual boxes,
\bea 
S(total) \eq S_{max} \cr \cr
S_{max}(total) \eq 2S_{max} \cr \cr
\rm Negentropy \it &=& S_{max}
\eea
In other words the high degree of correlation between the microstates gives rise to  negentropy, a resource which in principle can be exploited to do a significant amount of work.

\bn

\bf Exercise: \rm
Design a machine to extract work from this setup.

 Then patent it.

\bn

\section{Uncomplexity}

\subsection{The Auxiliary System}

Instead of using operator complexity I'm going to work with state complexity for the rest of this lecture. State complexity is the number of gates that it takes to go from a simple unentangled state of $K$ qubits to the state in question. 

Physical states can be described as normalized vectors in the Hilbert space of $K$ qubits with the phase modded out. In other words they are points on $CP(2^K - 1).$ The evolution of a state is represented by an ``auxilliary system" \url{https://arxiv.org/pdf/1701.01107.pdf}---a fictitious classical system moving  on $CP(2^K - 1).$

One way of representing  a state is by the coefficients in the expansion of the state in some basis---say the computational basis. Thus a state of $K$ qubits is labeled by $2^K$ complex numbers,
\be 
|\psi\ra \to (\alpha_1, \alpha_2,......\alpha_{2^K})
\label{psi}
\ee
The $\alpha$ can be thought of as  coordinates of a point in the phase space of the Auxiliary system.
Since the dimension of the space is $2^K$ we can roughly picture the system as a classical collection of $2^K$ particles. That's more than enough to do statistical mechanics. 

As I explained in Lecture I, the entropy of this \it auxiliary \rm
system is the average complexity of an ensemble of states. The second law of complexity (analogous to the second law of thermodynamics) states that the complexity tends to increase
\url{https://arxiv.org/pdf/1701.01107.pdf} . In fact the second law of complexity is the second law of thermodynamics applied to the auxiliary system.
As in the case of thermodynamics, when the number of degrees of freedom becomes large the tendency for complexity to increase becomes overwhelming.

Starting with a simple unentangled state and evolving it for an exponential time will bring the state to  complexity equilibrium (maximal complexity) analogous to thermal equilibrium. Complexity equilibrium is a far more thorough mixing of degrees of freedom than thermal equilibrium and takes a much longer time to be established. Being \it out of complexity equilibrium \rm  is much more subtle than being out of thermal equilibrium. Thermal equilibrium is about local easily measured correlations: complexity equilibrium is about extremely subtle global correlations which are very hard to detect.
A system can easily be in thermal equilibrium and still very far from complexity equilibrium.

The parallel with entropy  suggests that we define a resource analogous to negentropy. The term that  Adam Brown and I coined for this resource is \it uncomplexity. \rm We might have called it Negcomplexity but we didn't. Some people have suggested the term simplicity, but a state can have lots of uncomplexity and still be very complex.

 Uncomplexity is defined by analogy with Negentropy, 
\be 
\rm Uncomplexity \it = \CC_{max} - C
\ee
where for a system of $K$ qubits the maximum state complexity is $2^K.$ Just as negentropy is the room for entropy to grow, uncomplexity is the room for complexity to grow. 
The idea is that uncomplexity is a resource for doing some kind of  work. In this case work does not refer to lifting a weight, but of doing directed computation, i.e.,
computation with a goal. 

\subsection{Combining Auxiliary Systems}

In thermodynamics it is important to understand how systems combine, as for example, when we combined two boxes of gas.  The question arises: What does it mean to combine two auxiliary systems? For example suppose $\CA$ is the auxiliary system corresponding to some quantum system $\CQ$. Is there a quantum system whose auxiliary is two copies of $\CA$? The first thing that comes to mind is to simply double the quantum system from $\CQ$ to two copies $\CQ \cup \CQ.$
However the number of coordinates describing the Auxiliary system for $\CQ \cup \CQ$ is not twice the number describing the original system $\CQ$. It is the square of that number.
Evidently doubling the quantum system is not doubling the auxiliary system.

In fact the right thing to do if we want to double the number of components in \ref{psi} is to add just one qubit. Let the computational basis for the additional qubit be $|0\ra, |1\ra.$
Any state can be written in the  form,
\be 
|\Psi\ra = |0\ra |\psi_0\ra + |1\ra |\psi_1\ra .
\ee
This obviously doubles the number of components in \ref{psi},
\be 
|\Psi\ra \to (\alpha_1, \alpha_2,......\alpha_{2^K}, \beta_1, \beta_2,......\beta_{2^K}),
\ee
Thus doubling the number of coordinates describing the auxiliary system.
If we think of the coefficients as coordinates of fictitious particles, then the number of particles is doubled by adding
the extra qubit!

With this in mind let us build an analogy with the two boxes of gas in the previous section. First consider a state,
\be 
|\Psi\ra = |0\ra|max_0\ra + |1\ra|max_1\ra
\label{Psi1}
\ee
where $|\max_{0,1}\ra$ are maximally complex states of $K$ qubits. Furthermore  assume that $|\max_{0}\ra$ and $|\max_{1}\ra$ have maximal relative complexity (see lecture I). This latter condition is the anolog of the two gasses in figure \ref{2box-uncor} being completely uncorrelated.

The maximum complexity of states $|\Psi \ra$ is $2^{K+1} = 2\times 2^K$. It is also true that if the relative complexity of $|\max_{0}\ra $ and $|\max_{1}\ra$ is maximal then the actual complexity of $|\Psi\ra$ is maximal, i.e., $2^{K+1}$. It follows that the uncomplexity of \ref{Psi1} is zero. If indeed uncomplexity is a necessary resource for some purpose, then the state \ref{Psi1} is useless for that purpose.

\bn

Next consider a similar state except that $|\max_1\ra$ and $|\max_2\ra$ are identical. In this case the single qubit is unentangled with the other $K$ qubits: it is said to be  ``clean." 
The $\alpha_i$ and $\beta_i$ are equal to each other and we have a situation which is analogous to  the one in figure \ref{2box-cor} where the two boxes were in identical microstates. 
Each box of gas in figure \ref{2box-cor} was in equilibrium, but the combined system was far from equilibrium. 

It is very easy to see that if $|max_1\ra = |max_2\ra$ then the complexity of the combined system of $(K+1)$ qubits is the same as the complexity of either component, i.e., $2^K$. 
Evidently, by adding a  single qubit to a maximally complex state we have doubled the maximal complexity but have not changed the actual complexity. Thus the addition of  a single clean qubit  restores the uncomplexity to the same value that one box of gas would have if it were in a simple unentangled state.

\section{Uncomplexity as a Resource}

In discussing computational work it is helpful to separate what can be done by reversible (unitary) operations from what can be done by applying measurements. Starting with a maximally complex state, or any other state, the complexity can be eliminated by  measuring  all Pauli $Z$ operators. The outcome will be a unentangled product state which can be brought to the state $|000...0\ra$ by applying no more than $K$ single qubit gates. This would have the effect of restoring the uncomplexity to its maximal value\footnote{Notice that we did the same thing by adding a single clean qubit.}.

But measurements are not reversible operations, and necessarily dump heat into the environment. By computational work I will mean the part of a protocol that can be achieved reversibly. For example we may want to reversibly transform the state from some input to a target state that has some information that we can later extract by making a measurement. The computational work by definition is associated with the \it reversible transformation only, \rm and not with the \it final measurement.
\rm

For most purposes maximally complex pure states cannot be distinguished from maximally mixed density matrices. Both will give random results for almost all measurements. Since a reversible operation on a maximally mixed state does nothing, maximally mixed and maximally complex states are useless as a starting point for doing computational work.

Thus to do computational work we need some uncomplexity, i.e., some separation between the actual complexity of a computer, and maximal possible complexity $2^K.$ Uncomplexity is therefore a necessary resource, but it is not generally sufficient. What I am going to show you, by a concrete example, is how adding a single clean  qubit to a maximally complex system of $K$ qubits restores the ability to do a great deal of computational work. 

\section{The Power of One Clean Qubit}

That a resource analogous to negentropy can be vastly increased by merely adding a single qubit seems very counterintuitive. If I told you that by adding a single molecule to a gas of $10^{30}$ molecules in thermal equilibrium, it  would allow you to extract a large amount of useful work you would probably dismiss the claim as nonsense. You would of course be right. When I first thought about the fact that a single qubit can restore an exponential amount of uncomplexity I was skeptical about its meaning. So as I always do when I am confused about quantum information I went to Patrick Hayden. I asked him: Can adding one qubit to a computer significantly increase its power?

Patrick listened to my story and was less dismissive than I thought he would be. He said that it reminded him of something called \it one clean qubit computation. \rm \ After some Googling I found the original paper ``On the Power of One Bit of Quantum Information" by Ray Laflamme  (the same as in Gregory-Laflamme)  and Emanuel Knill (quant-ph/9802037). What I'm going to explain is a slightly modified version of their method.

%Let's begin with $K$ qubits in a Haar-random (maximally complex) state $|\max\ra$ with no uncomplexity. As I explained, such a state is essentially useless for  reversible directed computation. Now let us add a single clean  qubit in the state $|0\ra.$ This does not change the complexity of the state but it does double the maximum complexity. By adding the clean qubit the uncomplexity jumps to $2^K.$ 

To illustrate the power that we gain from a single clean qubit, let's consider the problem of computing the trace of a unitary operator in $SU(2^K).$ There are a number of problems that reduce to computing such a trace. Unfortunately computing the trace of a $2^K \times 2^K$ matrix is very hard for the simple reason that it means adding $2^K$ numbers. 

I will call the unitary whose trace is to be computed $G.$ $G$ itself has some complexity which is defined by the minimal number of gates satisfying 
\be 
G = g_n \ g_{n-1}....g_1.
\label{G=gg}
\ee
The object is to start with the maximally complex state of $K$ qubits, add a single qubit,
\be 
|max\ra |0\ra,
\ee
and then transform it to a state that has some accessible information about $\Tr G$. It would also be highly preferable to do it in a way that does not depend of the particular maximally entangled state $|max\ra$ since describing such a state is itself exponentially difficult.

\subsection{The Protocol}
Here is the circuit that accomplishes the task:
\begin{figure}[H]
\begin{center}
\includegraphics[scale=.2]{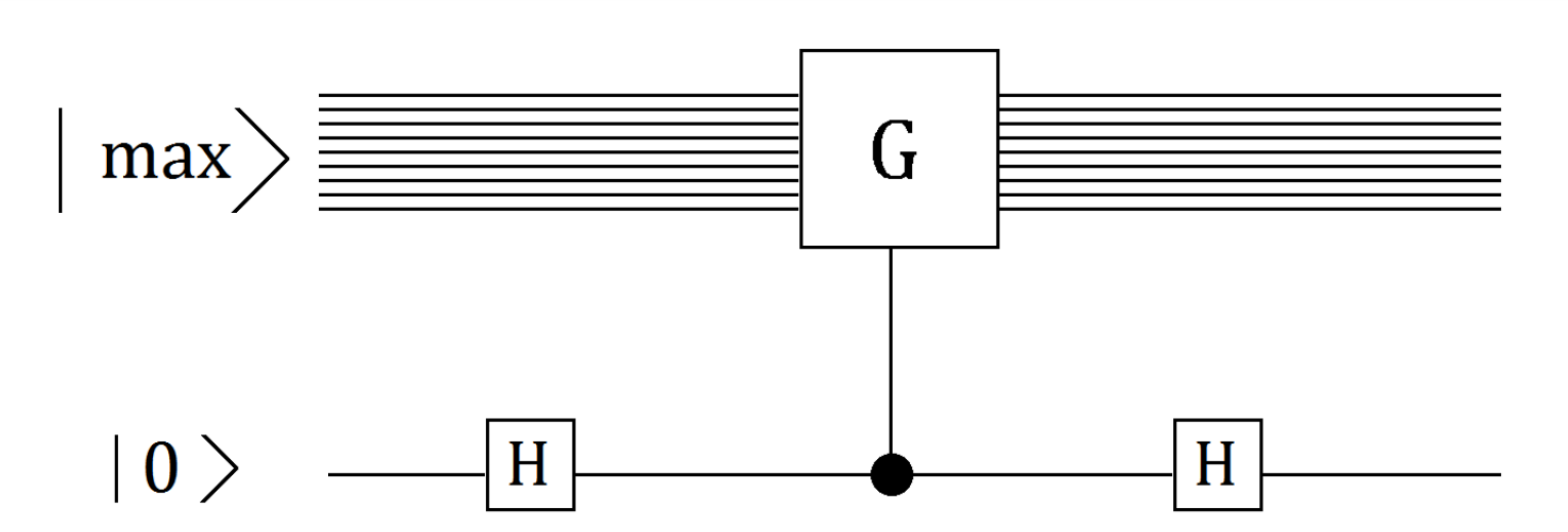}
\caption{}
\label{Clean1}
\end{center}
\end{figure}
Let me explain figure \ref{Clean1}. The initial state is,
\be 
|\Psi\ra  = |max\ra \otimes |0\ra
\ee
The single-qubit gates $H$ are Hadamard gates, which have the following action,
\bea
H |0\ra &=& \frac{1}{\sqrt 2} \big{(}|0\ra + |1\ra\big{)} \cr
H |1\ra &=& \frac{1}{\sqrt 2} \big{(}|0\ra - |1\ra\big{)}
\eea
The protocol begins with a Hadamard acting on the clean qubit to transform it to $\big{(}|0\ra + |1\ra\big{)}$.

Next, a ``controlled" $G$ operation acts. The dot at the bottom indicates a controlled operation: if the clean qubit is in the state $|0\ra$ then the unitary operator $G$ acts on the remaining qubits; if the clean qubit is in state $|1\ra$ then $G$ does not act and $|max\ra$ is left undisturbed. The last element of the circuit is another Hadamard.

It is straightforward to follow the circuit and see that the final state is,
\bea 
|\Psi'\ra &=&{ \frac{1}{2} } \big{(} G+1 \big{)} \otimes |0\ra \cr \cr
 &+& 
\frac{1}{2} \big{(} G-1 \big{)} \otimes |1\ra
\eea
Finally we compute the expectation value of the $Z$ Pauli operator for the clean qubit (which by the way, is no longer clean).
\be 
\la Z \ra = \la max| \left( \frac{G^{\dag}+G}{2}\right) |max\ra
\ee

We now make use of the fact that $|max\ra$ is maximally complex. A maximally complex state is Haar-random, which allows us to replace the expectation value by the normalized trace,

\be 
\la Z \ra = \Tr \left( \frac{G^{\dag}+G}{2}\right) 
\ee

Thus the circuit transforms the initial state to a final state that indeed has information about $\Tr G.$ Of course that is not the end of the story; we still have to measure $Z$ and repeat the procedure enough times to determine the $\la Z \ra$ but that is not part of the reversible process that we define to be the computational work.

It is also possible to determine $ \Tr \left( \frac{G^{\dag}-G}{2}\right) $ by a similar method which I will leave to you.

\begin{figure}[H]
\begin{center}
\includegraphics[scale=.2]{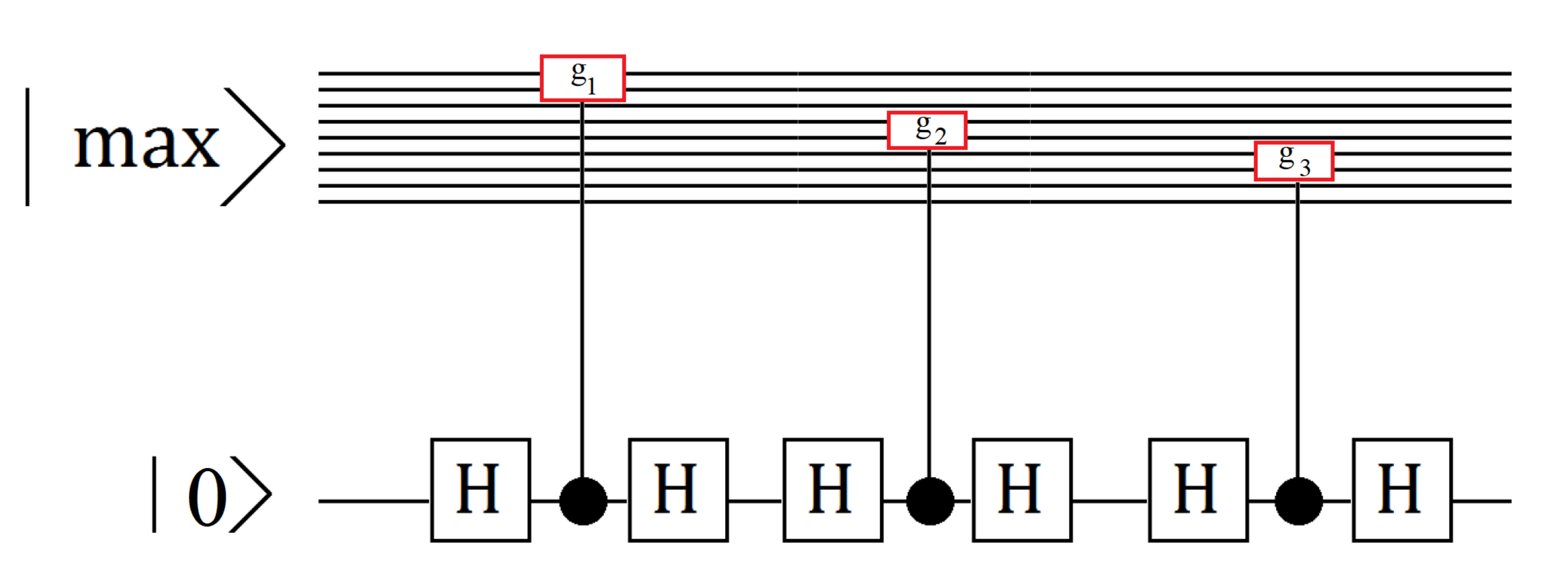}
\caption{}
\label{Clean2}
\end{center}
\end{figure}

One might complain that applying $G$ is not a \kl \ operation, but that is easy to repair assuming that $G$ can be written as a product of gates as in \ref{G=gg}. In that case the circuit in figure \ref{Clean1} can also be decomposed into local gates. This is shown in figure \ref{Clean2}.

The circuit can be simplified by using $H^2 =1$ to eliminate all the Hadamards except for the ones at the beginning and end.
I will leave you to ponder this circuit and show that it is the same as the original circuit in figure \ref{Clean1}. 

\subsection{Expending Uncomplexity and Negentropy}

The uncomplexity of $|max\ra$ is zero but by adding the clean qubit we bought ourselves an uncomplexity $2^K.$ But the action of the circuit will inevitably increase the complexity by roughly the number of applied gates. Assuming that \ref{G=gg} is the minimal description of $G$ then we can expect the increase of complexity to  be of order $n$ which is also the complexity of $G$. To put it another way, the application of the circuit decreases the uncomplexity from $2^K$ to $ \big( 2^K - \CC(G) \big)$. In other words the computational work in computing $\Tr G$ expends an uncomplexity  equal to the complexity of $G$.

In this example uncomplexity is not the only resource we made use of. A maximally complex state is also a state of maximum entropy which in this case is $S = K \log 2.$ The addition of the clean qubit does not change the entropy, but after the circuit has acted the clean qubit is no longer clean. This implies that the entropy increases but only by an amount less than or equal to one bit. Thus we have also expended Negentropy, but no more than one bit.

One bit of negentropy can be stored in a single qubit, but obviously a single qubit cannot  by itself be used to calculate $\Tr G$. Evidently in this case we needed both the large uncomplexity and a tiny bit of negentropy to do the computation.

\section{Spacetime and Uncomplexity}

Now I want to return to the gravitational aspects of complexity, or more to the point, the gravitational aspects of uncomplexity. 

\bn

\subsection{CA }
In these lectures I haven't spoken yet about the  the complexity-action correspondence  (D. Carmi, R. C. Myers and P. Rath,
\url{https://arxiv.org/pdf/1612.00433.pdf}
  \ \  A. Brown, D. A. Roberts, L. Susskind, B. Swingle and Y. Zhao,
  \url{https://arxiv.org/pdf/1509.07876.pdf} .
 That's a whole subject of its own but it's fairly well known. I will limit the discussion here to the  basic facts that I'll need to make contact with uncomplexity.

Let's begin with the idea of a Wheeler-DeWitt patch. I'll restrict the discussion to one-sided AdS black holes. Pick a time $t$ and consider all co-dimension $1$ surfaces anchored to the boundary at that time. The WdW patch is simply the union of all such surfaces. It's boundaries include light-like surfaces and   possibly pieces of space-like singularities. 

\begin{figure}[H]
\begin{center}
\includegraphics[scale=.4]{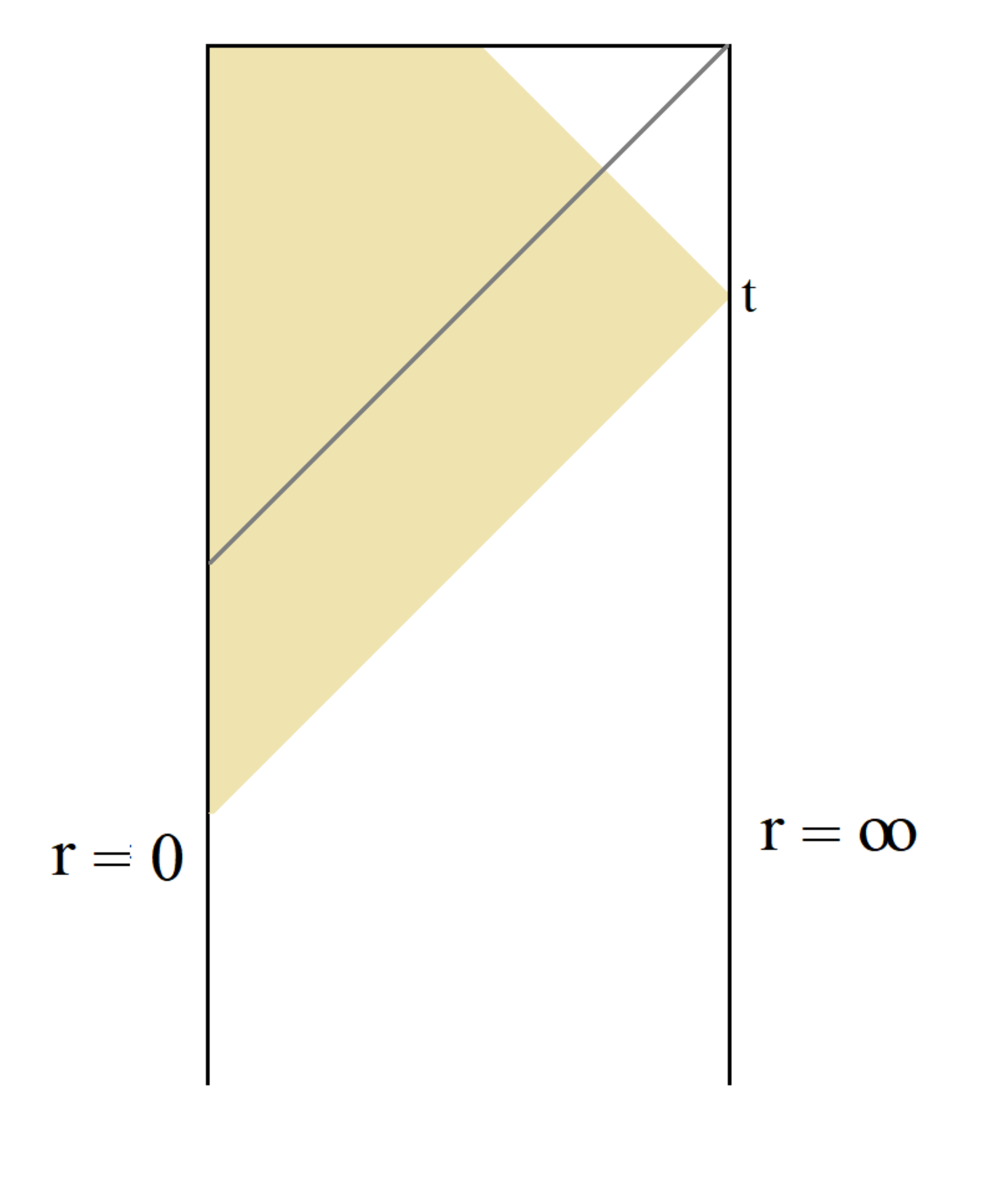}
\caption{}
\label{S0}
\end{center}
\end{figure}

The WDW patch consists of a portion outside the horizon and a portion  behind the horizon. The CA correspondence is that the complexity of the quantum state is the action Einstein-Hilbert action of the WDW patch. In the cases we will be interested in the action is simply proportional to the space-time volume of the WDW patch. The numerical factor relating the two will not concern us.

Generally the space-time volume diverges near the boundary, but the part that corresponds to the complexity of the black hole only involves the region behind the horizon. Thus we identify complexity with the space-time volume of the portion of WDW behind the horizon.

\subsection{Geometric Interpretation of Uncomplexity}
 
\begin{figure}[H]
\begin{center}
\includegraphics[scale=.2]{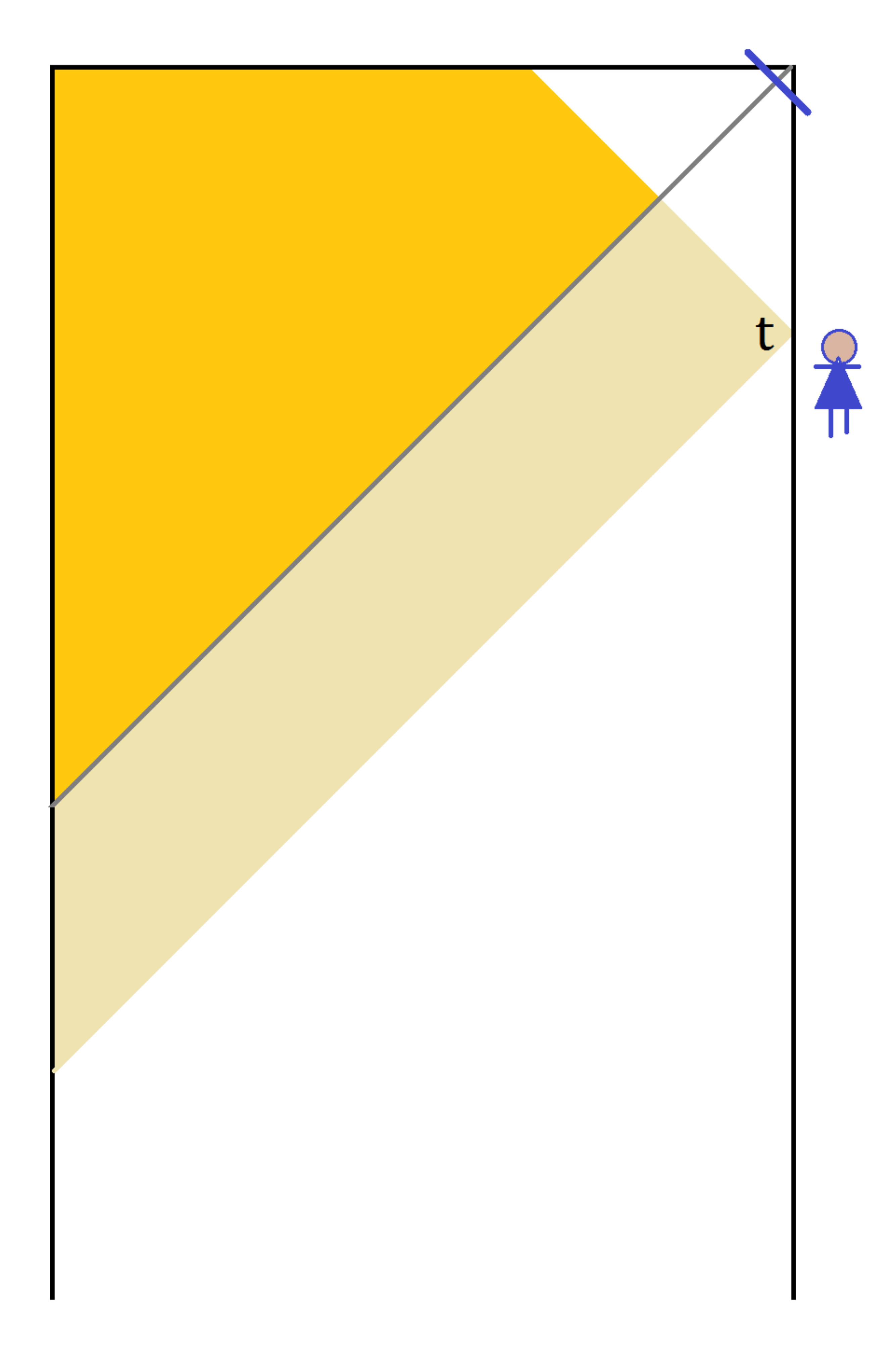}
\caption{}
\label{S1}
\end{center}
\end{figure}

Figure \ref{S1} illustrates a one-sided AdS black hole formed by the collapse of some matter. In the upper right corner the small slash-mark indicates the exponential time at which the complexity becomes maximal. In Lecture II  I explained the reasons for thinking that a breakdown of classical GR and possibly the occurrence of firewalls might take place at this  time.
The figure also shows Alice at time $t$ who is contemplating a jump into the black  hole. She would like to know how much spacetime volume is available to her to explore.

Finally the figure also shows the WDW patch at time $t$, and in particular the portion of the WDW patch behind the horizon.

Let's consider the question: What is the uncomplexity of the quantum state of the black hole at time $t$? To compute the uncomplexity we must first determine the complexity. According to the complexity-action duality the complexity is proportional to the action of the portion of the WDW patch behind the horizon, in other words the spacetime volume of the darker orange region. 

\begin{figure}[H]
\begin{center}
\includegraphics[scale=.2]{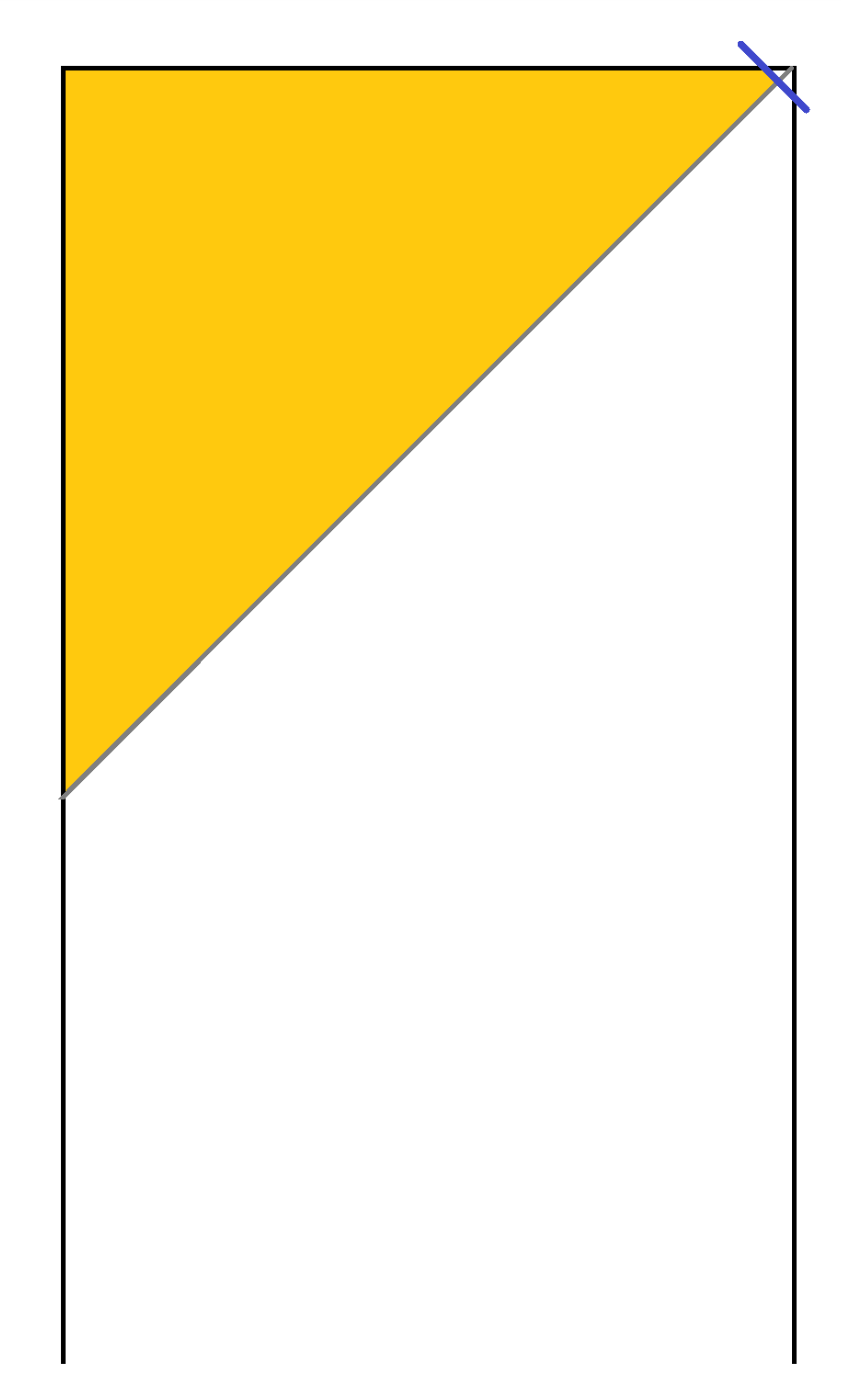}
\caption{}
\label{S2}
\end{center}
\end{figure}

Next, to compute the uncomplexity we must know the maximum possible complexity. That's given by the complexity evaluated at the time of the slash. The relevant spacetime volume is shown in figure \ref{S2}.

\begin{figure}[H]
\begin{center}
\includegraphics[scale=.2]{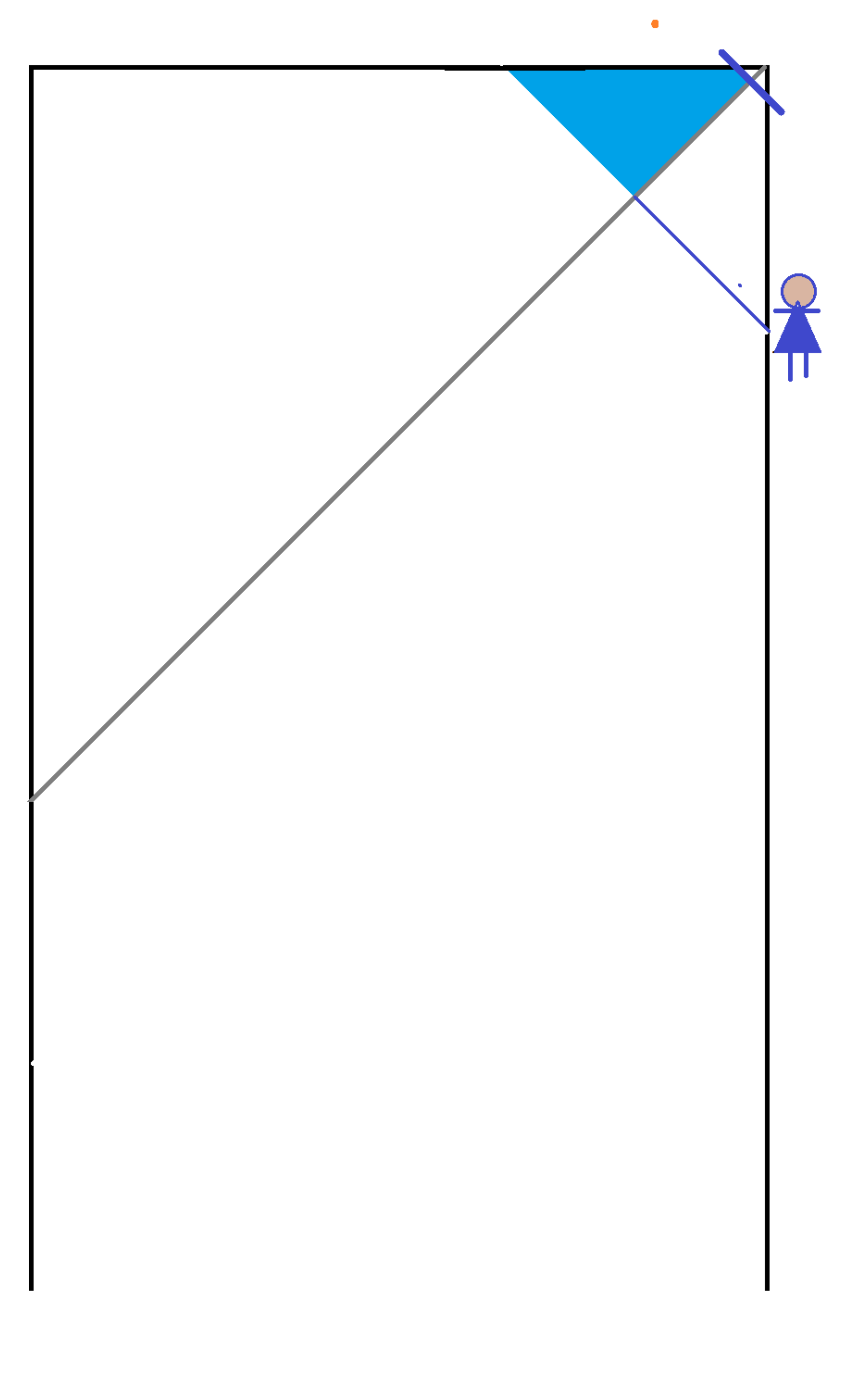}
\caption{}
\label{S3}
\end{center}
\end{figure}

The uncomplexity is of course the difference between the maximum complexity and the actual complexity at time $t$,
\be 
\rm Uncomplexity \it = \CC_{max} - \CC(t).
\ee
which is the spacetime volume shown in blue in figure \ref{S3}.

What is the meaning of the blue region in figure \ref{S3}?
It is the total available  spacetime  that Alice can visit if she jumps in after time $t$. Thus we find an elegant geometric meaning to  uncomplexity as a kind of spacetime resource available to  observers who jump into the black hole after time $t$.

\bn

Of course Alice can do a lot better if she happens to have a few thermal photons to spare. For each clean thermal photon that she throws in, the uncomplexity increases by an exponential amount. The mechanism was explained in Lecture II where we saw that a thermal photon can clear out any firewalls for an exponential time. 

\bn

Thus we see the remarkable power of one clean qubit in two ways, one involving computation and one involving the elimination of firewalls behind the horizon.

\part{Conclusion}

To me, the idea that computational complexity might  govern the interior geometries of black holes was startling. Complexity, especially quantum complexity, is such a subtle and finespun quantity that ordinarily it is impossible to measure or even to compute. What makes complexity  so tenuous is that it is defined as a global minimum, whereas in most physical  contexts  local minima are the relevant things. This is especially concerning when we consider the very long time behavior of a black hole.

Although I have not stressed it in these lectures, relative complexity can be thought of in geometric terms  (Dowling, Nielsen  \url{https://arxiv.org/pdf/quant-ph/0701004.pdf}) as the length of the shortest geodesic between two points $U$ and $V$, the geometry  being defined by a certain right-invariant complexity metric. The evolution of a system $U(t)$ sweeps out a geodesic, which for some period of time is the shortest geodesic connecting the identity to $U(t).$ Eventually a cut-locus \url{https://en.wikipedia.org/wiki/Cut_locus_(Riemannian_manifold)} will be reached, at which point a shorter geodesic will emerge. An example is a torus with incommensurate cycles. Starting at a  point that we may call $I$ we may move along a geodesic forever without coming back to the same point. The length of the geodesic, measured from $I$  will grow forever. But once we pass the cut-locus, a shorter geodesic will suddenly emerge. The original geodesic continues on, and is completely smooth at the cut-locus, but it is no longer the global minimum. We saw just this type of behavior  in figure \ref{cut} in Lecture I where shorter paths materialized when the complexity reached its maximal value.

That complexity governs the geometry of black hole interiors has sometimes been greeted with skepticism just because of the global character of complexity. The argument is often expressed by saying that there is no physical reason for the geometry to suddenly make a transition at the cut-locus since the Hamiltonian evolution of a black hole is completely smooth. To put it another way, physics is governed by local stationary points.

\begin{figure}[H]
\begin{center}
\includegraphics[scale=.4]{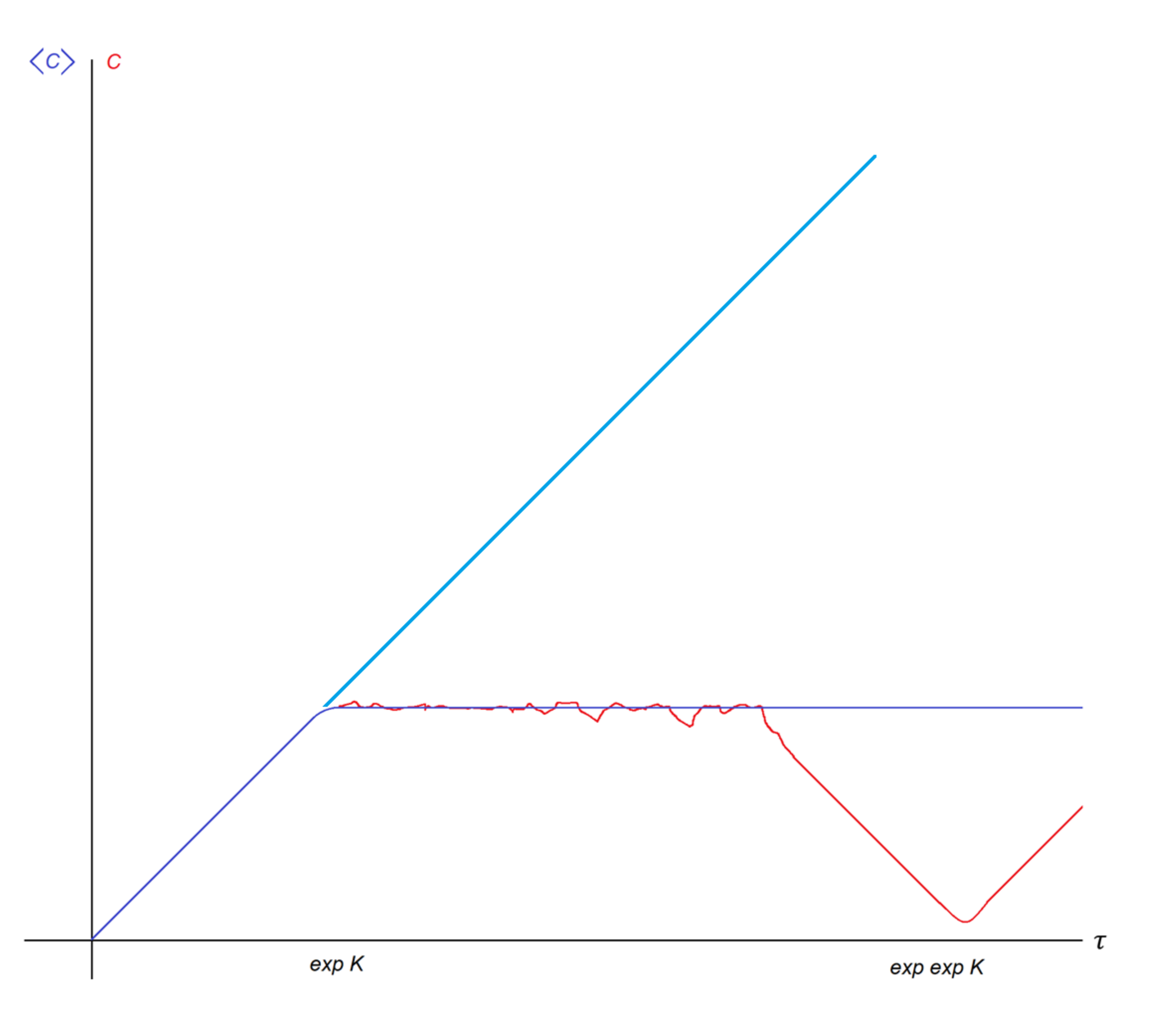}
\caption{}
\label{wrong}
\end{center}
\end{figure}

Consider figure \ref{2nd-law}. The transition from the linearly growing behavior to complexity equilibrium occurs sharply---as it would at a cut-locus---even though the Hamiltonian evolution of the system is perfectly smooth. If we suppose that the geometry of the wormhole is insensitive to such transitions,  the growth of the wormhole would continue forever. 
The light blue extrapolation in figure \ref{wrong} illustrates what the growth of a wormhole volume might look like in that case, while the complexity would follow the original curve in figure \ref{2nd-law}.

To see why this must be wrong, let's consider the extreme situation in which the complexity executes a quantum recurrence at the doubly exponential time $\exp \exp S.$  The obvious interpretation of the V-shaped portion of the red curve   is that a white-hole black-hole bounce has taken place, as in figure \ref{Foliated-BTZ2}. Down at the bottom of the V the system has returned to a state very close to the thermofield-double. The TFD is a state in which the wormhole length vanishes. It would not be hard to confirm that  the wormhole is short. A simple diagnostic would be the correlation function of fields at opposite ends of the wormhole; a short wormhole would indicate a large correlation; a very long wormhole would indicate a very small correlation. The fact that a quantum recurrence occurred implies that the correlation would be large.

Therefore we must conclude that somewhere between $t=0$ and $t=exp{ \ exp{ \ S}  \ }$ the derivative of the  wormhole length must have turned around. The natural place for that to happen is when the system reaches complexity equilibrium, i.e., at the cut locus.

I'll conclude these lectures with some philosophical remarks that I think apply not only to my own lectures but to the entire  PiTP 2018 series, especially to the lectures about entanglement, chaos,  SYK,  and complexity. They represent my own views and other lecturers may or may not agree with them.

\bn

For some time we have been moving toward the view that there are no independent laws of gravity, only laws of quantum mechanics applied to certain kinds of complex systems. The laws of gravity are emergent, and moreover they are not really laws; they are statistical tendencies. The trend goes all the ways back to the  laws of black holes as formulated by Bardeen, Carter, and Hawking in 1972. In particular the second law of black holes states that the total area of all horizons increases with time. This follows from the Einstein field equations which we usually think of as the absolute laws of gravity. But following Bekenstein, we now know that  the second law of black holes is nothing but the second law of thermodynamics: total entropy increases. However the second law is not an absolute law. It is a statistical tendency---a very strong tendency---but nonetheless a tendency.

Scrambling, chaos, and operator growth 
( D. Roberts, D. Stanford and A. Streicher,
 \url{https://arxiv.org/pdf/1802.02633.pdf} )
are very general phenomena seen in most  quantum systems with large number of degrees of freedom. It now seems that they are connected with the Newtonian law that the momentum of an object increases as it falls, in particular as it falls  toward a horizon ( \url{https://arxiv.org/pdf/1802.01198.pdf} 
  \url{https://arxiv.org/pdf/1804.04156.pdf}).
Scrambling and operator growth are phenomena that apply to all complex quantum systems, but again they are statistical tendencies. 

Finally, in these lectures I discussed the growth of spacetime behind horizons---another consequence of Einstein's equations. It has much in common with the growth of horizon area, but unlike that phenomenon it is not connected with the ordinary second law of thermodynamics. Instead it is a consequence of the second law of quantum  complexity which is  also a statistical tendency.

All quantum systems exhibit these tendencies to a greater or lesser extent, but only special quantum systems exhibit the  detailed kind of locality that we identify with Einstein gravity. In the context of AdS/CFT only special CFTs give rise to bulk locality on scales much smaller than the AdS radius. Those systems require very strongly coupled large-$N$ gauge theories.  Most quantum field theories cannot be extrapolated to large coupling without encountering  phase transitions or worse. Only supersymmetric QFTs have  analyticity properties that ensure against such breakdown at large coupling. Thus, to my knowledge, only super-theories have sub-AdS locality. That includes not only AdS/CFT but also BFSS  matrix theory   \url{https://arxiv.org/pdf/hep-th/9610043.pdf}. 

Of course the ``real world"  is not even approximately supersymmetric, yet it is very local on microscopic scales. This is a big puzzle that nothing in my lectures, or the other PiTP lectures address.

De Sitter space presents an even thornier puzzle. DS does not have the kind of time-like boundary that underpins gauge-gravity duality. The tools from AdS/CFT do not seem to apply in their currently understood form, possibly including the the connection between complexity and spacetime geometry. Nevertheless we live in a de Sitter world in which gravity is perfectly healthy.  
Evidently we are missing some  important things which this  PiTP program not touched on. Sooner of later we will need to get out of the comfort zone and move past AdS and supersymmetry.

\end{document}